\documentclass[aastex63,numberedappendix]{emulateapj}
\usepackage{rotating}
\usepackage{amsmath, color, natbib,graphicx,tablefootnote,subfigure}
\usepackage{graphics,subfigure,hyperref,natbib,float}
\usepackage[rightcaption]{sidecap}
\newcommand{\CH}[1]{\colhead{#1}}

\newcommand\ii{{\sc ii}}
\newcommand\iii{{\sc iii}}
\newcommand{\CM}{\checkmark}
\newcommand{\W}{$\lambda$}
\usepackage[dvipsnames]{xcolor}

\shorttitle{CHAOS IV}
\shortauthors{Berg et al.}
\defcitealias{berg15}{B15}
\defcitealias{croxall15}{C15}
\defcitealias{croxall16}{C16}

\begin{document}  
\title{CHAOS IV: Gas-Phase Abundance Trends from the First Four CHAOS Galaxies}
\author{Danielle A.\ Berg$^{1,2}$, 
            Richard W.\ Pogge$^{1,2}$,  
            Evan D.\ Skillman$^{3}$, 
            Kevin V.\ Croxall$^{4}$, 
            John Moustakas$^{5}$,
            Noah S.J. Rogers$^{3}$,
            Jiayi Sun$^{1}$}
            
\affil{   $^1$Department of Astronomy, The Ohio State University, 140 W 18th Ave., Columbus, OH, 43210; berg.249@osu.edu \\
           $^2$Center for Cosmology \& AstroParticle Physics, The Ohio State University, 191 West Woodruff Avenue, Columbus, OH 43210 \\
           $^3$Minnesota Institute for Astrophysics, University of Minnesota, 116 Church St. SE, Minneapolis, MN 55455 \\        
           $^4$Expeed Software, 100 W Old Wilson Bridge Rd Suite 216, Worthington, OH 43085 \\
           $^5$Department of Physics \& Astronomy, Siena College, 515 Loudon Road, Loudonville, NY 12211 }

\email{berg.249@osu.edu}


\begin{abstract}
The chemical abundances of spiral galaxies, as probed by \ion{H}{2} regions across their disks, 
are key to understanding the evolution of galaxies over a wide range of environments. 
We present LBT/MODS spectra of 52 \ion{H}{2} regions in NGC3184 as part of the 
CHemical Abundances Of Spirals (CHAOS) project. 
We explore the direct-method gas-phase abundance trends for the first four CHAOS galaxies, using 
temperature measurements from one or more auroral line detections in 190 individual \ion{H}{2} regions. 
We find the dispersion in $T_e-T_e$ relationships is dependent on ionization, 
as characterized by $F_{\lambda5007}/F_{\lambda3727}$, and so recommend ionization-based 
temperature priorities for abundance calculations. 
We confirm our previous results that [\ion{N}{2}] and [\ion{S}{3}] provide the most robust 
measures of electron temperature in low-ionization zones, while [\ion{O}{3}] provides reliable 
electron temperatures in high-ionization nebula. 
We measure relative and absolute abundances for O, N, S, Ar, and Ne. 
The four CHAOS galaxies marginally conform with a universal O/H gradient, 
as found by empirical IFU studies when plotted relative to effective radius.
However, after adjusting for vertical offsets, we find a tight universal N/O gradient of 
$\alpha_{\rm N/O}=-0.33$ dex/$R_e$ with $\sigma_{tot.} = 0.08$ for $R_g/R_e < 2.0$, 
where N is dominated by secondary production. 
Despite this tight universal N/O gradient, the scatter in the N/O--O/H relationship is significant.
Interestingly, the scatter is similar when N/O is plotted relative to O/H or S/H. 
The observable ionic states of S probe lower ionization and excitation energies than O, which might 
be more appropriate for characterizing abundances in metal-rich \ion{H}{2} regions.
\end{abstract}

\keywords{galaxies: abundances - galaxies: spiral - galaxies: evolution - galaxies: individual (NGC~3184) - galaxies: ISM - ISM: lines and bands}


\section{INTRODUCTION}\label{sec:intro}

The history of a galaxy can be traced by the abundances of heavy elements,
as they are produced and accumulated as successive generations of stars 
return their newly synthesized elements to the interstellar medium (ISM). 
In spiral galaxies, ISM abundance studies are dominated by the disk, where the
majority of their star formation occurs, and are typically characterized 
by negative radial gradients of oxygen and nitrogen abundances \citep[e.g.,][]{pagel81,garnett87,zaritsky94}.
The abundance gradients across the disks of spiral galaxies provide essential 
observational constraints for chemical evolution models of galaxies, 
and support the inside-out growth theory of galaxy disk formation.

Emission lines originating from \ion{H}{2} regions provide an excellent probe of 
the gas-phase abundances and, thus, the radial metallicity gradients in disk galaxies.
Further, \ion{H}{2} regions, which are ionized by recently-formed massive stars that carry the 
same chemical signature from the gas in which they were formed,
allow us to measure the cumulative chemical evolution of the present-day ISM. 

Galaxy surveys conducted with integral field unit (IFU) spectrographs are 
spatially resolving large numbers of low redshift galaxies
\citep[e.g.,][]{sanchez12,bryant15,bundy15}
and intermediate-redshift galaxies are being targeted using ground-based infrared spectrographs
\citep[e.g., $z\sim2-3$ lensed or stacked galaxies;][]{erb10,shapley15,steidel14,rigby15,berg18}.
In the future, these studies will enable us to answer important questions that impact our 
understanding of galaxy formation and evolution, such as the importance of metallicity gradients
over cosmic time, the magnitude of azimuthal variations, and integrated light versus resolved studies.
However, presently, most of these studies must use abundance correlations with strong emission-lines
to interpret their data (strong-line methods), and so are inherently limited
by the large uncertainties associated with the calibrations of these methods 
\citep[up to 0.7 dex in absolute abundance;][]{kewley08,moustakas10}.
Until we can truly understand the abundances of the local spiral galaxies and improve 
our calibration toolset, we cannot be completely confident in our measures from IFU 
studies or of the chemical evolution of galaxies at high redshift.

Many studies have used multi-object spectroscopy to attempt to directly 
measure the nebular physical conditions and abundances and map out their 
trends across the disks of spiral galaxies.
However, because direct measurements of gas-phase abundances via one of the ``direct" methods
(i.e., auroral or recombination lines) have long been prohibitively expensive 
in terms of telescope time, the majority of these studies are limited to 
first order trends using a dozen or fewer abundance detections per galaxy.
This challenge motivated the CHemical Abundances Of Spirals 
\citep[CHAOS;][]{berg15} project: a large database of high quality \ion{H}{2} region 
spectra over a large range in abundances and physical conditions in nearby spiral galaxies. 
These spectra provide direct abundances, estimates of temperature stratification and their 
corresponding corrections to lower absolute abundances, and allow calibrations based on 
observed abundances over expanded parameter space rather than photoionization models. 

While the absolute abundance scale of \ion{H}{2} regions is still a topic of debate
\citep[see, for example, the discussion of the Abundance Discrepancy Factor in][]{bresolin16},
the CHAOS survey is building a large sample of direct abundances, observed and analyzed uniformly,
allowing us to characterize the possible systematics of the direct method.
To date, CHAOS has increased, by more than an order-of-magnitude, the number of \ion{H}{2} 
regions with high-quality spectrophotometry to facilitate the first detailed 
direct measurements of the chemical abundances in a sample of nearby disk galaxies.
So far, results for individual galaxies have been reported for 
NGC~628 (M74) in \citet[][hereafter, B15]{berg15},
NGC~5194 (M51a) in \citet[][hereafter, C15]{croxall15}, and 
NGC~5457 (M101) in \citet[][hereafter, C16]{croxall16}.
Here we present new direct abundances for NGC~3184 and, combined with past results,
present the first analyses of a sample of four CHAOS galaxies, totaling 190 \ion{H}{2} 
regions with measured auroral line based temperatures.

The paper is organized as follows.
In Section~2 we briefly review the CHAOS data, including the spectroscopic observations 
(\S~2.1), reductions (\S~2.2), and emission line measurements (\S~2.3).
Section~3 details the nebular electron temperature and density measurements,
recommended ionization-based temperature priorities, 
as well as the abundance determinations.
Radial abundance trends for the first four CHAOS galaxies are reported in Section~4,
beginning with radial O/H and S/H abundances in \S~4.1 and \S~4.2, respectively.
In \S~4.3 we propose a universal secondary N/O gradient.
We discuss secondary drivers of the observed abundance trends in Section~5, namely
azimuthal variations (\S~5.1), surface density relationships (\S~5.2), and
effective yields (\S~5.3).
Section~6 examines abundance trends with metallicity for the CHAOS sample, where
$\alpha$/O and N/O trends are discussed in \S~6.1 and \S~6.2, respectively.
Finally, we focus on N/O trends in Section~7.
We discuss the production of N/O in spiral galaxies in \S~7.1 and consider sources 
of scatter in the N/O--O/H relationship in \S~7.2.
A summary of our results is provided in Section~8.


\begin{deluxetable*}{lcccc}
\tablewidth{0pt}
\tablecaption{Adopted Properties of CHAOS Galaxies}
\tablehead{
\CH{Property} 				& \CH{NGC~628}	& \CH{NGC~5194}	& \CH{NGC~5457}	& \CH{NGC~3184}}
\startdata
{R.A.}					& {01:36:41.75} 	& {13:29:52.71} 	& {14:03:12.5}		& {10:18:16.86}		\\ 
{Decl.}					& {15:47:01.18}		& {47:11:42.62}		& {54:20:56}		& {41:25:26.59}		\\ 
{Type}					& {SA(s)c}			& {SA(s)bc pec}	& {SAB(rs)cd}		& {SAB(rs)cd}		\\ 
{Redshift}					& {0.00219}		& {0.00154}		& {0.00080}		& {0.00198}		\\ 
{Adopted D (Mpc)}			& {$7.2^1$}	    	& {$7.9^2$}	    	& {$7.4^3$}		& {11.7}$^4$ 		\\ 
{Inclination (deg.)}			& {5$^5$}			& {$22^{6}$}		& {$18^{7}$}		& {16}$^{8}$		\\ 
{P.A. (deg.)}				& {12$^{9}$}		& {$172^{7}$}		& {$39^{7}$}		& {179}$^{8}$		\\ 
{${m_B}$ (mag)}			& {10.01}	 		& {9.08}			& {7.99}			& {10.44}	    		\\ 
{log $M_\star$ ($M_\odot$)} 	& 10.0			& 10.5			& 10.4			& 10.2	     		\\ 
{$v_{\rm flat}$ (km s$^{-1}$)}	& 200 			& 210			& 210           		& 200	        		\\ 
{$R_{25}$ (arcsec)}			& {315.0$^{9}$}		& {$336.6^{9}$}		& {$864.0^{10}$}	& {222.0$^{9}$} \smallskip \\ 
\multicolumn{5}{l}{CHAOS-Derived Properties:}\\	
{$R_e$ (arcsec)}			& {95.4}			& {94.7}			& {197.6}			& {93.2}        		\\ 
{$R_g$ Coverage ($R_e$)}	& 2.3  			& 3.4     			& 4.6    			& 2.0      			\\ 
{$T_e$ Regions$^a$} 		& {$45^{11}$}   		& {28$^{12}$}  		& {72$^{13}$}  		& {30$^{14}$} 		   
\enddata
\tablecomments{
Adopted properties for the current sample of CHAOS galaxies:
NGC~628, NGC~5194, NGC~5457, and NGC~3184.
Rows 1 and 2 give the RA and Dec of the optical center in units of 
hours, minutes, seconds, and degrees, arcminutes, arcseconds respectively.
The RAs, Decls, galaxy type (Row 3) and redshifts (Row 4) are taken from 
the NASA/IPAC Extragalactic Database (NED). 
Adopted distances, inclinations, and position angles are given in Rows 5--7.
Rows 8--10 list B-band magnitude \citep{devaucouleurs91}, stellar mass, and $v_{\rm flat}$ of each galaxy.
Stellar masses were determined using the integrated 3.6 $\mu$m flux in \citet{dale09} and 
rotation speed is adopted from the simple flat rotation curve reported in \citet{leroy13}.
Rows 11 and 12 give the optical radius at the $B_{25}$ mag arcsec$^{-2}$ and 
the half-light radius, as determined in this work (see Appendix~\ref{sec:A1} for details), 
of the system in arcseconds, respectively.
Row 13 provides the radial coverage of the CHAOS observations in units of $R_e$.
Finally, the number of \ion{H}{2} regions with direct auroral-line temperature 
measurements from [\ion{O}{3}], [\ion{N}{2}], or [\ion{S}{3}] are tabulated in Row 14.\\
References: 
(1) \citet{vandyk06}; (2) \citet{baron07}; (3) \citet{ferrarese00}; (4) \citet{bose14}; 
(5) \citet{shostak84}; (6) \citet{colombo14}; (7) \citet{walter08}; 
(8) \citet{jimenez-donaire17}; (9) \citet{egusa09}; (10) \citet{kennicutt03b}; 
(11) \citetalias{berg15}; (12) \citetalias{croxall15}; (13) \citetalias{croxall16}; (14) this work. \\
$^a$Only regions with $T_e$[\ion{O}{3}], $T_e$[\ion{S}{3}], or $T_e$[\ion{N}{2}] are tallied here. }
\label{tbl1}
\end{deluxetable*}


\section{NEW CHAOS SPECTROSCOPIC OBSERVATIONS OF NGC~3184}\label{data}

\subsection{Optical Spectroscopy}
All CHAOS observations are obtained following a consistent methodology,
but here we highlight details specific to new observations of NGC~3184.
Optical spectra of NGC~3184 were obtained during March 2012 and January 2013
using the Multi-Object Double Spectrographs 
\citep[MODS,][]{pogge10} on the Large Binocular Telescope (LBT). 
The spectra were acquired with the MODS1 unit as the MODS2 spectrograph 
was not available at the time of the observations.  
We obtained simultaneous blue and red spectra using the G400L (400 lines mm$^{-1}$, R$\approx$1850) 
and G670L (250 lines mm$^{-1}$, R$\approx$2300) gratings, respectively.  
This setup provided broad spectral coverage extending from 3200 -- 10,000 \AA.  
Multiple fields were targeted in order to maximize the number of \ion{H}{2} regions with auroral 
line detections, i.e., [\ion{S}{2}] \W\W4068,4076, [\ion{O}{3}] \W4363, [\ion{N}{2}] \W5755,  
[\ion{S}{3}] \W6312, and [\ion{O}{2}] \W\W7320,7330.
Individual field masks, cut to target 17--25 \ion{H}{2} regions simultaneously, 
were observed for six exposures of 1200s, or a total integration time of 2-hours per field.

\begin{figure*}
\begin{center}
\includegraphics[scale=0.35, trim=0mm 20mm 0mm 20mm, clip]{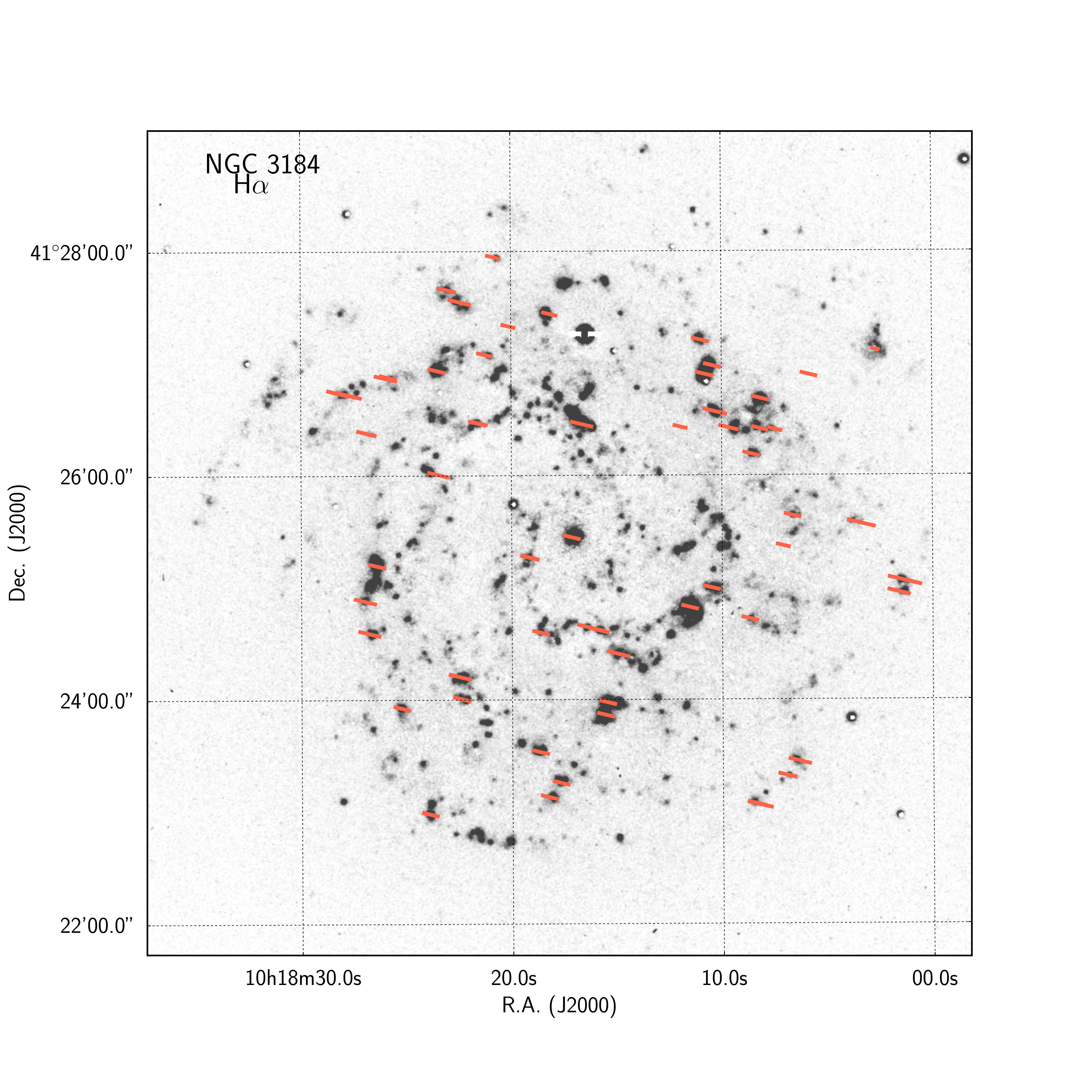}
\caption{Continuum-subtracted H$\alpha$ SINGS image of NGC~3184 \citep{munoz-mateos09}.
The footprints of CHAOS slits are overlaid in light red, representing the 
Field 1, 2, and 3 slit positions observed at the LBT. 
The slit positions targeted \ion{H}{2} regions, although not always centered in order to maximize
effective usage of mask real estate and obtain background within the slit. 
See Table~\ref{tbl2} for more details.}
\label{fig1}
\end{center}
\end{figure*}

Targeted \ion{H}{2} regions in NGC~3184, as well as 
alignment stars, were selected based on archival broad-band and H$\alpha$ imaging 
from the SINGS program \citep{kennicutt03b,munoz-mateos09}.  
Slits were cut to be 1\arcsec\ wide by a minimum of $\sim$10\arcsec\ long, to cover the extent 
of individual \ion{H}{2} regions, and extended to utilize extra space for sky.  
Slits were placed on relatively bright \ion{H}{2} regions across the entirety of the disk 
with the goal of  ensuring that both radial and azimuthal trends in the abundances could be investigated.
The locations of the slits for each of the three MODS fields observed in NGC~3184
are shown in Figure~\ref{fig1}.

We refer to the locations of the observed \ion{H}{2} regions in NGC~3184 as offsets, in 
right ascension and declination, from the center of the galaxy (see Table~\ref{tbl3} in Appendix~\ref{sec:A1}).  
The observations were obtained at relatively low airmass ($\lesssim1.2$).
Furthermore, slits were cut close to the median parallactic angle of the observing window 
for NGC~3184. 
The combination of low airmass and matching the parallactic angle minimizes flux lost due to 
differential atmospheric refraction between 3200 -- 10,000 \AA\ \citep{filippenko82}.

We report the new observations of NGC~3184 in Appendix~\ref{sec:A1}, while
details of previously reported observations can be found in 
\citetalias{berg15} for NGC~628, 
\citetalias{croxall15} for NGC~5194, and 
\citetalias{croxall16} for NGC~5457.
The adopted properties of these four galaxies are listed in Table~\ref{tbl1}.
Note that for NGC~628, NGC~5194, and NGC~5457 we report properties of these galaxies 
as adopted by the original CHAOS studies.
It may be of interest to some readers that since the time of the previous CHAOS studies,
updated (and likely more accurate) distances have been measured for NGC~628 and NGC~5194 by 
\citet{mcquinn17} and for NGC~5457 by \citet{jang17} using the tip of the red giant branch
method.
While many absolute properties change with galaxy distance, the results presented here are 
concerned only with relative abundance trends versus $R_e$ or $R_{25}$, and so are not affected 
by the updated distances.


\subsection{Spectral Reductions}\label{sec:reduct}

\begin{figure*}
\centering
	\includegraphics[scale = 1.85, trim = 5mm 0mm 5mm 0mm]{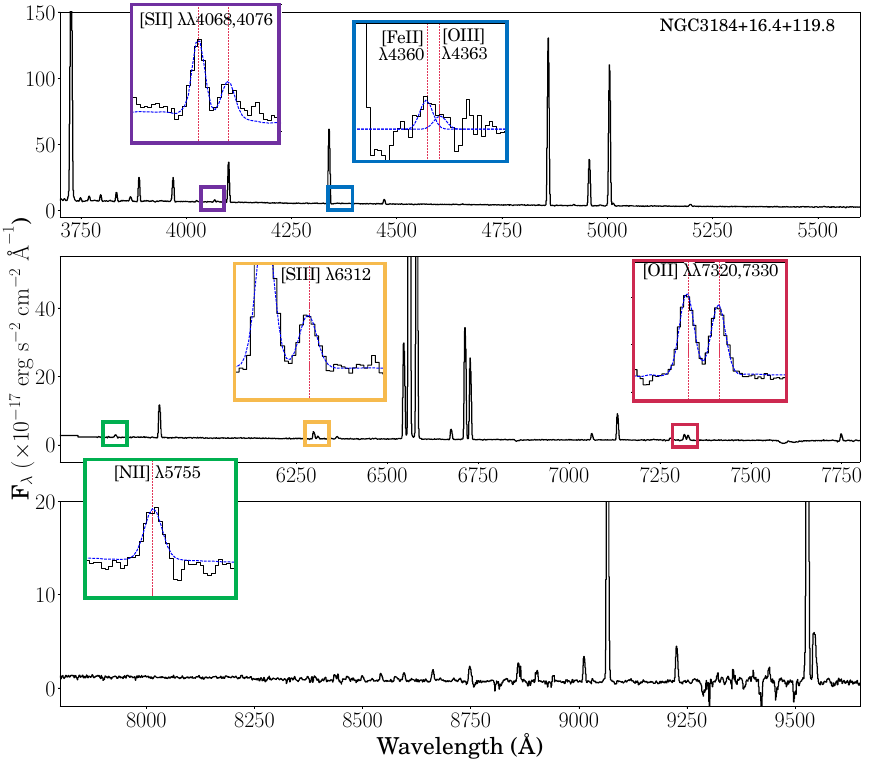}   
    \caption{Demonstration of a one dimensional spectrum taken with MODS1/LBT of region $+16.4+119.8$ in 
NGC~3184 with auroral line detections at a strength of 3$\sigma$ or greater.
The observed spectrum is plotted as a black line, with the model in blue. 
In the expanded windows, we mark and label the
five temperature-sensitive auroral emission line features used in this paper:
[S~\ii] $\lambda\lambda4068,4076$
[O~\iii] $\lambda4636$, 
[N~\ii] $\lambda5755$, 
[S~\iii] $\lambda6312$, and 
[O~\ii] $\lambda\lambda7320,7330$.
This spectrum lacks an [\ion{O}{3}] \W4363 detection as the majority of the emission in that region is actually due to a contaminating [\ion{Fe}{2}] line at \W4360 (see blue box).
Note that major telluric absorption features are not corrected for (see bottom panel). 
\vspace{18ex}}
\label{fig2}
\end{figure*}

For a detailed description of the data reduction procedures we refer the reader to \citepalias{berg15}.  
Here, we only note the primary points of our data processing.  
Spectra were reduced and analyzed using the beta-version of the MODS reduction pipeline
\footnote{\url{http://www.astronomy.ohio-state.edu/MODS/Software/modsIDL/}} which runs within the XIDL
\footnote{\url{http://www.ucolick.org/~xavier/IDL/}} reduction package.  
Given that the bright disks of CHAOS galaxies can complicate local sky subtraction, 
additional sky slits were cut in each mask that provided a basis for clean sky subtraction.
Continuum subtraction was performed in each slit by scaling the continuum flux from the sky-slit
to the local background continuum level.
One-dimensional spectra were then corrected for atmospheric extinction and flux calibrated based on 
observations of flux standard stars \citep{bohlin14}. 
At least one flux standard was observed on each night science data were obtained. 
An example of a flux-calibrated spectrum is shown in Figure \ref{fig2}. 


\subsection{Emission Line Measurements}\label{sec:iraf}
We provide a more detailed description of the adopted continuum modeling and line fitting 
procedures applied to the CHAOS observations in \citetalias{berg15}.  
Below, we only highlight the fundamental components of this process.  
We model the underlying continuum of our MODS1 spectra using the STARLIGHT\footnote{\url{www.starlight.ufsc.br}} 
spectral synthesis code \citep{fernandes05} in conjunction with the models of \citet{bruzual03}. 
Allowing for an additional nebular continuum, we fit each emission line with a Gaussian profile.  
We note that we have modeled blended lines (H7, H8, and H11 -- H14) in the Balmer series based 
on the measurements of unblended Balmer lines and the tabulated atomic ratios of \citet{hummer87},
assuming Case B recombination.  

We correct the strength of emission features for line-of-sight reddening using the relative 
intensities of the four strongest Balmer lines (H$\alpha$/H$\beta$, H$\gamma$/H$\beta$,
H$\delta$/H$\beta$).  
We report the determined values of E(B--V) in 
Table~4 of Appendix~\ref{sec:A1}.\footnote{We note that previous CHAOS papers also report
the E(B--V) reddening, but had incorrectly labeled this
quantity as c(H$\beta$).
}
We do not apply an ad-hoc correction to account for Balmer absorption as the lines 
were fit simultaneously with the stellar population models.  
The stellar models contain stellar absorption with an equivalent 
width of $\approx$1\,--\,2 \AA\ in the H$\beta$ line.
The uncertainty associated with each measurement is determined from measurements of the 
spectral variance, extracted from the two-dimensional variance image, uncertainty 
associated with the flux calibration, 
Poisson noise in the continuum, read noise, sky noise, flat fielding calibration error, 
error in continuum placement, and error in the determination of the reddening.  
We also include a 2\% uncertainty based on the precision of the adopted flux calibration 
standards \citep[][see discussion in Berg et al. 2015]{oke90}.

A few emission features required extra care, such as the intrinsically faint auroral lines
that are critical to this study.
As has been done with the previous CHAOS galaxies, we inspected the lines by-eye and 
measured the flux of each auroral line by-hand in the extracted spectra to confirm the fit.  
In cases where these measurements were in disagreement, 
we adopted the by-hand measurement.  
This was most common for the [\ion{N}{2}] \W5755 line which falls near the 
wavelength region affected by the dichroic cutoff of MODS and the ``red bump"
Wolf-Rayet carbon features.
Additionally, we have updated our line fitting code to include the [\ion{Fe}{2}] \W4360
emission feature, which may significantly contaminate [\ion{O}{3}] \W4363 line
measurements at high metallicities \citep[12+log(O/H) $>$ 8.4;][]{curti17}.

Finally, the [\ion{O}{2}]~\W\W3726,3729 doublet is blended for all observations
due to the moderate resolution of MODS.
However, two components are apparent in the doublet profile for the majority of 
spectra, and are therefore modeled using two Gaussian profiles.
The reported [\ion{O}{2}] \W3727 fluxes represent the total flux in the doublet.

The reddening-corrected emission line intensities measured from \ion{H}{2} regions in 
NGC~628, NGC~5194, and NGC~5457 have been previously reported in \citetalias{berg15},
\citetalias{croxall15}, and \citetalias{croxall16}, respectively.
For the NGC~3184 observations reported here, the reddening-corrected line intensities
are listed in Table~4 of Appendix~\ref{sec:A1}.


\begin{figure*}
\begin{center}
	\includegraphics[scale = 0.75, trim = 0mm 0mm 0mm 0mm, clip]{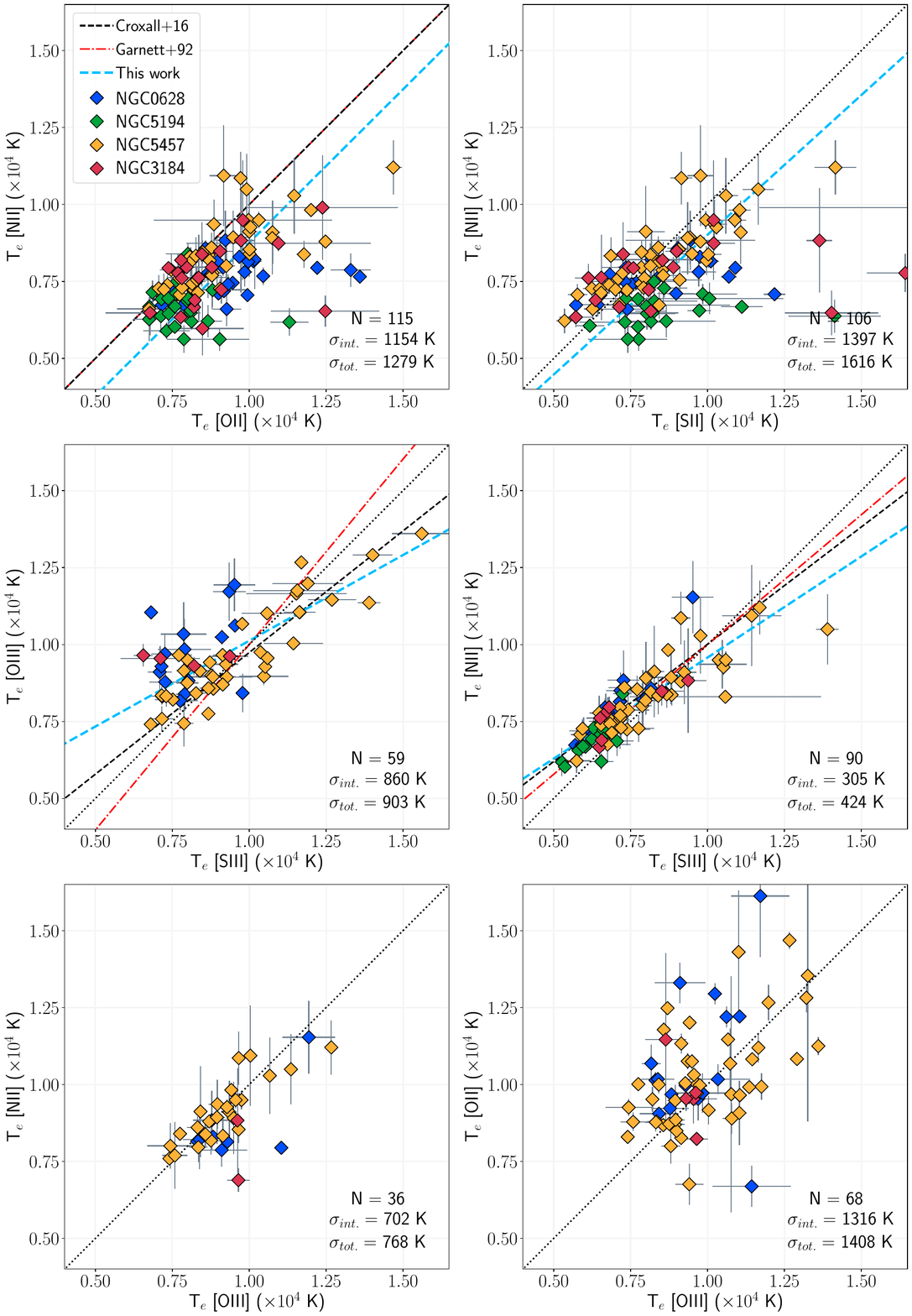}
\caption{Comparing temperature relationships for different ions from all four CHAOS galaxies.
  The black dotted lines assume a one-to-one relationship, red dot-dashed lines are the
  photoionization model relationships from \citet{garnett92}, black dashed lines are the 
  updated empirical relationships from \citetalias{croxall16}, and blue dashed lines are 
  the best linear fits to the data from all four CHAOS galaxies.
  The top panels compare measures of the low-ionization zone temperatures. 
  {\it Top left}: $T_e$[\ion{N}{2}] versus $T_e$[\ion{O}{2}], showing large scatter, and
  {\it top right}: $T_e$[\ion{N}{2}] versus $T_e$[\ion{S}{2}], offset significantly from one-to-one.
  The middle panels compare to the intermediate-ionization temperature, $T_e$[\ion{S}{3}],
  revealing the scattered $T_e$[\ion{O}{3}] versus $T_e$[\ion{S}{3}] trend ({\it left})
  and the tight correlation between T[\ion{N}{2}] and T[\ion{S}{3}] ({\it right}).
  The bottom panels show further comparisons to the high-ionization temperature, $T_e$[\ion{O}{3}].
  The T[\ion{N}{2}] versus T[\ion{O}{3}] trend ({\it left}) is relatively well behaved,
  but has few points, whereas T[\ion{O}{2}] versus T[\ion{O}{3}] ({\it right}) is a scatter plot.
  We adopt the \citetalias{croxall16} relationships, given in Equations~$1-3$, for this work.}
\label{fig3}
\end{center}
\end{figure*}


\section{Direct Gas-Phase Abundances}

\subsection{Electron Temperature and Density Determinations}\label{sec:temden}
The combined sensitivity and large wavelength coverage of CHAOS observations allows 
electron temperature and density measurements from multiple ions.
The temperature-sensitive auroral-to-nebular line ratios most commonly observed 
in the CHAOS spectra are
[\ion{S}{2}] \W\W4068,4076/\W\W6717,6731; 
[\ion{O}{3}] \W4363/\W\W4959,5007;
[\ion{N}{2}] \W5755/\W\W6548,6584;
[\ion{S}{3}] \W6312/\W9069,9532; and
[\ion{O}{2}] \W\W7320,7330/\W\W3727,3729.
To account for possible contamination by atmospheric absorption of the red [\ion{S}{3}] lines,
we follow our practice in \citetalias{berg15} of upward correcting the weaker 
of the two lines by the theoretical ratio of \W9532/\W9069 = 2.47.
Assuming a three-zone ionization structure, these measurements probe 
the physical conditions throughout the nebula, and allow for the comparison 
of multiple measures in the low-ionization zone.
We use the ratio of the [\ion{S}{2}] \W\W6717,6731 emission lines as a 
sensitive probe of the nebular electron density in typical \ion{H}{2} 
regions ($10^{1.5}<n_e ({\rm cm}^{-3}) <10^{3.5}$).
In order to compare the first four CHAOS galaxies in a uniform, consistent manner,
we recalculate the nebular temperatures and densities 
adopting the atomic data reported in Table~4 of \citetalias{berg15} 
and using the observed temperature- and density-sensitive line ratios 
with the P{\sc y}N{\sc eb} package in {\sc python} 
\citep{luridiana12, luridiana15}.

\subsubsection{Temperature Relationships}
It is common practice to use temperature-temperature ($T_e - T_e$) 
relationships derived from photoionization models to infer the temperatures 
in unobserved ionization zones.
The relationships of \citet[][hereafter, G92]{garnett92} are a typical choice; 
however, significant updates in atomic data 
(especially for [\ion{S}{3}] and [\ion{O}{2}]; see Figure~4 in \citetalias{berg15})
have occurred since the time of that work and so new relationships are warranted.

In \citetalias{croxall16}, we obtained temperature measurements from one or more
auroral lines in 74 \ion{H}{2} regions in M101, the largest number in a single galaxy
to date.
These data used the updated atomic data recommended in \citetalias{berg15}, and
provided a large dataset of well measured temperatures from multiple ions 
that allowed us to empirically determine new $T_e - T_e$ relationships:
\begin{small}
\begin{eqnarray}
	T_e{[\rm{NII}]} = (0.714\pm0.142)\times T_e{[\rm{OIII}]} + (2.57\pm1.25),\\ \label{eqn1}
  	T_e{[\rm{SIII}]} = (1.312\pm0.075)\times T_e{[\rm{NII}]} - (3.13\pm0.58), \\ \label{eqn2}
	T_e{[\rm{SIII}]} = (1.265\pm0.140)\times T_e{[\rm{OIII}]} - (2.32\pm1.35), \label{eqn3}
\end{eqnarray}
\end{small}
where temperatures are in units of $10^4$ K.

Using the combined data from the first four CHAOS galaxies, we compile a 
sample of 190 individual \ion{H}{2} regions with multiple auroral line measurements.
Of these regions, 175 have $T_e$[\ion{O}{3}], $T_e$[\ion{S}{3}], or $T_e$[\ion{N}{2}].
In Figure~\ref{fig3} we compare these data to the $T_e - T_e$ relationships 
of G92 (red dot-dashed lines) and \citetalias{croxall16} (black dashed lines).
For reference, the line of equality is shown as a dotted black line.
We recognize that these are simple $T_e-T_e$ relationships; in the future we will
use the full CHAOS dataset to explore more complicated $T_e-T_e$ relationships,
for example, accounting for the effects of ionization discussed below.

For each set of variables, we determine the best fit $T_e - T_e$ relationship
using a Bayesian linear regression.
Specifically, we use the code python {\sc linmix}\footnote{https://github.com/jmeyers314/linmix}, 
which is an implementation of the linear mixture model algorithm developed 
by \citet{kelly07} to fit data with uncertainties on two variables, 
including explicit treatment of intrinsic scatter.
Intrinsic scatter, $\sigma_{int.}$, is due to real deviations in the physical 
properties of our sources that are not completely captured by the variables considered.
By introducing an additional term representing the intrinsic scatter to the weighting of 
each data point in the fit, we can determine the median of the normally-distributed 
intrinsic random scatter about the regression.
The calculated total and intrinsic scatters, $\sigma_{tot.}$ and $\sigma_{int.}$ 
respectively, as well as the number of regions used in the fit, 
are presented in Figure~\ref{fig3}.

The top two panels of Figure~\ref{fig3} compare temperature measurements 
that characterize the low-ionization zone.
On the left, we use the 115 regions with both [\ion{N}{2}] and [\ion{O}{2}] 
measurements in our sample, and find a best fit of
$T_e$[\ion{N}{2}]$= [T_e$[\ion{O}{2}]$ - (1.203\pm1.144)]/(1.004\pm0.150)$.
As expected, the overall trend follows a one-to-one relationship within the 
limits of the uncertainties, but with both large total ($\sigma_{tot.} = 1280$ K)
and intrinsic ($\sigma_{int.} = 1150$ K) scatters.
While equal temperatures are expected from photoionization models,
the data tend to be shifted toward higher $T_e$[\ion{O}{2}].
This is true for the majority of the sample, which is clustered within
$1000-2000$ K of the equality relationship, but especially for the 
more extreme outliers that offset up to roughly 5000 K.

We note that dielectronic recombination can contribute to the observed [\ion{O}{2}] 
emission, especially \W\W7320,7330, in more metal-rich nebulae \citep[e.g.,][]{rubin86}.
The magnitude of the effect increases strongly with decreasing temperature 
(increasing metallicity) but depends on the electron density.
To this end, \citet{liu01} showed that recombination can play an important role 
in exciting both the [\ion{O}{2}] \W\W7320,7330 and [\ion{N}{2}] \W5754 auroral 
lines in the higher-density gas of planetary nebulae ($>10^3$ cm$^{-3}$).
These authors showed that this effect leads to overestimated [\ion{O}{2}]-
and [\ion{N}{2}]-derived electron temperature measurements.
However, we show below that $T_e$[\ion{N}{2}] is well behaved with respect to
$T_e$[\ion{S}{3}] which implies that the recombination
contribution must be small at the low densities of our nebulae.
Thus our data are consistent with previous reports of systematically 
larger $T_e$[\ion{O}{2}] than $T_e$[\ion{N}{2}] measurements 
\citep[e.g.,][]{esteban09,pilyugin09,berg15} that cannot be accounted 
for by recombination processes, and so we do not favor 
[\ion{O}{2}] as a reliable low-ionization zone temperature indicator.
We reserve further analysis for the complete CHAOS sample,
where we will revisit the reliability of [\ion{O}{2}] as a diagnostic and 
investigate the effects of sky contamination, recombination, and reddening.

In the top right panel of Figure~\ref{fig3}, we compare 
[\ion{N}{2}] and [\ion{S}{2}] using the [\ion{S}{2}]
temperatures presented in \citetalias{croxall16}, plus newly derived
values for NGC~628, NGC~5194, and NGC~3184, 
comprising a sample of 106 regions.
As expected for two ions that probe similar low-ionization gas, 
the best fit is consistent with equality as
$T_e$[\ion{N}{2}]$=[T_e$[\ion{S}{2}]$ - (0.072\pm1.392)]/(1.101\pm0.180)$.
Again, the intrinsic scatter accounts for the majority of the total scatter;
however, the large deviations observed indicate that observational uncertainties 
still play a large role at high [\ion{S}{2}] temperatures.


\begin{SCfigure*}
	\includegraphics[scale = 0.3125, trim = 5mm 200mm 5mm 200mm, clip]{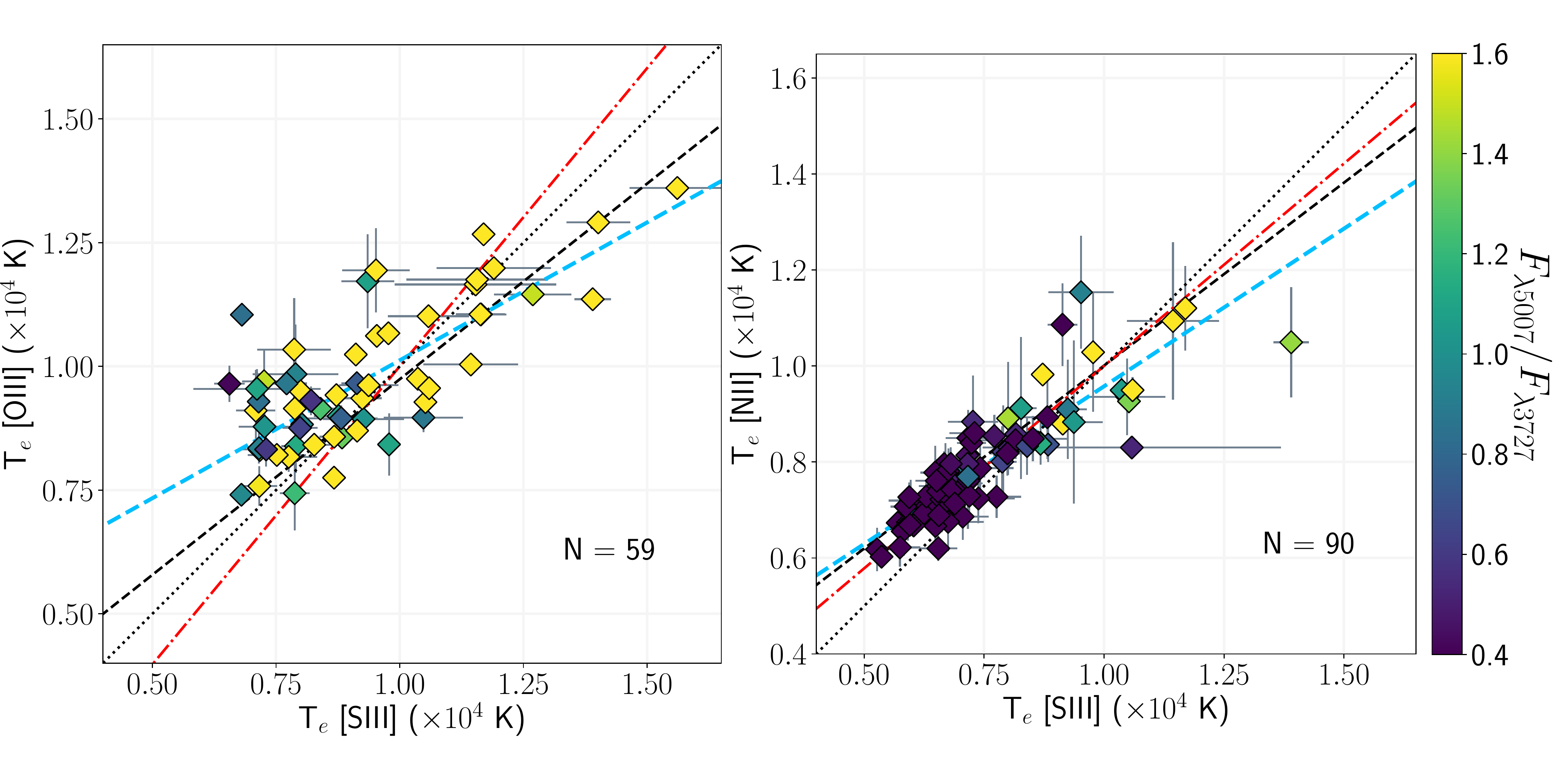}
\caption{
$T_e$[\ion{O}{3}] versus $T_e$[\ion{S}{3}] ({\it left})
and $T_e$[\ion{N}{2}] versus $T_e$[\ion{S}{3}] ({\it right}) for all four CHAOS galaxies
color-coded by the reddening-corrected [\ion{O}{3}] \W5007/[\ion{O}{2}] \W3727 flux ratios.
The tight correlation between $T_e$[\ion{N}{2}] and $T_e$[\ion{S}{3}] seen for the dark blue / purple points 
promotes the use of these low- and intermediate-ionization zone temperatures for low-ionization 
\ion{H}{2} regions (low $F_{\lambda5007}$/$F_{\lambda3727}$).
However, comparing the high-ionization yellow points in the two plots suggests 
it is better to use the high-ionization zone temperature, $T_e$[\ion{O}{3}],  
for \ion{H}{2} regions with high $F_{\lambda5007}$/$F_{\lambda3727}$.
\\
\\} 
\label{fig4}
\end{SCfigure*}


In the middle two panels of Figure~\ref{fig3}, we examine the relationship
between the intermediate-ionization zone, characterized by [\ion{S}{3}],
with both the high-ionization zone ([\ion{O}{3}]; left) 
and low-ionization zone ([\ion{N}{2}]; right).
In the middle left panel, we find the best fit to the $T_e$[\ion{O}{3}]$-T_e$[\ion{S}{3}]
relationship is in good agreement with \citetalias{croxall16},  
but diverges from G92 for the hottest regions observed:
$T_e$[\ion{S}{3}]~$= (1.795\pm0.067)\times T_e$[\ion{O}{3}] $- (8.167\pm1.122)$.
Previous studies have reported large discrepancies between
$T_e$[\ion{O}{3}] and $T_e$[\ion{S}{3}] and significant scatter in their relationship
\citep[e.g.,][]{hagele06,perez-montero06,binette12,berg15}.
The $T_e$[\ion{O}{3}]$-T_e$[\ion{S}{3}] relationship for our sample of 59 regions 
is no exception, with a significant scatter of $\sigma_{tot.} = 900$ K that can be
attributed almost entirely to intrinsic scatter ($\sigma_{int.} = 860$ K).
Given the large number of outliers presented in both our sample and the literature,
we reiterate and stress the finding of \citetalias{berg15} that $T_e$[\ion{O}{3}] 
alone is less reliable than $T_e$[\ion{S}{3}] or $T_e$[\ion{N}{2}] for abundance 
calculations in metal-rich \ion{H}{2} regions. 

\citet{curti17} cautioned of the potential contamination of the temperature-sensitive
[\ion{O}{3}] \W4363 line by the neighboring [\ion{Fe}{2}] \W4360 line.
This effect is especially prominent at abundances of 12+log(O/H)$>8.4$, 
where the [\ion{Fe}{2}] line increases in strength and the [\ion{O}{3}] \W4363
line becomes faint due to the decreasing \ion{H}{2} region temperature.
Because \citet{curti17} study used stacks of integrated galaxy light spectra in
their study, the source of the [\ion{Fe}{2}] \W4360 emission is difficult to trace; however,
as a precaution we have added the \ion{Fe}{2} emission feature to our line fitting code 
so that the [\ion{Fe}{2}] \W4360 and [\ion{O}{3}] \W4363 lines are simultaneously fit and deblended,
and have inspected the fits by-eye (see \S~\ref{sec:reduct}). 
In fact, we do not measure $T_e$[\ion{O}{3}] in any very metal rich 
\ion{H}{2} regions in CHAOS and so do not find any significant [\ion{Fe}{2}] 
contamination affecting our $T_e$[\ion{O}{3}] measurements.
For instance, [\ion{Fe}{2}] \W4360 emission is seen in the blue inset window of the spectrum 
shown in Figure~\ref{fig2}.
However, [\ion{O}{3}] \W4363 was not strong enough to be identified as a detection 
and so the high-ionization zone temperature was inferred from $T_e$[\ion{S}{3}] 
and not affected by the [\ion{Fe}{2}] contamination.

In the middle right panel of Figure~\ref{fig3} we plot $T_e$[\ion{N}{2}] versus $T_e$[\ion{S}{3}].
Similar to the trend reported in \citetalias{berg15}, we find a very tight correlation,
especially for the coolest, most metal-rich regions typical of CHAOS (with $T_e < 10^4$ K).
The best fit line (blue) to the 90 regions is 
$T_e$[\ion{S}{3}]~$=~(1.522\pm0.042)\times~T_e$[\ion{N}{2}]$- (4.576\pm0.463)$,
in agreement with the relationship of \citetalias{croxall16} 
(black dashed line) and about which the dispersion is quite small: $\sigma_{tot.} = 420$ K.
The \citetalias{croxall16} relationship is also very similar to the G92 relationship,
where differences (seen in both bottom panels) are likely due to changes
in the adopted [\ion{S}{3}] atomic data.

Finally, we compare the low- and high-ionization zones in the bottom two panels of Figure~\ref{fig3}.
On the left, the relationship between the low-ionization zone $T_e$[\ion{N}{2}] and
the high-ionization zone $T_e$[\ion{O}{3}] is reasonably well behaved, but has too
few data points to analyze further.
On the other hand, the $T_e$[\ion{O}{2}]$-T_e$[\ion{O}{3}] plot shows a cloud of
scattered points that is difficult to characterize.

Significant [\ion{O}{3}] \W4363, [\ion{N}{2}] \W5755, and/or [\ion{S}{3}] \W6312
detections are measured in 30 regions in NGC~3184, resulting in direct oxygen
abundance measurements.
The electron temperatures and densities characterizing each \ion{H}{2}
region observed in NGC~3184 are reported in Table~5 in Appendix~\ref{sec:A1}.

\subsubsection{Ionization-Based Temperature Priorities}
CHAOS has proven highly successful at measuring significant detections of 
both [\ion{N}{2}] \W5755 and [\ion{S}{3}] \W6312, demonstrating the utility 
of these lines in metal-rich \ion{H}{2} regions.
Given the robust $T_e$[\ion{N}{2}]$-T_e$[\ion{S}{3}] relationship demonstrated for 
the 90 \ion{H}{2} regions with simultaneous detections,
our results further endorse the recommendation of \citetalias{berg15} to prioritize 
these two temperature indicators. 
However, it is interesting that the $T_e$[\ion{N}{2}]$-T_e$[\ion{S}{3}]
relation shows a notable increase in dispersion for $T_e > 10^4$ K, whereas 
the dispersion in the $T_e$[\ion{O}{3}]$-T_e$[\ion{S}{3}] relationship seems
to settle down in that same $T_e$ regime.

Recently, \citet{yates19} measured a large range of $T_e$[\ion{O}{3}]/$T_e$[\ion{O}{2}]
ratios spanning significant temperature (and, due to its inverse dependence, metallicity) 
parameter space from a sample of 130 \ion{H}{2} regions and integrated-light galaxies.
They postulate that deviations from equal temperatures are rooted in the ionization 
structure of the nebulae, where O$^{++}$-dominated nebulae have cooler [\ion{O}{3}] temperatures
and O$^{+}$-dominated nebulae have cooler [\ion{O}{2}] temperatures.
Because the relative flux of the [\ion{O}{3}] \W5007 and [\ion{O}{2}] \W3727 emission lines are 
dependent on the number of oxygen ions in the O$^{++}$ relative to O$^{+}$ state, we can
use the [\ion{O}{3}] \W5007/[\ion{O}{2}] \W3727 ratio as a proxy for O$^{++}$/O$^+$, 
or the ionization structure.

In Figure~\ref{fig4} we reproduce the $T_e$[\ion{O}{3}]$-T_e$[\ion{S}{3}]
and $T_e$[\ion{N}{2}]$-T_e$[\ion{S}{3}] relationships from Figure~\ref{fig3},
but with the points color-coded by their reddening-corrected 
[\ion{O}{3}] \W5007/[\ion{O}{2}] \W3727 flux ratios.
As expected, low ionization \ion{H}{2} regions (low 
$F_{\lambda5007}$/$F_{\lambda3727}$;
dark blue/purple points) show the tightest correlation between the low- and intermediate-ionization zone
temperatures ($T_e$[\ion{N}{2}] versus $T_e$[\ion{S}{3}]) and high ionization \ion{H}{2} regions 
(high $F_{\lambda5007}$/$F_{\lambda3727}$; yellow points) show the tightest correlation
between high- and intermediate- ionization zone temperatures ($T_e$[\ion{O}{3}] versus 
$T_e$[\ion{S}{3}]).
Motivated by these dispersion-ionization correlations, we recommend simple, yet improved, 
ionization-based temperature priorities below.

While few $T_e$[\ion{O}{3}] detections were found in the first CHAOS paper examining NGC~628,
many more detections were added with the addition of NGC~5457, revealing the utility of 
$T_e$[\ion{O}{3}] at
high $T_e$ and high ionization (high $F_{\lambda5007}$/$F_{\lambda3727}$).
Therefore, we prefer a $T_e$[\ion{O}{3}] measurement for high ionization nebulae that are 
dominated by the O$^{++}$ zone and a $T_e$[\ion{N}{2}] measurement for low ionization nebulae
that are predominantly O$^+$, where $T_e$[\ion{S}{3}] is used in the absence of a [\ion{N}{2}] \W5755 
detection.
In order to apply this rubric, we define a high (low) ionization nebula criteria
of $F_{\lambda5007}$/$F_{\lambda3727} >$ ($<$) 1.25.
This division was chosen based on a statistical analysis of the $T_e$[\ion{O}{3}]-based
oxygen abundance dispersion with $F_{\lambda5007}$/$F_{\lambda3727}$ using
data from the \citetalias{croxall16} study of M101 and the \citet{rosolowsky08} study of M33, 
where dispersion was minimized for $F_{\lambda5007}$/$F_{\lambda3727} > 1.25$.
The details of this analysis will be presented in \citet{Berg20}.


\subsection{Abundance Determinations}
We calculate absolute and relative abundances using the P{\sc y}N{\sc eb} package in {\sc python}, 
assuming a five-level atom model \citep{derobertis87}, the atomic data reported in Table~4 of 
\citetalias{berg15}, and the temperatures determined 
from the [\ion{O}{3}], [\ion{S}{3}], and/or [\ion{N}{2}] measured temperatures
in conjunction with $T_e - T_e$ scaling relationships.
We showed in Section~\ref{sec:temden} that our electron temperature results
for the first four CHAOS galaxies are consistent with the \citetalias{croxall16} 
$T_e - T_e$ relationships, therefore, we use Equations~$1-3$ to determine the
temperatures of unmeasured ionization zones. 
Further, the dispersion in our measured $T_e - T_e$ relationships correlates with the average
ionization of the nebulae, as represented by the $O_{32} = F_{\lambda5007}/F_{\lambda3727}$ ratio.

\begin{figure}
\begin{center}
	\includegraphics[scale = 0.57, trim = 24mm 200mm 10mm 25mm, clip]{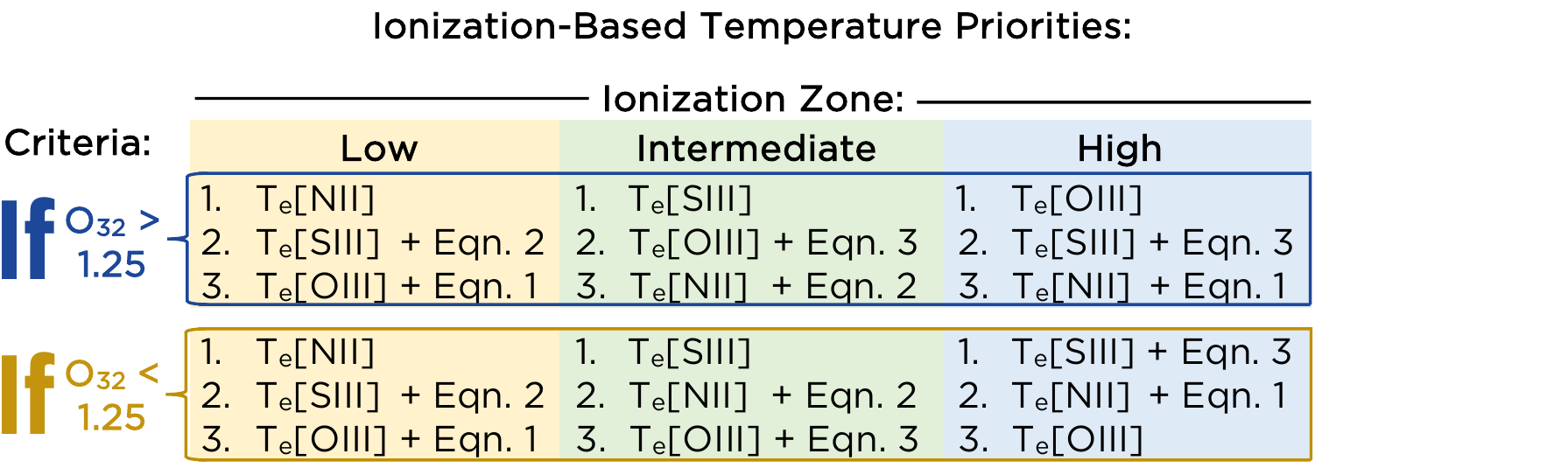}
	\caption{
	Updated temperature prioritization for different ionization zones from the CHAOS data.
	The priorities are to be used to select the first measured ion temperature from each ordered list
	and are split into two separate decision trees based on the $O_{32} = F_{\lambda5007}/F_{\lambda3727}$ ratio,
	which is used to determine the average ionization of an \ion{H}{2} region.}
  \label{fig5}
\end{center}
\end{figure}

We adopt the ionization-based temperature prioritization depicted in Figure~\ref{fig5}.
Specifically, if all three ionic temperatures are measured and the average ionization of the nebula 
is relatively high ($O_{32} > 1.25$), we prioritize $T_e$[\ion{N}{2}] for the low-ionization zone,
$T_e$[\ion{S}{3}] for the intermediate-ionization zone, and $T_e$[\ion{O}{3}] for the high-ionization zone.
If instead the average ionization of the \ion{H}{2} region is relatively low  ($O_{32} < 1.25$),
we adopt the measured low- and intermediate-ionization zone temperatures as before, 
but instead use $T_e$[\ion{S}{3}] in combination with Eqn.~3 to infer the high-ionization
zone temperature.
The justification of this choice is the large dispersions for high-ionization points 
in the $T_e-T_e$ relations shown in Figure~\ref{fig4}, with the result that we have less 
confidence in \W4363 in this regime (see discussion in \S~4.2).
In the absence of a measurement of the appropriate ionization-zone temperature,
temperatures should be inferred from the next preferred ion measured (following the
ordering in Figure~\ref{fig5}) in combination with the $T_e - T_e$ relationships from
Equations~$1-3$.

While the ionization-based temperature prioritizations presented here offer an improvement
to temperature-based abundance determinations, we note two caveats.
First, it is best to have independent measurements of the temperature in each ionization
zone to reduce the reliability on relationships from photoionization modeling.
Second, there are inherent, systematic uncertainties remaining due to the nominal assumption that 
\ion{H}{2} region structures can be simply divided into three 1D ionization zones when 
the reality is much more complicated.

\subsubsection{Oxygen Abundances}\label{sec:oabund}
We adopt the ionization-based temperature prioritization recommended in 
Figure~\ref{fig5} in order to determine the abundances of the first four CHAOS 
galaxies in a uniform, homogeneous manner.
Ionic abundances relative to hydrogen are calculated using:
\begin{equation}
    \frac{N(X^i)}{N(\rm{H}^+)} = \frac{I_{\lambda(i)}}{I_{\rm{H}\beta}}\frac{j_{\rm{H}\beta}}{j_{\lambda(i)}},
\end{equation}
where the emissivity coefficient, $j_{\lambda(i)}$, is sensitive to the 
adopted temperature.

The total oxygen abundance is calculated as the sum of the O$^+$/H$^+$ 
and O$^{++}$/H$^+$ ion fractions.
While emission from O$^{+3}$ is negligible in typical star-forming regions,
some oxygen might be in O$^0$ phase for the moderate-to-low ionization 
parameters characteristic of the CHAOS data \citep[$-2.5 <$ log $U < -4.0$; see, 
for example, Figure~5 in][]{berg19}.
In the current work, we can estimate the typical contribution to the oxygen 
abundance by O$^0$ emission using the [\ion{O}{1}] \W6300 feature, which can be 
distinguished from the [\ion{O}{1}] \W6300 night sky line at the distances of 
our sample and the resolution of MODS.
For our sample, the average $I$([\ion{O}{1}]\W6300)/$I$(H$\beta) = 0.022$,
corresponding to an O$^0$/(O$^0 + $O$^+ +$O$^{++}$) fraction of 3\%.
This means that, on average, the oxygen abundance may be underestimated by 
only $\Delta$O/H$<0.02$ dex due to missing O$^0$/H$^+$ contributes.
Given that possible contributions from O$^0$ are typically less significant 
than the uncertainties on the oxygen abundance measurements, O$^0$/H$^+$ 
is not included in our oxygen abundance determinations, consistent
with previously published CHAOS data.

The total oxygen abundances for our NGC~3184 sample are reported in Table~5
of Appendix~A, noting that neither O$^0$ nor contributions from dust
\citep[also typically $<0.1$ dex;][]{peimbert10,pena-guerrero12} are included.
Additionally, given that the abundances reported in previous CHAOS works were
not derived with methodology consistent with Figure~\ref{fig5}, we re-derive 
the abundances for NGC~628, NGC~5194, and NGC~5457 in order to compare our sample
in a uniform manner.
Since both NGC~628 and NGC~5194 were analyzed following the methodology laid out
in \citetalias{berg15} and both had very few [\ion{O}{3}] \W4363 detections,
their results were not significantly modified.
\citetalias{croxall16}'s study of NGC~5457, on the other hand, generally 
prioritized [\ion{O}{3}]-derived temperatures for the purpose of comparing 
to $T_e$[\ion{O}{3}]-based abundances in the literature.
The total and relative abundances for NGC~628, NGC~5194, and NGC~5457 used in 
this work are report in Table~6 in Appendix~~\ref{sec:A2}.


\begin{figure}
\begin{tabular}{r}
  \includegraphics[scale = 0.36, trim = 225mm 20mm 0mm 5mm, clip]{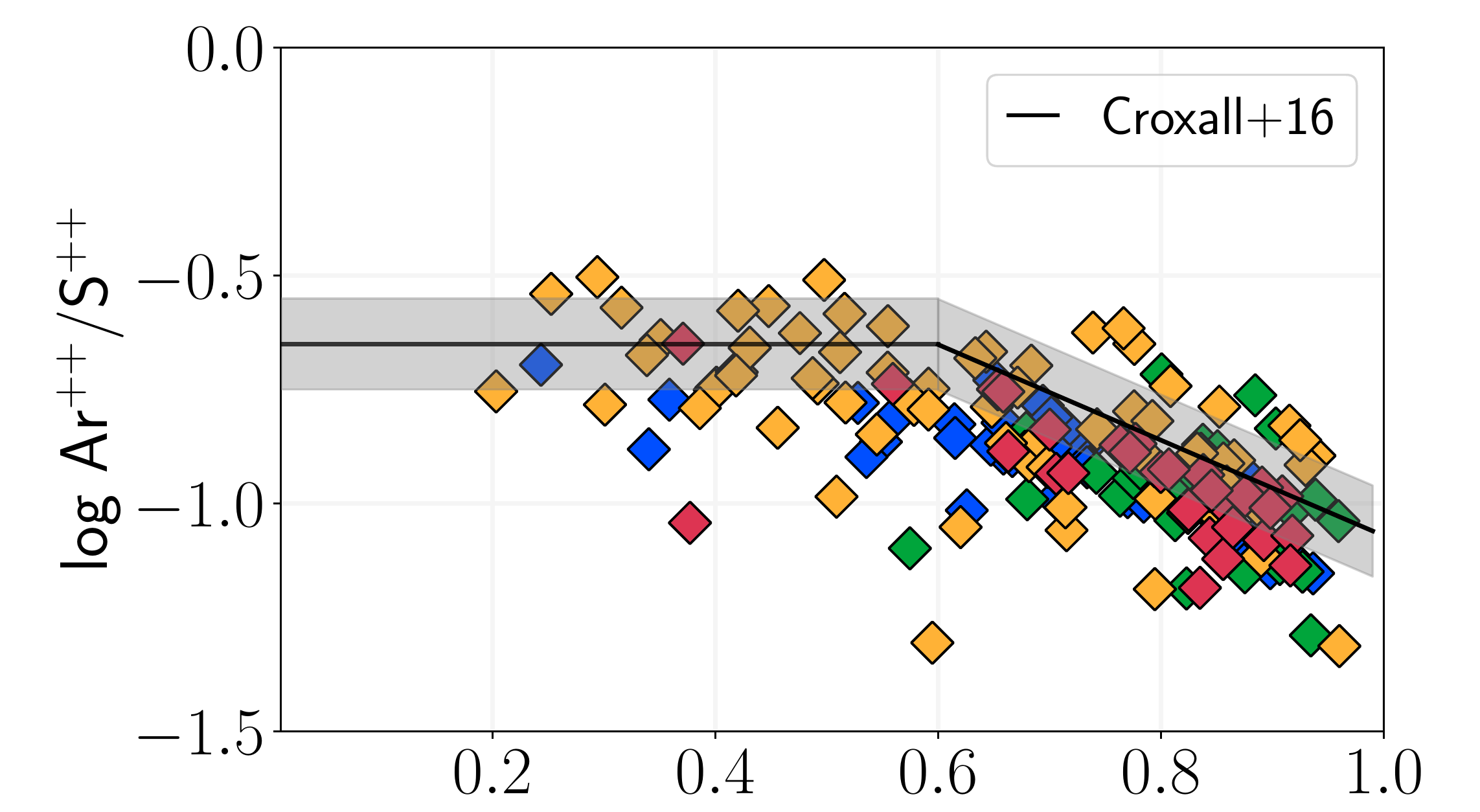} \\
  \includegraphics[scale = 0.36, trim = 0mm 20mm 225mm 5mm, clip]{CHAOSIV_ArICF_B19.pdf} \\
  \includegraphics[scale = 0.36, trim = 0mm 10mm 0mm 5mm, clip]{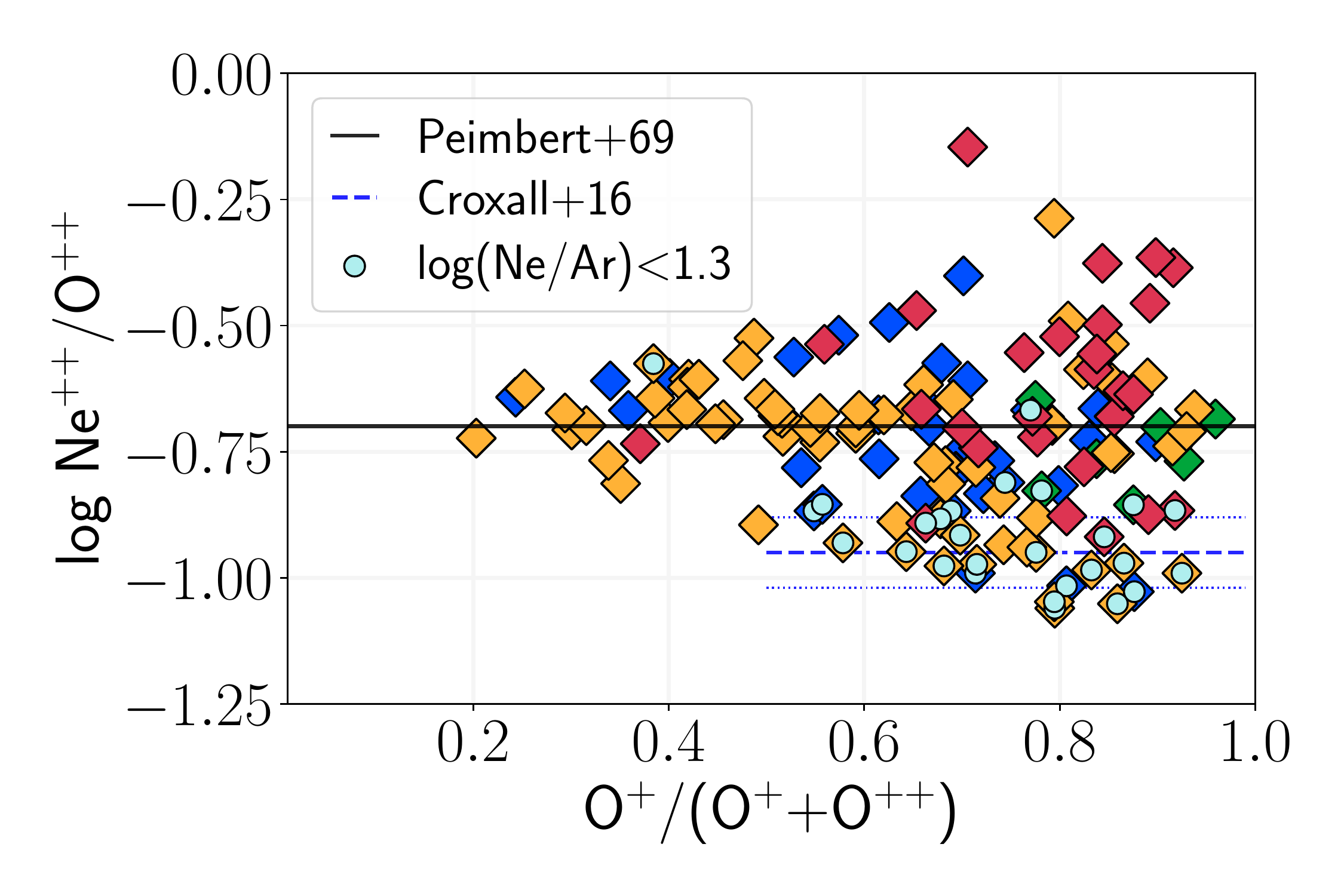} 
\end{tabular}
\caption{
Three of the ionization correction factors considered in this work versus
low-ionization fraction of oxygen, O$+$/(O$^{+} +$O$^{++}$).
In the top panel, we plot the Ar$^{++}$/S$^{++}$ ICF relationship 
introduced by \citetalias{croxall16}.
We note that NGC~3184 seems to deviate from the relationship to lower 
Ar$^{++}$/S$^{++}$ values with decreasing ionization.
In the middle panel, we plot the Ar$^{++}$/O$^{++}$ ratio and corresponding 
ICF(Ar) from \citet{thuan95}.
The NGC~3184 observations align well with this relationship.
In the bottom panel, we plot the Ne$^{++}$/O$^{++}$ ratio,
revealing two populations at low ionization (see also \citetalias{croxall16}).
We also consider regions with low log(Ne/Ar) ratios ($<1.3$; light blue circles),
which largely correspond to the low Ne$^{++}$/O$^{++}$ points.}
\label{fig6}
\end{figure}  


\subsubsection{Nitrogen Abundances}\label{sec:nabund}
We also observe significant N, S, Ar, and Ne emission lines in our spectra
that allow us to determine their relative abundances. 
However, when emission lines from prominent ionization stages are absent in the 
optical, their abundance determinations require an ionization correction factor (ICF)
to account for the unobserved ionic species. 
For nitrogen, we employ the common assumption that N/O = N$^+$/O$^+$,
such that the ICF(N) = (O$^+$ $+$ O$^{++}$)/O$^+$ \citep{peimbert67}.
While the O$^+$ ionization zone overlaps both N$^+$ and N$^{++}$, 
N/O = N$^+$/O$^+$ benefits from comparing two ions in the same temperature zone, 
and \citet{nava06} found this assumption valid within a precision of roughly 10\%.
We report the ionic, total, and relative N abundances for NGC~3184 in Table~5 in Appendix~A.
We also list the ICF, where the uncertainty is solely a propagation of the emission line uncertainties.

\subsubsection{Sulfur Abundances}\label{sec:sabund}
For sulfur, both S$^{+}$ (10.36--22.34 ev) and S$^{++}$ (22.34--34.79 eV) 
span the O$^{+}$ zone (13.62--35.12 eV), as the transitions from
S$^{++}$ to S$^{+3}$ and O$^{+}$ to O$^{++}$ are nearly coincident.
We note that the low ionization energy of S$^+$ means that [\ion{S}{2}]
emission can originate from outside the \ion{H}{2} regions ($E \leq 13.59$ eV),
and, therefore, caution must be used when interpreting these lines.
While we do not currently correct for such diffuse ionized gas in CHAOS,
the high-ionization of our nebulae ensure that the this gas only constitutes a small fraction.

In high-ionization nebulae, S$^{+3}$ (34.79--47.22 eV) lies in the O$^{++}$ zone (35.12--54.94 ev).
To account for the unseen S$^{+3}$ ionization state 
we employ the ICF from \citet{thuan95} for high-ionization \ion{H}{2} 
regions characterized for O$^+$/O$\leq 0.4$,
where the total O is assumed to be O = O$^+ +$O$^{++}$.
However, because the metal-rich \ion{H}{2} regions of CHAOS are typically
cooler and moderate-ionization, we follow the recommendation of 
\citetalias{croxall16} and adopt ICF(S) = O/O$^{++}$
(or simply S/O = (S$^+ +$S$^{++}$)/O$^{+}$) 
for O$^+$/O$ > 0.4$ \citep[see, also,][]{peimbert69}.
The resulting ICFs and ionic, total, and relative S abundances for NGC~3184 
are tabulated in Table~5 in Appendix~\ref{sec:A1}.
The uncertainty on the ICF(S) is a propagation of the emission line uncertainties
for O$^+$/O$ > 0.4$ and 10\% of the ICF(S) in the case of O$^+$/O$\leq 0.4$ \citep[see][]{thuan95}.

\subsubsection{Argon Abundances}\label{sec:arabund}
In the case of argon, only the Ar$^{++}$ ionization state is observed in the 
majority of CHAOS optical spectra, but the ionization potentials of 
O$^+$ (13.62--35.12 eV) and O$^{++}$ (35.12--54.94 ev) encompass portions of 
Ar$^+$ (15.76--27.63 eV), Ar$^{++}$ (27.63--40.74 eV), and Ar$^{+3}$ (40.74--59.81 eV).
While ratios of sulfur and oxygen ions relative to Ar$^{++}$ have both 
been used individually in the past to trace unseen argon ions, \citetalias{croxall16} 
found that the low-ionization regions of the CHAOS NGC~5457 sample 
are not well represented by either.
Instead, \citetalias{croxall16} corrected for the decrease in Ar$^{++}$/S$^{++}$ 
seen in low-ionization nebula by adopting a linear correction to Ar$^{++}$/S$^{++}$: 
log(Ar$^{++}$/S$^{++}) = -1.049\times$(O$^+$/O)$- 0.022$, for O$^+$/O $\geq 0.6$.
For higher ionization nebulae, Ar$^{++}$/S$^{++}$ was uncorrelated
with O$^+$/O and so a constant value of log(Ar$^{++}$/S$^{++}) = -0.65$ was 
assumed, similar to \citet{kennicutt03a}.

The log(Ar$^{++}$/S$^{++}$) correction from \citetalias{croxall16} is shown 
in the top panel of Figure~\ref{fig6}. 
The previously reported trend of decreasing Ar$^{++}$/S$^{++}$ with increasing 
O$^+$/O is reproduced, but with more dispersion in the updated ionic abundances,
especially for NGC~5457 -- the data it was derived for.
We find that all four CHAOS galaxies follow just as well the Ar$^{++}$/O$^{++}$-based 
ICF of \citet{thuan95} over the full range in O$^+$/O probed by the sample.
Given that three of the galaxies seem to be systematically offset from the Ar$^{++}$/S$^{++}$
relation, we choose to apply the ICF(Ar) from \citet{thuan95}, 
which has an uncertainty of 10\%, to all four CHAOS galaxies.
The differences between the updated ion fractions and those measured in 
\citetalias{croxall16} support the finding by \citet{yates19} and this work 
that ionization plays an important role in the temperature and metallicity 
determinations of an \ion{H}{2} region.
We list the resulting Ar abundances in Table~5 of Appendix~\ref{sec:A1}.
            
\subsubsection{Neon Abundances}\label{sec:neabund}
Neon is similar to argon in that only one ionization state is typically observed, 
Ne$^{++}$ (40.96--63.45 eV). 
Therefore, we use the ICF suggested by \citet{peimbert69} and \citet{crockett06} to 
correct for the unobserved Ne$^+$ ions (21.57--40.96 eV): ICF(Ne) = O/O$^{++}$,
where standard propagation of errors is used to determine the uncertainty.
Then, Ne/O = Ne$^{++}$/O$^{++}$.
Just as \citetalias{croxall16} reported a bifurcation in the Ne$^{++}$/O$^{++}$ 
values of NGC~5457, we see a similar downward dispersion for low ionization 
(O$^+$/O $> 0.5$) in the bottom 
panel of Figure~\ref{fig6} for our four-galaxy sample \citep[see, also,][]{kennicutt03b}.
Interestingly, we also note an upturn to high Ne$^{++}$/O$^{++}$ values for
some low-ionization nebulae.

The unseen Ne$^+$ (21.56--40.96 eV) partially overlaps with both the O$^+$ and O$^{++}$ ionization zones.
This means that a significant fraction of Ne likely lies in the Ne$^+$ state,
especially for the moderate-ionization nebulae observed by CHAOS.
This results in underestimated total Ne abundances in low- to moderate-ionization nebulae,  
a well-known issue with the classical ICF(Ne) \citep{torres-peimbert77,peimbert92}.
\citet{garcia-rojas13} observed a similar trend in the Ne/Ar ratios of planetary nebulae,
where low-ionization targets appeared Ne-poor and Ar-rich.
Interestingly, many of the low Ne$^{++}$/O$^{++}$ CHAOS points also exhibit the lowest
values of log(Ar/Ne), which are plotted as light blue circles in Figure~\ref{fig6}.

Using the average Ar/Ne ratio of the CHAOS sample as a guide, we apply a Ne/Ar
correction that is normalized to the average value for low-ionization regions 
(O$^{+}$/O $> 0.5$) and update the Ne/O values (see Section~6.1).
Overall, this correction seems to pull the regions with low-ionization Ne/O abundances up,
while regions with suspiciously low Ar/O abundances in NGC~5457 are adversely affected.
The resulting Ne abundances are reported in Table~5 of Appendix~\ref{sec:A1}.
While this updated ICF(Ne) is clearly not perfect, these relationships are illuminating 
and suggest that a more sophisticated ICF is needed to fully correct the total Ne abundance.
A more in depth discussion of the analysis of the CHAOS ICFs can be found in \citetalias{croxall16}.


\section{Radial Abundance Trends}


\begin{figure}
\epsscale{1.0}
  \centering
  \includegraphics[scale = 0.2225, trim = 6mm 5mm 0mm 0mm, clip]{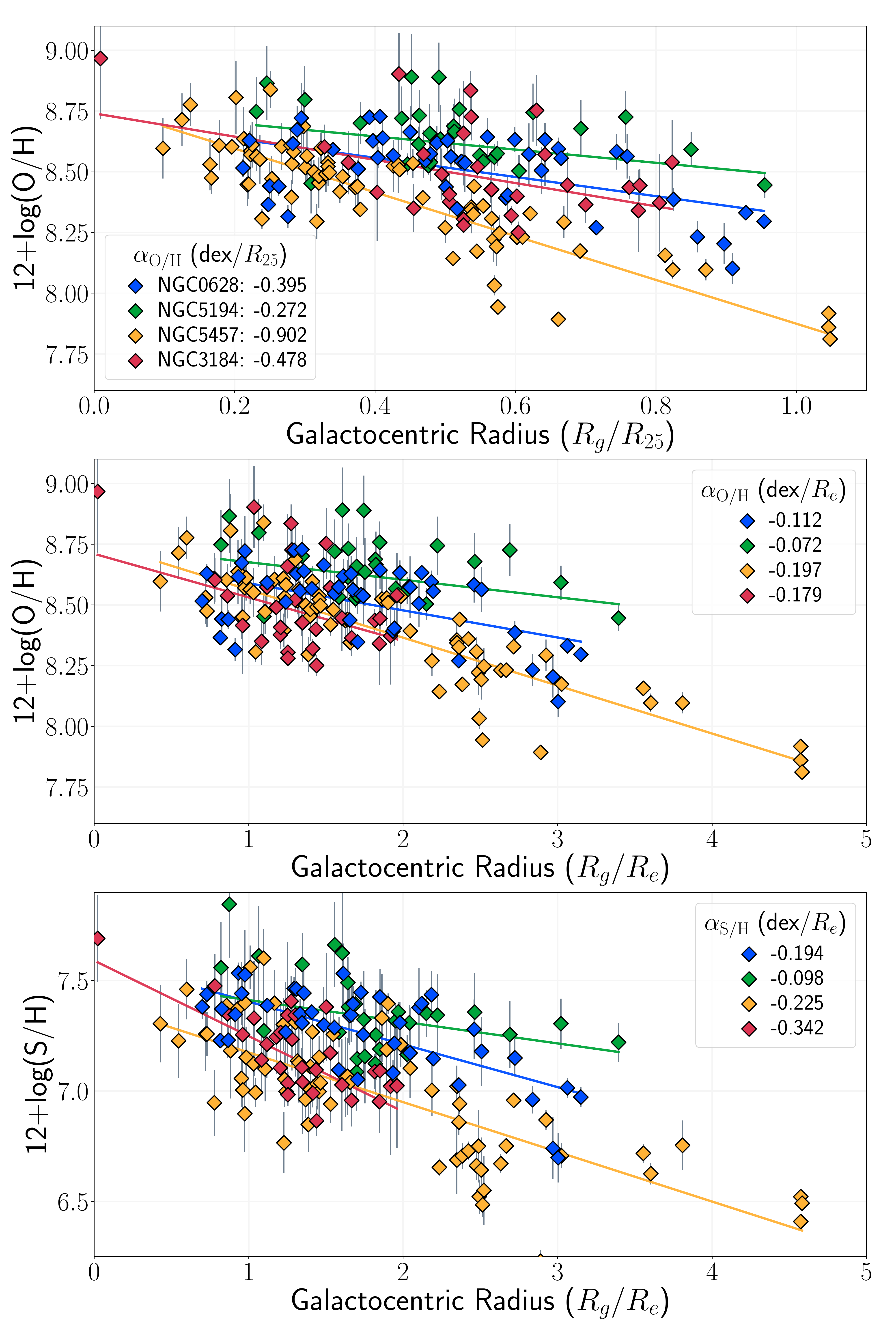}
  \caption{
  Abundance trends plotted vs. galactocentric radius for the first four CHAOS galaxies:
  the O/H gradient normalized to the $R_{25}$ radius of each galaxy ({\it top}),
  the O/H gradient normalized to each galaxy's effective radius, $R_e$ ({\it middle}),
  and the S/H gradient normalized to $R_e$ ({\it bottom}).
  We note that the O/H gradients appear to be no more ordered when plotted relative to $R_e$,
  as originally proposed by \citet{sanchez14}, as they show individual differences in
  their slopes and dispersions regardless of the radial normalization.
  Interestingly, S/H gradients show a similar ordering as O/H.
  See Table~\ref{tbl1} for properties of the CHAOS galaxies.}
  \label{fig7}
\end{figure}  


\subsection{Radial Oxygen Abundance Gradients}
In the past, studies of radial abundance trends have used both a variety 
of methods to characterize abundance and to normalize the galactocentric radius
to show significant variations in the gradients of different galaxies
\citep[e.g.,][]{zaritsky94,moustakas10}.
However, many of these studies have relied on abundance measurements in
just a handful of \ion{H}{2} regions per galaxy.
More recently, abundance trends have been studied in large numbers of \ion{H}{2}
regions using integral field unit (IFU) spectroscopy of individual galaxies.
Using empirical oxygen abundances determined from CALIFA IFU spectra, \citet{sanchez14} 
found a universal O/H gradient with a characteristic slope of 
$\alpha_{\rm O/H} = -0.10\pm0.09$ dex$/R_e$ over $0.3 < R_g/R_e < 2.0$ for 306 galaxies, 
whereas \citet{sanchez-menguiano16} report a shallower slope of 
$\alpha_{\rm O/H} = -0.075$ dex$/R_e$, with $\sigma=0.016$ dex for 122 face-on spiral galaxies.
However, the recent study of 102 spiral galaxies using VLT/MUSE IFU spectra by 
\citep{sanchez-menguiano18} found a distribution of slopes with an
average of $\alpha_{\rm O/H} = -0.10\pm0.03$ dex$/R_e$.
These authors find that radial gradients are steepest when the presence of an inner
drop or an outer flattening is also detected in the radial profile, and point to radial
motions in shaping the abundance profiles.

While IFU studies have greatly expanded our understanding of abundance gradients,
they have thus far relied on strong-line abundance calibrations, and therefore 
have systematic uncertainties \citep[e.g., see reviews from][]{kewley08,maiolino19}.
CHAOS now allows us to compare radial abundance trends using large numbers
of direct abundance measurements in \ion{H}{2} regions.
We display the O/H abundances derived in Section~\ref{sec:oabund} for the 
four CHAOS galaxies in Figure~\ref{fig7} as a function of galactocentric radius. 
Because the locations of individual \ion{H}{2} regions are known 
with high precision relative to one another, we consider only the 
uncertainties associated with oxygen abundance here. 
We plot the galactocentric radius relative to the isophotal ($R_{25}$) and effective ($R_e$)
radii of each galaxy in the top and middle panels of Figure~\ref{fig7}, respectively.
Because there is no visual evidence for an outer-disk flattening in the 
O/H gradient in the coverage of the CHAOS sample, we characterize the O/H 
gradient in each galaxy with a single, Bayesian linear regression using
the python {\sc linmix} code (solid lines).
Parameters of the resulting fits are given in Table~2.\looseness=-2

Comparing the individual O/H gradients in Figure~\ref{fig7}, 
there are apparent differences in both the O/H versus $R_g/R_{25}$ and O/H
versus $R_g/R_e$ gradients in the top and middle panels, respectively. 
While the gradients align more closely when plotted versus the effective radius ($R_e$),
the gradients of individual galaxies are still uniquely distinct.
The four CHAOS galaxies have a range of slopes of 
$-0.20 < \alpha_{\rm O/H}$ (dex$/R_e$) $< -0.07$.
Because the high-quality direct abundances of the CHAOS sample allow us to 
better constrain the unique gradient of an individual galaxy, 
we are seeing tangible gradient differences, even amongst just 4 galaxies, 
but within the dispersion seen for the large CALIFA samples of strong-line abundances. 
In this sense, the CHAOS data are demonstrating that O/H versus $R_g/R_e$ gradients 
are {\it not} uniformly behaved.

NGC~5194 presents the largest deviation from the typical CHAOS slope, where its nearly flat
slope has been attributed to interactions with its companion, NGC~5195, resulting in radial
migration and mixing of the interstellar gas \citepalias[see discussion in][]{croxall15}.
However, even when we only consider the three non-interacting spiral galaxies in our sample,
we find tangible differences in the O/H abundance gradients and dispersions of individual CHAOS 
galaxies.
The varying coefficients of the best-fit gradients characterizing the CHAOS galaxies 
(tabulated in Table~\ref{tbl2}) show that 
detailed direct abundance measurements reveal a range in the chemical evolution of individual galaxies. \looseness=-2


\subsection{Radial Sulfur Abundance Gradients}
Sulfur abundances can be an extremely useful tool, 
particularly in the absence of oxygen abundance information.
Notably, sulfur abundances only require a limited wavelength coverage of
$\sim$\W4850--\W9100 (but better if coverage extends to $\sim$\W9600) to ensure 
measurement of all the necessary inputs to a direct abundance:
(i) reddening correction (from H$\alpha$/H$\beta$ and the Paschen lines),
(ii) density (from [\ion{S}{2}] \W6717/\W6731), 
(iii) temperature (from [\ion{S}{3}] \W6312/\W9069),
(iv) S$^+$ (from [\ion{S}{2}] \W\W6717,6731), and 
(v) S$^{++}$ (from [\ion{S}{3}] \W\W9069,9532).
Surveys with limited blue wavelength coverage \citep[e.g., MUSE;][]{bacon10} 
may therefore be able to take advantage 
sulfur's utility and measure direct abundance trends in the absence of the blue oxygen lines. 

Prompted by the importance of S as a temperature indicator, and the expectation of 
alpha-elements that S and O abundances should trace one another, we explore the S/H 
gradients of the CHAOS galaxies in the bottom panel of Figure~\ref{fig7}.
As before, we fit Bayesian linear regression models and report the results in Table~\ref{tbl2}.
The S/H and O/H gradients of our galaxies are all consistent within the 
uncertainties, with the interesting exception of NGC~628.
These fits suggest that S/H abundances provide an alternative direct measurement
of a galaxy's metallicity gradient.
S/H abundances may also be easier to measure in moderate- to 
metal-rich \ion{H}{2} regions where [\ion{S}{3}] \W6312 is significantly 
detected more often than [\ion{O}{3}] \W4363.
However, it is important to note that S/H abundances have the disadvantage of requiring an 
ICF for the unseen S$^{+3}$ and thus, are generally considered inferior to O/H abundances.
Typically, in the CHAOS sample, the correction for S$^{+3}$ is less than 20\%, 
but it can get as high as 80\%, so caution is warranted.

Why does sulfur seem to behave so well for the CHAOS sample?
While the dominant observable ionic states of O in the CHAOS spectra, 
O$^+$ and O$^{++}$, probe the full ionization range of \ion{H}{2} region 
nebulae, our data largely consist of moderate-ionization nebulae.
Our regions have O$^+$/O ionization fractions that are typical
of the more metal-rich \ion{H}{2} regions in spiral galaxies, and this combination
produces regions that are both more moderate ionization and have cooler temperatures.
Given this, it is perhaps not surprising that $T_e$[\ion{S}{3}] characterizes the CHAOS data so well.
At the typically higher metallicities of the CHAOS regions, the nebula are generally 
lower-excitation and so have large S$^{++}$ fractions (i.e., S$^{++}$ is the dominant ionization zone).
To be quantitative, given the excitation energy of [\ion{S}{3}] \W6312 (3.37 eV), 
a temperature of $T_e\sim7000$ K is required for 1\%\ of the electrons to excite [\ion{S}{3}].
This temperature is well matched to the majority of our \ion{H}{2} regions, 
which have temperature measurements of $6000$ K$ < T_e < 8000$ K.
On the other hand, the excitation energy of [\ion{O}{3}] \W4363 (5.35 eV) requires
a much hotter nebular temperature of $T_e\sim11000$ K for 1\%\ of electrons 
to excite [\ion{O}{3}].
In these typically moderate-ionization nebula, not only is O$^{++}$ a sub-dominant ion, 
but the relatively low electron temperature of the gas will rarely
excite to the upper level of O$^{++}$ from which \W4363 is emitted. 
In contrast, the observable ionic states of S in the CHAOS spectra (S$^+$, S$^{++}$) 
probe the lower ionization zones ($\lesssim35$ eV) that are dominant in the majority 
of metal-rich \ion{H}{2} regions.


\subsection{Radial N/O Abundance Gradients: \\
A Universal N/O Relationship}\label{sec:UNO}
\begin{figure}
\epsscale{1.0}
  \centering
  \includegraphics[scale = 0.24, trim = 10mm 25mm 0mm 40mm, clip]{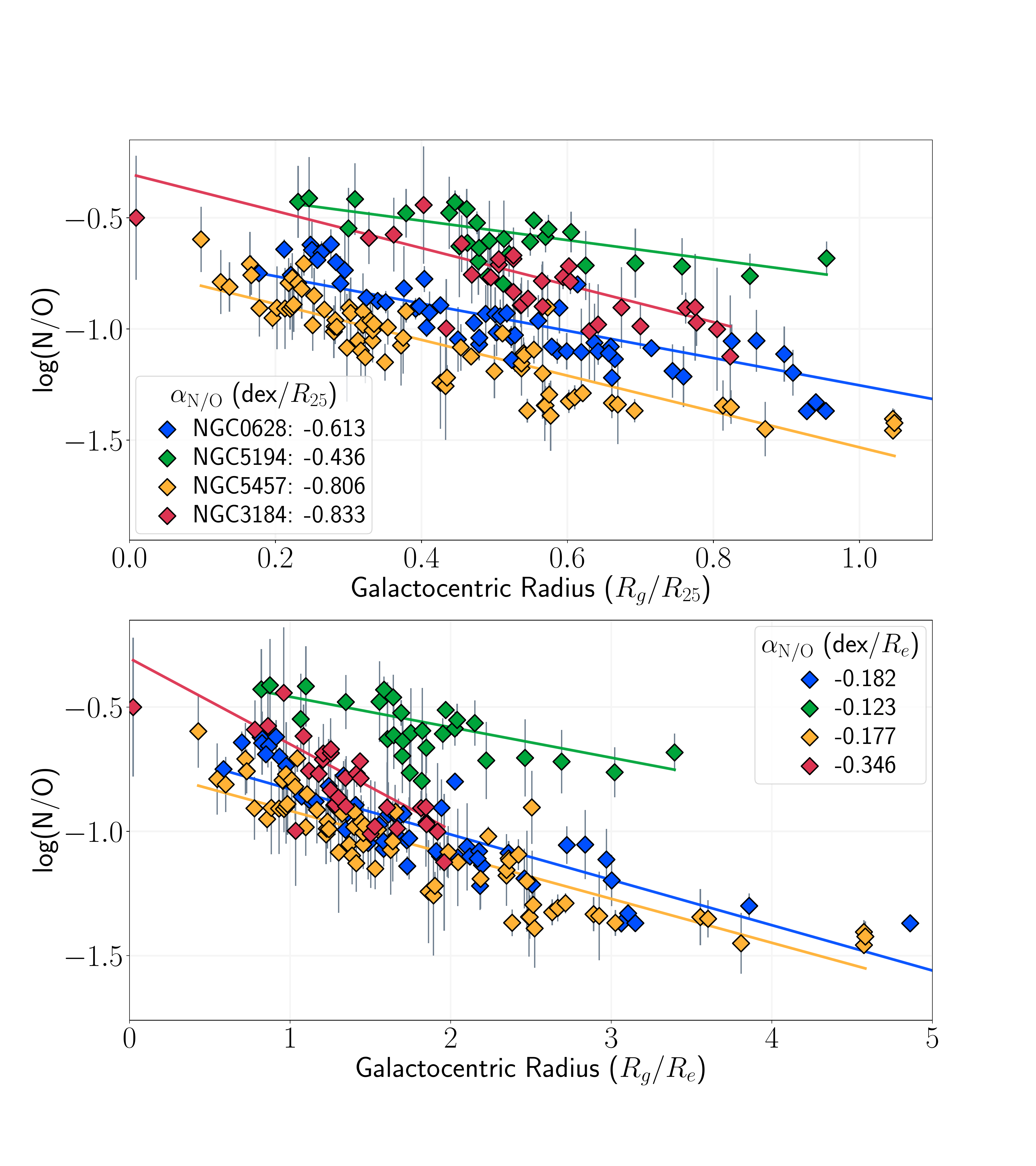}
  \caption{
  N/O abundance plotted vs. galactocentric radius for CHAOS galaxy sample presented here.
  {\it Top:} similar to Figure~\ref{fig6}, the N/O trend is normalized to the $R_{25}$ radius for each galaxy.
  {\it Bottom:} the N/O gradient relative to effective radius, $R_e$.
  NGC~5194 appears as a clear outlier when plotted in this way.}
  \label{fig8}
\end{figure}

The N/O abundances for the four CHAOS galaxies are presented in Figure~\ref{fig8}.
Galactocentric radii are normalized to the isophotal radius, $R_{25}$, of each galaxy 
in the top panel and to the effective radius, $R_e$, in the bottom panel.
Once again we analyze gradients of galaxies by comparing their 
individual Bayesian linear regression fits (solid lines).
Interestingly, when trends in N/O versus $R_g/R_{25}$ are considered as a single, 
linear relationship as was done with O/H in Section~4.2, all four galaxies appear 
to have similar gradients, only offset from one another.
Additionally, as noted in previous CHAOS papers, the N/O relationships are more tightly 
ordered with radius than the O/H gradients, presented by smaller dispersions.
On the other hand, when the N/O trends are normalized by their effective radius 
(bottom panel), three of the four galaxies (NGC~628, NGC~5457, and NGC~3184) shift
to lie nearly on top of one another, while NGC~5194 emerges as an outlier once again. \looseness=-2


\begin{figure*}
  \centering
  \includegraphics[scale = 0.3, trim = 8mm 5mm 0mm 0mm, clip]{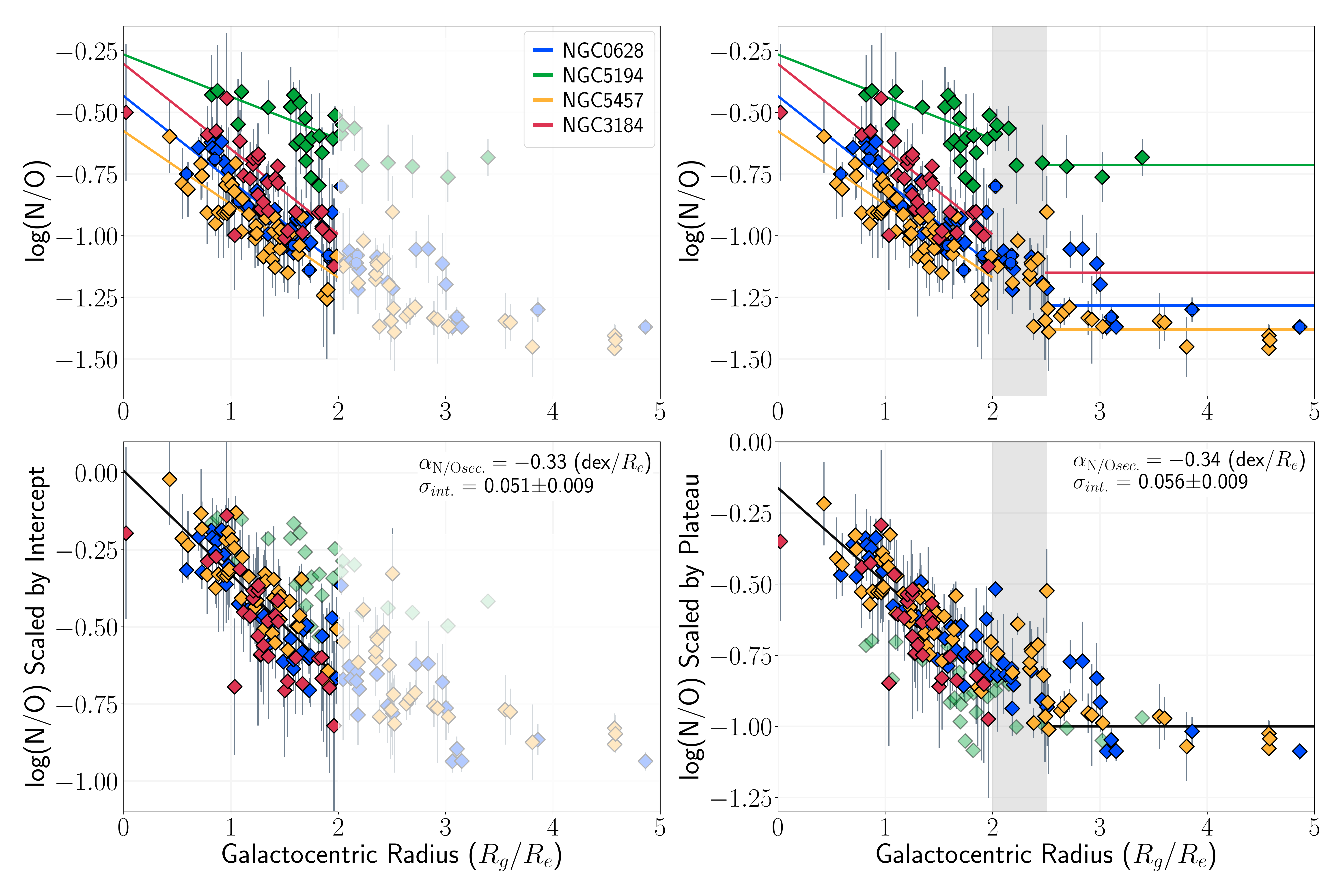}
  \caption{ 
  {\it Top left:} N/O versus galactocentric radius of the CHAOS sample with separate fits to N/O 
  for inner disks ($R_e/R_g < 2.0$).
  {\it Bottom left:} N/O trends of each galaxy are offset by the y-intercept of their fits above,
  producing a remarkably tight N/O gradient for the three non-interacting galaxies.
  {\it Top right:} Considering the full gradient, data in the outer disk ($R_e/R_g > 2.5$) appears 
  to form a flatter trend and so we shade the potential transition grey.
  As an illustrative exercise, a variance-weighted average plateau is fit for $R_e/R_g > 2.5$ for each 
  galaxy (and assumed for NGC~3184 based on the extrapolated fit for $R_g/R_e < 2.25$).
  {\it Bottom right:} N/O trends are normalized by the average outer disk N/O value of each galaxy,
  again revealing a universal N/O gradient for $R_e/R_g < 2.0$.
  If these trends are physical, then the outer flat trend may be the primary N plateau set
  by the galaxy's SFH and the inner gradient a primary$+$secondary N trend that transitions
  near $2.0 < R_e/R_g < 2.5$.
  Data sets with larger radial coverage are needed to test this prediction.}
  \label{fig9}
\end{figure*}  


We further investigate the similarities of the CHAOS N/O gradients by 
comparing them over the same radial extent.
Limited by the coverage of NGC~3184, we refit the N/O gradient of 
the $R_g/R_e < 2.0$ inner disks of the CHAOS galaxies with a Bayesian
linear regression model and plot them as solid lines in the top left-hand 
panel of Figure~\ref{fig9}.
Now, three of the four galaxies have trends that run parallel to one another:
all have very tight trends with slopes of $\alpha_{\rm N/O}=-0.3$ dex/$R_e$ 
and dispersions of $\sigma < 0.06$ dex (see Table~\ref{tbl2}).
Given that the inner disk radial gradients decline more steeply for N/O than O/H,
these trends are indicative of secondary nitrogen.

In order to isolate the secondary N/O trend of the CHAOS sample, 
we remove the offset between galaxies by subtracting their 
individual y-intercept offsets.
The resulting scaled N/O versus O/H relationships are shown in the bottom 
left-hand panel of Figure~\ref{fig9}, where a tight secondary N/O relationship 
emerges that characterizes the entire CHAOS sample well.
Given the relatively flat gradient of NGC~5194 in the top left-hand panel of 
Figure~\ref{fig9}, we fit the secondary N/O relationship excluding NGC~5194 
(denoted by the semi-transparent green points) in the bottom left-hand panel 
of Figure~\ref{fig9}.
The Bayesian linear regression reports a slope of $\alpha_{\rm N/O}=-0.33$ dex/$R_e$, 
with a very small total dispersion of $\sigma = 0.08$ dex.

It is remarkable that a simple shift produces such a tight secondary N/O gradient
for these three galaxies, and indicates that a physical origin may be responsible.
A common interpretation of N/O trends owes vertical offsets to differences
in individual star formation histories (SFHs) that set the primary N/O plateau
\citep[e.g.,][]{henry00}.
Given the limited disk coverage of the CHAOS sample, it is difficult to determine
the primary N/O plateau that is expected at large radii (low metallicity). 
However, we can explore the existing data in the outer disk as an illustrative exercise.
Using NGC~5457 as our best and largest dataset for exploring radial trends,
we note that the N/O trend is approximately flat for $R_g/R_e>2.5$, and so
adopt $2.0<R_g/R_e<2.5$ as the transition from primary to secondary N production
(gray-shaded band).
In the upper right-hand panel of Figure~\ref{fig9}, we fit a 
weighted average to the N/O values for $R_g/R_e > 2.5$.
For NGC~3184, no N/O measurements exist for $R_g/R_e > 2.5$, 
and so a (toy-model) plateau was assumed based on the value of the 
extrapolated secondary relationship at the transition radius.

In the bottom right-hand Figure~\ref{fig9} we apply a second scaling method.
We normalize the individual N/O relationships by their corresponding plateaus 
and once again see a tight secondary N/O relationship emerges
that characterizes the inner disk of the CHAOS sample well.
Fitting a Bayesian linear regression to the three non-interacting galaxies, 
we find a slope of $\alpha_{\rm N/O}=-0.34$ dex/$R_e$ and $\sigma = 0.08$,
equal to the slope determined using a y-intercept offset.
Once again, we find remarkable consistency of the N/O gradient slopes, 
regardless of the offset method used, suggesting a universal N/O gradient.
The agreement between the bottom two panels of Figure~\ref{fig9}
may be indicative of a break near $2.0<R_g/R_e<2.5$ and a transition to a 
flatter gradient for $R_g/R_e>2.5$.
We currently do not have sufficient data coverage of the outer CHAOS disks,
but more radially-extended data sets will be able to test this break/plateau prediction.
Coefficients for the secondary N/O fits are tabulated in Table~\ref{tbl2}.

If the slope of N/O versus radius is simply dependent on metallicity, then a universal 
N/O gradient like the one depicted in Figure~\ref{fig9} can be interpreted as 
resulting directly from the nucleosynthetic yields of the stars producing it.
In yield models, the integrated N yield is dominated by intermediate mass stars and
increases with increasing metallicity, while the oxygen yields from massive stars
decrease with increasing metallicity.
Further, the small observed scatter about this relationship could result from 
the fact that we are observing regions of star-formation with differing average burst 
ages, and the majority of N is produced around 250 Myr after the burst onset, whereas the 
massive stars producing oxygen have main-sequence lifetimes of only a few Myr
(see discussion in Section~7).


\section{Secondary Drivers of \\
Abundance Trends}

Even with the precise abundance gradients of spiral galaxies afforded by the CHAOS project,
many open questions remain regarding metallicity gradients in disk galaxies.
Here we explore possible environment effects through azimuthal variations
and surface density profiles. \looseness=-2

\subsection{Azimuthal Variations}
Beyond simple gradients in spiral galaxies, other patterns in the spatial distribution of
metals in the ISM may be key to understanding the redistribution of recently synthesized
products. 
While some processes happen on relatively short timescale, such as local oxygen production 
from massive stars \citep[$< 30$ Myr;][]{pipino09} and \ion{H}{2} region mixing on sub-kpc scales
($< 100$ Myr), the timescale for differential rotation to chemically homogenize an 
annulus of the ISM is much longer \citep[$\sim$1 Gyr; see, e.g.,][]{kreckel18}.
Further, the fate of metals after they are produced is unclear, as the spatial and temporal 
scales on which oxygen enriches the ISM occurs are poorly known.
Therefore, azimuthal inhomogeneities are expected and can inform us
about asymmetric processes occurring in the disk. 

\citet{ho17} studied the azimuthal variations in the oxygen abundance gradient of 
the nearby, strongly-barred, spiral galaxy NGC~1365 as part of the TYPHOON program, 
finding O/H to be lower, on average, by 0.2 dex downstream from the spiral arms.
Given the strong correlation with spiral pattern, these authors find that the observed
abundance variations are due to the mixing and dilution processes driven by the spiral
density waves.
On the other hand, the TYPHOON program has also reported a much smaller magnitude of 0.06 dex
azimuthal variations for the unbarred spiral galaxy NGC 2997 \citep{ho18}.

We test for azimuthal variations in the CHAOS sample by examining the offset 
in direct abundance from each galaxy's average gradient for O/H and N/O
as a function of both radius and position angle with in the disk.
We find no evidence of systematic azimuthal variations in the direct abundance 
CHAOS sample of unbarred spiral galaxies explored here.
However, while CHAOS observations span broad radial and azimuthal coverage, 
region selection is biased to the highest surface-brightness \ion{H}{2} regions, 
and so may not include the faint inter-arm coverage needed to unveil these subtle trends.


\begin{figure*} 
  \centering
    \includegraphics[scale=0.30,trim = 0mm 0mm 0mm 0mm, clip]{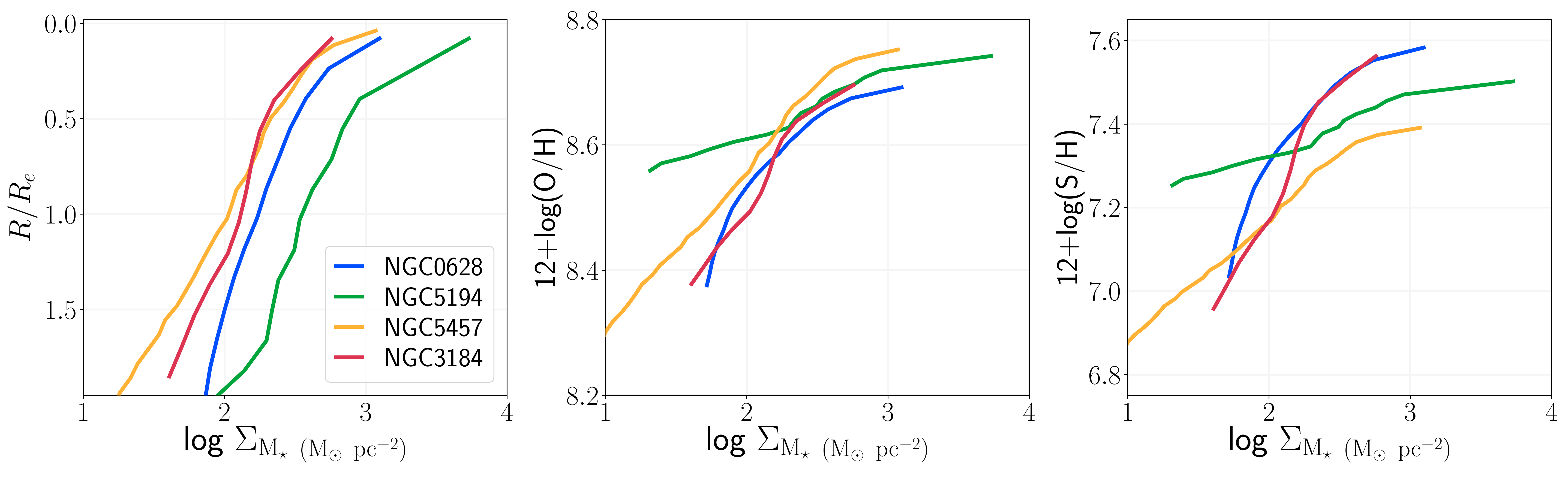}
  \caption{
  Stellar mass surface density trends for the four CHAOS galaxies relative to 
  galactocentric-radius (left panel), oxygen abundance (middle panel), and sulfur
  abundance (right panel).
  While NGC~5194 has abnormally flat abundance trends, the oxygen abundance of the three 
  non-interacting galaxies closely correlates with stellar mass surface density.}
  \label{fig10}
\end{figure*} 


\begin{figure*} 
  \centering
    \includegraphics[scale=0.215,trim = 0mm 0mm 0mm 0mm, clip]{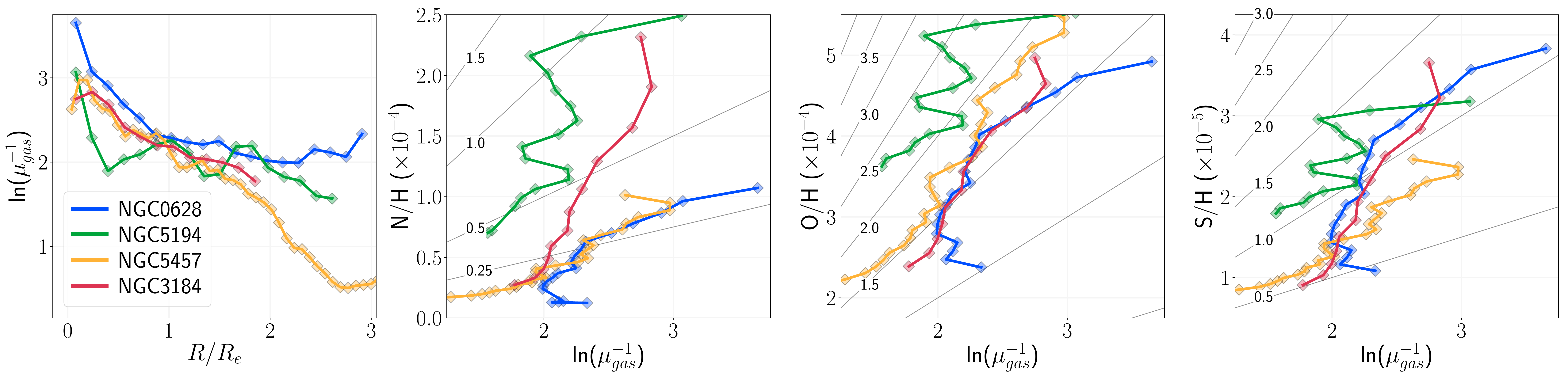}
  \caption{
  Trends of the logarithm of the inverse of the gas fraction for the CHAOS sample.
  The radially-averaged profiles are similar for all four galaxies (first panel), 
  whereas abundance trends for N/H (second panel), O/H (third panel), and S/H 
  (last panel) show more significant variations.
  For the abundance versus inverse gas fraction trends, lines of constant elemental effective 
  yield are drawn, where the yield labels are in that same units as the y-axis
  ($\times10^{-4}$, $\times10^{-4}$, and $\times10^{-5}$ for N/H, O/H, and S/H, respectively).  }
  \label{fig11}
\end{figure*} 


\subsection{Surface Density Relationships}
A fundamental relationship of global galaxy evolution is the luminosity-metallicity 
relationship, which includes spiral disk galaxies \citep[e.g.,][]{garnett87,vila-costas93,zaritsky94}.
This relationship typically refers to the total or average metallicity of a galaxy,
but what does this mean for the abundance gradients in individual spiral galaxies?
While several recent studies support a characteristic oxygen abundance gradient for the
main disk of spiral galaxies \citep[e.g.,][]{sanchez14,sanchez-menguiano18},
\citet{belfiore17} reported an increasing oxygen abundance slope (dex/$R_e$) 
with stellar mass for SDSS-IV MaNGA \citep{bundy15} galaxies with M$_\star < 10^{10.5}$ M$_\odot$.
However, in a study of 49 local star-forming galaxies, \citet{ho15} found that 
metallicity gradients expressed in terms of the isophotal radius ($R_{25}$) did not 
correlate with either stellar mass or luminosity,
but rather increase with decreasing total stellar mass when expressed in terms of dex/kpc
\citep[see, also,][]{pilyugin19}.
Alternatively, \citet{pilyugin19} concluded in their study of MaNGA galaxies
that oxygen abundance is governed by a galaxy's rotational velocity.
Despite these works, no clear evidence has emerged to conclusively determine
the dependence of abundance gradients on basic galaxy properties or halo
properties (e.g., rotational velocity).

Locally, the oxygen abundance trends of spiral galaxies have also been observed to 
correlate with stellar mass surface density 
\citep[e.g.,][]{mccall82,edmunds84,ryder95,garnett97}.
In Figure~\ref{fig10}, we examine the stellar mass surface density profiles for
the CHAOS galaxies (see Appendix~\ref{sec:A3} for details).
The left panel shows the typical trend of decreasing stellar mass surface density
as you move further out in the disk, but with NGC~5194 having a slightly elevated 
density of stars compared to the others.
In the middle and right panels, we plot the local surface mass-metallicity relationship
for O/H and S/H, respectively.
Similar to the global relationship \citep[see, e.g.,][]{tremonti04}, local metallicity 
measurements also increase with mass surface density and plateau at high mass values.
This local trend is especially tight for the three non-interacting CHAOS galaxies.

The metallicity-surface density relationships in Figure~\ref{fig10} may reflect 
fundamental similarities in the evolution of non-barred, non-interacting spiral galaxies.
For example, \cite{ryder95} argues for a galaxy evolution model that includes self-regulating
star formation, where energy injected into the ISM by newly-formed stars inhibits 
further star formation.
These models were able to successfully reproduce the observed correlations between surface 
brightness and SFR \citep{dopita94} and surface mass density \citep[e.g.,][]{phillips91,ryder95,garnett97}.
The current work supports these ideas that stellar mass, gas mass, and SFR surface density are  
fundamental and interdependent parameters that govern the chemical evolution of spiral galaxies.
A more thorough investigation of the dependence of metallicity on local properties 
with be conducted in the future with the entire CHAOS sample.

\subsection{Effective Yields}
In a simple closed-box model, assuming instantaneous
recycling of stellar nucleosynthetic products and no gas flows,
chemical evolution is solely a function of the gas fraction, $\mu_{\rm gas}$:
$Z = y\cdot{\rm ln}(\mu^{-1})$,
where $Z$ is the metallicity and $y$ is the metal yield.
Inverting this equation, one can measure the effective yield, $y_{\rm eff}$,
given the observed metallicity, $Z_{\rm obs}$, and gas fraction:
\begin{equation}
	y_{\rm eff} = \frac{Z_{\rm obs}} {{\rm ln}(\mu_{\rm gas}^{-1})} \label{eqn4}.
\end{equation}

In Figure~\ref{fig11}, we plot the radially-averaged inverse gas fraction trends 
for the CHAOS sample (see Appendix C for the sources of the gas distributions).
While the inverse gas fractions steadily decrease with increasing radius for all
four galaxies (left panel), plotting abundance versus inverse gas fractions
reveals different effective yield trends (three right panels).
Nonetheless, the trends appear to be the most ordered for O/H and S/H, 
with similar slopes amongst the three non-interacting galaxies.
The less ordered trends for N/H may then be revealing the effects of varying gas flows in each galaxy
and the time effects of production in lower mass stars.
Further, this picture is consistent with the result from theoretical models based on stochastically 
forced diffusion that most scatter in observed abundance gradients ($\sim0.1$ dex) is due to stellar 
feedback and gas velocity dispersion \citep{krumholz18}.

Following Equation~\ref{eqn4}, these plots of abundance versus the inverse gas fraction
trace the effective yield of the relevant element.
The true yield is a function of stellar nucleosynthesis, but the effective yield (slope of 
$Z$-ln($\mu_{gas}^{-1}$) plots) will be altered from this value by gas inflows and outflows.
In this context, the similar slopes in O/H and S/H versus ln($\mu_{gas}^{-1}$) are indicative of
a closed-box effective yield of both oxygen and sulfur, whereas the O/H and S/H trends of 
NGC~5194 diverge as expected for gas flows associated with interacting galaxies.
According to Figure~\ref{fig11}, the CHAOS galaxies generally follow slopes of 
$0.5-1.25\times10^{-5}$ for sulfur and $1.5-2.0\times10^{-4}$ for oxygen, which corresponds to
$y_{\rm eff}$(O)$ = 0.006-0.008$ assuming Z $=0.02$ Z$_\odot$ and 12+log(O/H)$_\odot=8.69$
\citep{asplund09}.
These $y_{\rm eff}$(O) values are consistent with the range of effective oxygen yields 
measured for spiral galaxies by \citet{garnett02}, spanning 0.0033--0.017.
We note that the effective yield values \citet{garnett02} found for NGC~628 and NGC~5194 
are higher than our own, but this difference is largely accounted for by the offset in 
the measured abundance scales for these two galaxies.


\section{Abundance Trends With Metallicity}\label{sec:z}


\begin{figure*}
\centering
  \begin{tabular}{cc}
    \includegraphics[scale = 0.35, trim = 0mm 0mm 5mm 0mm, clip]{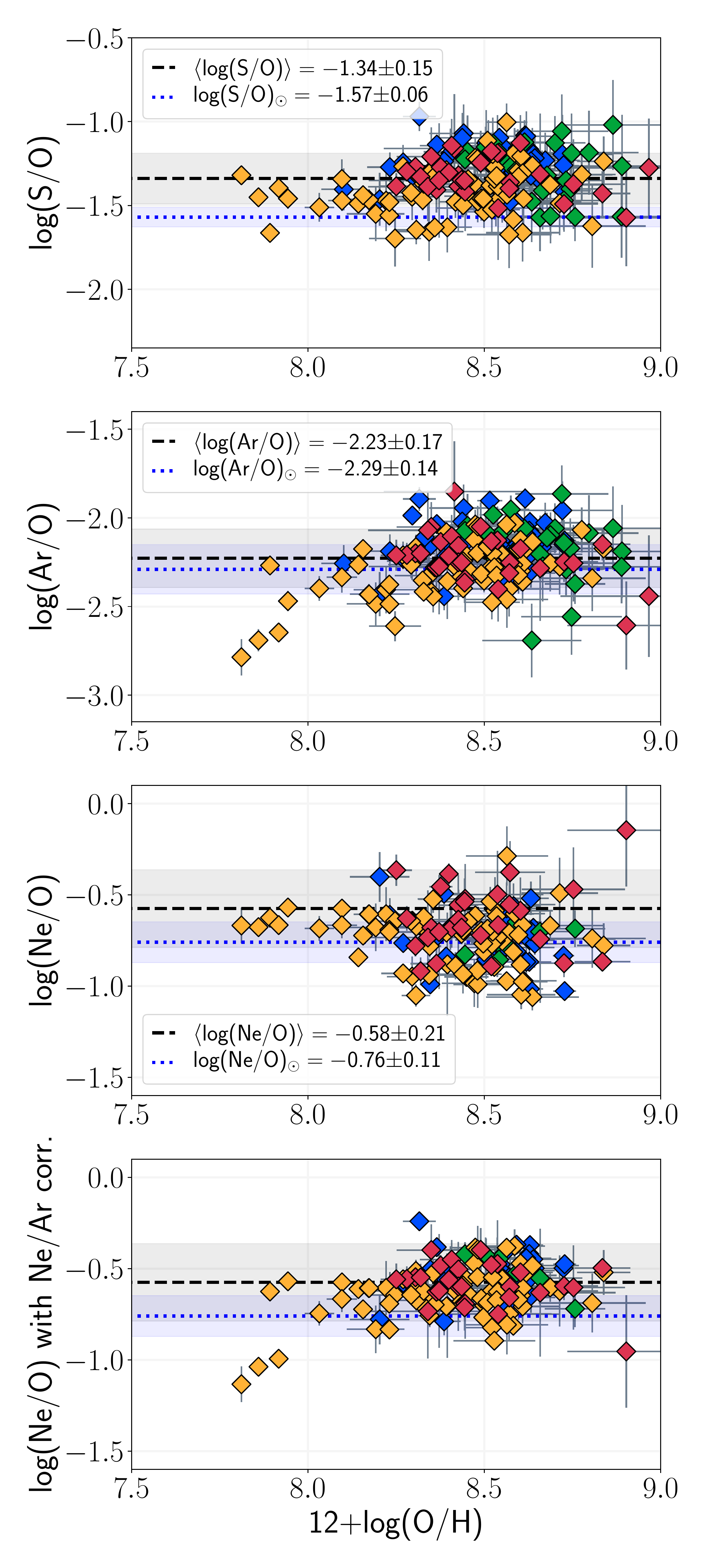} &
    \includegraphics[scale = 0.35, trim = 0mm 0mm 5mm 0mm, clip]{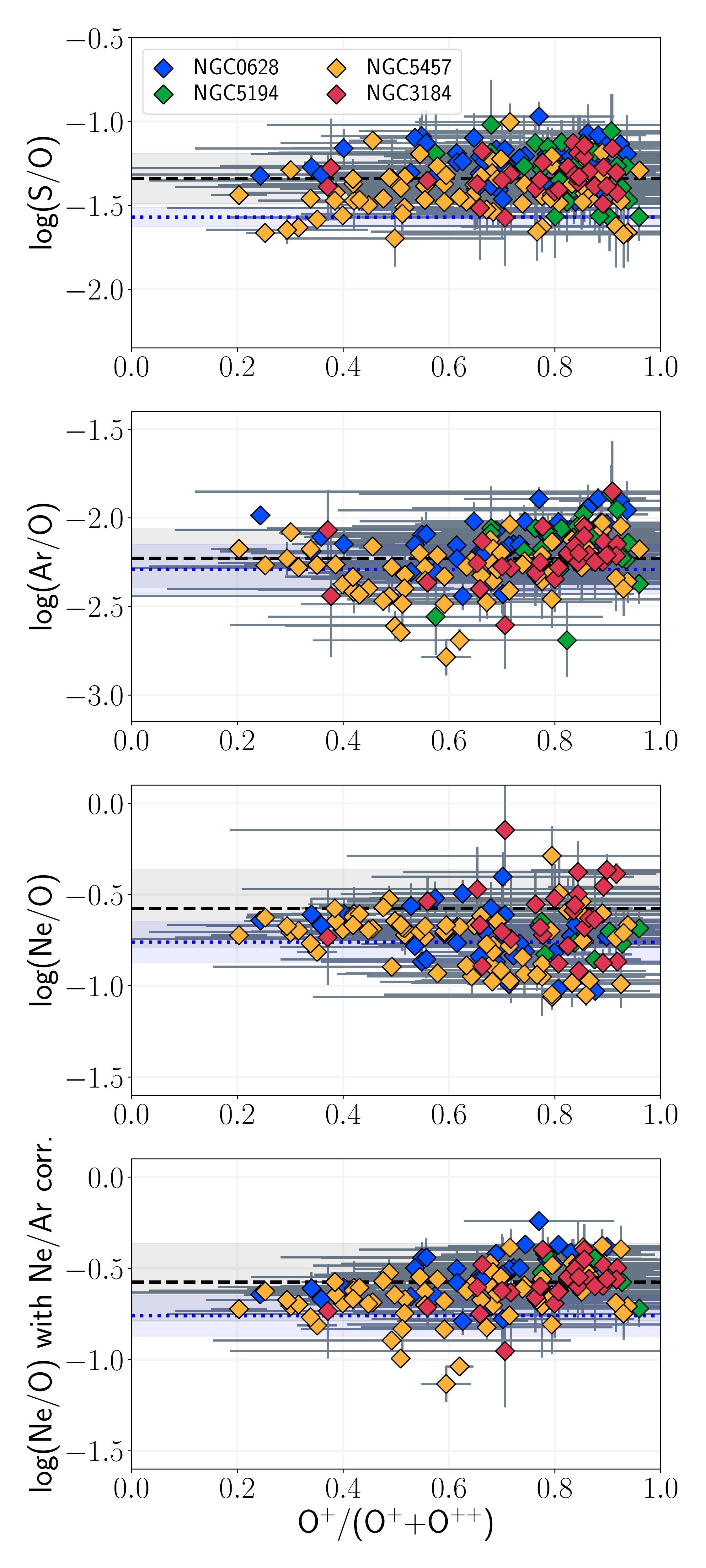}  
  \end{tabular}
  \caption{Alpha-element ratios for the CHAOS sample versus oxygen abundances
  (left) and ionization fraction (right).
  In each panel, the solar value and uncertainty from \citet{asplund09} is labeled and plotted as
  a blue dotted-line and blue shaded-band, respectively.
  The weighted-average and uncertainty of the CHAOS data is also given and denoted by a 
  black dashed-line and black shaded-band, respectively.
  The top two rows show the S/O and Ar/O ratios, both with relatively flat distributions.
  The Ar/O abundances for NGC~5457 were corrected by \citetalias{croxall16} 
  using the Ar$^{++}$/S$^{++}$ relation
  shown in Figure~\ref{fig6}, however, NGC~3184 uses the ICF(Ar) from \citet{thuan95}.
  The bottom two rows show the Ne/O ratio using the standard Ne$^{++}$/O$^{++}$
  ICF (third row) and further corrected for offsets in the Ne/Ar ratio (bottom row).}
  \label{fig12}
\end{figure*}


\subsection{Alpha/O Abundances}
Next, we turn our focus from abundance gradients to relative abundance
trends with O/H metallicity. 
In Figure~\ref{fig12} we plot the relative abundances of $\alpha$-elements.
In descending panel order we plot S/O, Ar/O, and Ne/O as a function of O/H (left side), 
where diamond points are color coded according to galaxy.

Stellar nucleosynthetic yields \citep[e.g.,][]{woosley95} indicate that 
$\alpha$-elements are predominantly produced on relatively short timescales 
by core-collapse supernovae (SNe; massive stars) explosions. 
The $\alpha$-element ratios in Figure~\ref{fig12} are, therefore, expected to 
be constant and so we plot the variance-weighted mean $\alpha$/O ratios of the 
CHAOS observations as black dashed lines in each panel.
The average values are denoted in the upper left corners
and can be compared to the solar values from \citet[][blue dotted line]{asplund09}.
The average CHAOS $\alpha$/O values are generally greater than solar,
but individual galaxies also show slight shifts from one another.

Relative to the constant relationship assumed in each panel of 
Figure~\ref{fig12}, the CHAOS observations visually show significant 
scatter and may also deviate in a systematic way. 
\citetalias{croxall16} discovered a significant population of low-ionization 
(high O$^{+}$/O) \ion{H}{2} regions in NGC~5457 with low Ne/O values.
A deeper exploration of the $\alpha$/O ratios in that work revealed a 
lack of previous observations in the low-ionization regime and challenges 
in finding an appropriate ICF to use.

Similar to \citetalias{croxall16}, in Section~\ref{sec:neabund} we found a large dispersion 
in the Ne$^{++}$/O$^{++}$ ratios of the CHAOS galaxies for low-ionization \ion{H}{2} regions. 
Additionally, many of these regions also exhibit exceptionally low values of log(Ar/Ne) 
(see Figure~\ref{fig6}).
This motivated us to apply a correction to the Ne/O abundances based on the offset in Ne/Ar from 
the average CHAOS value for low-ionization \ion{H}{2} regions (O$^{+}$/O $> 0.5$).
The updated Ne/O values, plotted in the bottom panel of Figure~\ref{fig12}, show a 
smaller dispersion around the mean sample value, but with a few significant NGC~5457 outliers.
While the proposed correction removes the bifurcation in Ne/O at low-ionization,
it seems to over-correct Ne/O abundance for the nebulae with discordantly low Ar/O abundances.

Following \citetalias{croxall16}, we further examine the $\alpha$/O dependence on ionization
by plotting our $\alpha$/O ratios for the four CHAOS galaxies versus O$^+$/O in the right 
column of Figure~\ref{fig12}.
For both Ar and S, there seems to be a small residual systematic dependence on ionization 
that is not adequately corrected for by \citetalias{croxall16} or other traditional ICFs.
In this case, the high-ionization \ion{H}{2} regions (O$^+$/O $< 0.5$) have 
S/O and Ar/O ratios that are generally under- and over-predicted, respectively, relative to 
the average, while the low-ionization \ion{H}{2} regions (O$^+$/O $> 0.5$) seem to be 
evenly dispersed about the mean.
In general, no simple corrections to the ICFs are yet apparent.
Instead, we will derive new ICFs for the CHAOS data using updated 
photoionization models in a future paper.  \looseness=-2


\subsection{N/O versus Metallicity}
Historically, N/O enrichment has been studied as a function of total oxygen 
abundance owing to the relative ease of integrated-light galaxy observations.
In this context, the observed scaling of nitrogen with oxygen has long been understood 
as a combination of primary nitrogen plus a linearly increasing fraction of secondary 
nitrogen that comes to dominate the total N/O relationship at intermediate metallicities 
\citep[e.g.,][]{vila-costas93,vanzee06a,berg12}.
Note that the scatter of the N/O--O/H relationship reported in previous studies is often
significantly larger than that of the CHAOS N/O radial gradients
\citep[e.g.,][]{vanzee06a,berg12}. \looseness=-2

\begin{figure*}
\begin{center}
\epsscale{1.0}
  \includegraphics[scale = 0.29, trim = 45mm 5mm 50mm 25mm, clip]{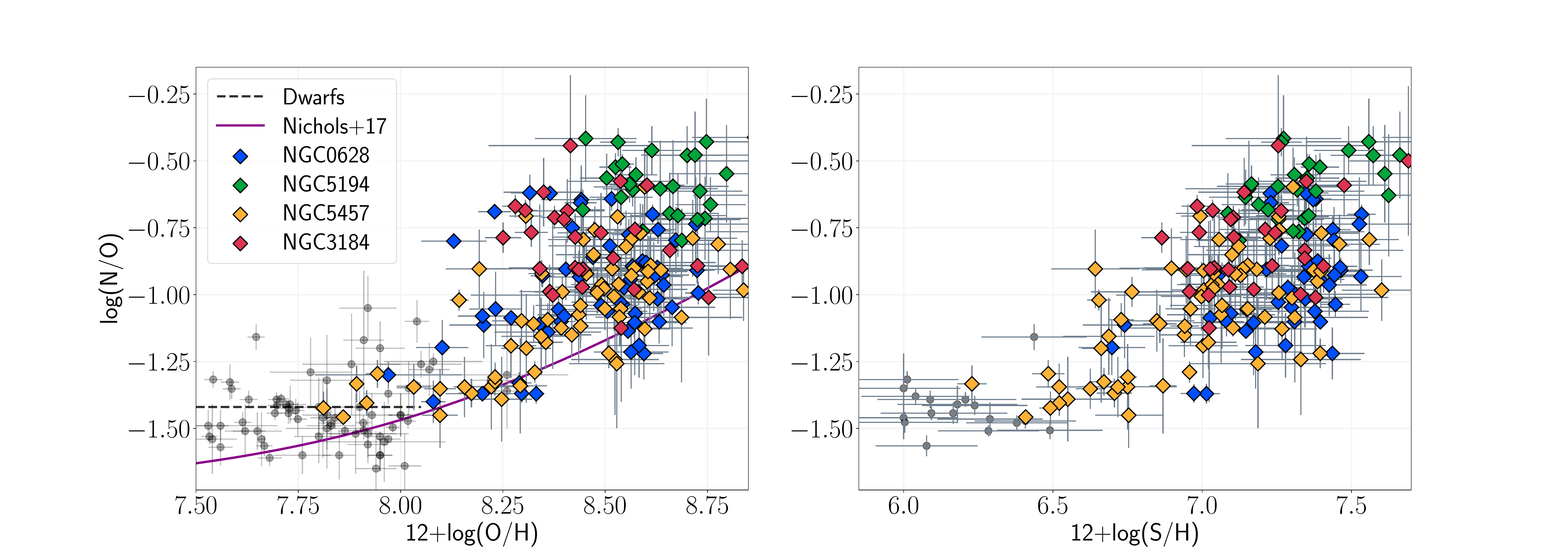} \\
  \caption{ 
  N/O versus O/H ({\it left}) and S/H ({\it right}) for the CHAOS galaxies (diamonds) 
  and local dwarf galaxies (grey circles).
  At low O/H or S/H, N is dominated by primary production and N/O is low 
  (grey dashed line is the average primary N/O plateau of dwarfs).
  At larger O/H or S/H, secondary N production begins to increase N/O.
  This is demonstrated by the empirical trend of stellar abundances (purple line),
  however, individual \ion{H}{2} regions show a large dispersion.}
  \label{fig13}
\end{center}
\end{figure*}

In Figure~\ref{fig13} we plot the N/O versus O/H values (left panel) and the
N/O versus S/H value (right panel) for the CHAOS galaxies.
For comparison, we also plot the empirical stellar N/O--O/H relationship from \citet{nicholls17}
and measured abundances for nearby metal-poor dwarf galaxies from \citet{berg19}, 
which should compose a primary N plateau at low O/H and S/H values. 
Despite the tight N/O radial gradients observed for individual CHAOS galaxies 
(see Figure~\ref{fig9}), large dispersion is seen in N/O when plotted versus O/H,
similar to previous N/O--O/H studies.
Guided by the stellar relationship (purple line), our data do follow the general trend
of low N/O due to primary nitrogen at low oxygen abundances, followed by increasing N/O,
presumably as secondary nitrogen becomes prominent, at larger O/H (12+log(O/H)$\gtrsim$8.2).
A similar trend is seen for N/O--S/H.
Yet, individual galaxies in our sample clearly occupy different regions on the N/O versus O/H 
and N/O versus S/H plots.
Interestingly, the collective trend of the four galaxies appears to produce a stronger
correlation between N/O with S/H than O/H.
However, significant scatter in seen for each galaxy, and the dispersions for the N/O--S/H 
and N/O--O/H relationships are consistent for each galaxy.


\section{Understanding The \\
Universal N/O Gradient}\label{sec:NO}


We now return to the universal N/O slope we found for the inner disks of CHAOS galaxies
in Section~4.3.
To understand the source of this trend, we must first 
understand how O and N are produced in these galaxies.
Despite the ease at which both O and N emission are observed, 
discovering the origin of N is far more complex than O.
Oxygen is primarily synthesized on short timescales by core-collapse SNe explosions 
of massive stars \citep[$M \gtrsim 9 M_\odot$, e.g.,][]{heger03}.
Nitrogen, on the other hand, is produced mainly by the CN branch of the CNO cycle,
which can occur in the H-burning layer of both massive stars and intermediate mass 
stars ($1 M_\odot < M < 9 M_\odot$).
The slowest step of the CNO cycle is the conversion of $^{14}$N to $^{15}$O, which results in a
pile up of $^{14}$N that can then be dredged-up by a convective layer.
In metal-poor gas, the seed O and C needed for the CNO cycle may come from a He-burning phase.
This path to N production is independent of the initial metal content of the star,
and so is referred to as ``primary" nucleosynthesis.
In more enriched gas at higher metallicities the CNO cycle increases N production 
proportional to the initial metal composition (O and C) of the star.
This type of N production is ``secondary" nitrogen owing to its dependence on 
the metallicity of the star in which it was synthesized.


\begin{figure*}
\epsscale{1.0}
  \centering
  \begin{tabular}{c}
  \includegraphics[scale = 0.98, trim = 48mm 72mm 25mm 82mm, clip]{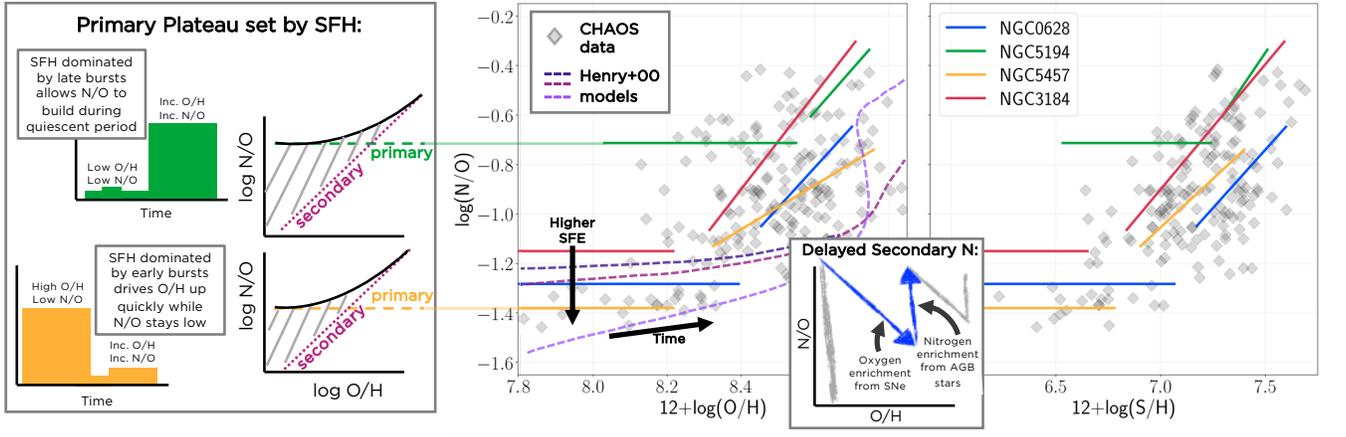}  
  \end{tabular}
  \caption{ 
  Schematic of how a galaxy's star formation history affects its overall N/O-O/H relationship.
  {\it Right:} average N/O versus O/H and S/H trends for the CHAOS sample.
  Secondary N slopes relative to S/H are similar for all four galaxies, however, 
  the trends are offset from one another both vertically and horizontally. 
  {\it Left:} two simple star formation histories and their resulting N/O-O/H trends are shown.
  On top, a star formation history with low star formation efficiency at early times allows 
  N/O to build up, raising the primary plateau and shifting the transition to secondary 
  dominance to the right.
  On the bottom, a star formation history with high star formation rate at early times 
  produces significant oxygen ahead of nitrogen production, setting a lower primary N/O 
  plateau and O/H transition to secondary N.
  In addition to these differences, the lower-right inset plot shows that the N/O ratios of 
  individual \ion{H}{2} regions are sensitive to time since the onset of the most recent burst,
  driving scatter from the average relationships.}
  \label{fig14}
\end{figure*}  


\subsection{Offsets Between Individual Galaxies}
A schematic of nitrogen production for the CHAOS galaxies is shown in Figure~\ref{fig14}.
The radial gradient fits to the N/O, O/H, and S/H relationships are combined to produce the
plotted N/O versus O/H relationship (middle panel) and N/O versus S/H (right panel) for each galaxy.
The progressively increasing N/O values at smaller galactocentric distance correspond to 
increasing O/H abundance, as is expected for secondary N production.  
This results in parallel secondary N/O slopes for the N/O--S/H trends in Figure~\ref{fig14},
and similar slopes in the N/O--O/H relationship for the three non-interacting galaxies.
However, the individual relationships are distinct in two ways.
First, each galaxy has a different primary plateau level, indicating large variations in 
their star formation histories, and, second, a different O/H transition value for when 
secondary N becomes important and turns the N/O curve upwards. 

\citet{henry00} found that chemical evolution models differing only by 
their assumed star formation efficiencies (SFEs) produced a range of primary N/O plateaus.  
We illustrate the effect of varying the SFE by over-plotting the \citet{henry00} 
constant SFR models, where efficiency has been varied by a factor of 25, 
on our N/O versus O/H data in Figure~\ref{fig14}.
For low SFRs, the build-up of oxygen is slow and on the order of the lag time 
before intermediate-mass stars begin ejecting nitrogen. 
This allows a high N/O plateau to be established at low oxygen abundances (darkest purple curve).
On the other hand, high star formation rates early in the star formation history (SFH) 
form a large number of massive stars that produce greater levels of oxygen ahead of 
N enrichment, establishing a lower plateau (lightest purple curve) and shifting the 
entire N/O--O/H trend in Figure~\ref{fig14} to the right towards greater O/H.
In between these scenarios, continuous star formation with roughly 250 Myr between bursts
will result in N and O increasing in lockstep, dependent on the elemental yields.
The coupling of the N/O plateau with galaxy SFH is also reported by 
cosmological hydrodynamical simulations of individual regions within 
spatially-resolved galaxies \citep{vincenzo18a}.
In these simulations, asymptotic giant branch (AGB) stars contribute significant
N at low O/H, but the exact value of the primary N/O plateau will vary 
from galaxy to galaxy according to the relative contributions from SNe and AGB 
stars, as determined by their galaxy formation time and SFH.

On the left-hand side of Figure~\ref{fig14}, we extend the highest N/O plateau 
from NGC~5194 (green) and the lowest N/O plateau from NGC~5457 (yellow).
Based on the above discussion, for the low N/O plateau of NGC~5457, we can put forth 
a star formation history scenario in which the star formation efficiency was high 
early in the galaxy's evolution, allowing oxygen to build up from many bursts of star 
formation before nitrogen was returned from longer-lived intermediate mass stars. 
Due to the higher level of nucleosynthetic products from massive stars, 
contributions from secondary nitrogen production may dominate 
over primary nitrogen production at relatively low O/H and S/H values.
On the other hand, the high N/O plateau of NGC~5194 could be due 
to a star formation history in which low star formation efficiency at early 
times allows nitrogen production, although delayed, to keep pace with oxygen 
and sulfur production and enrich the ISM. 
Here we assume low star formation efficiency to mean either constant, low star 
formation rates or long quiescent periods between bursts.
In this scenario, primary nitrogen production is the dominant mechanism until 
the galaxy reaches relatively high O/H.
Note, however, that this is a very simplistic model where N/O is changing monotonically;
in a hierarchical galaxy building scenario that may not be true.

In summary, the primary N/O plateau sensitively probes the SFH of a galaxy, 
rather than being set by the ratio of N to O yields, and explains the large range 
of plateau levels observed for spiral galaxies.
When this offset is accounted for, the N/O plateau then informs the primary
N production yields and the universal N/O gradient (see Figure~\ref{fig13})
is a direct probe of the secondary N yields of intermediate mass stars.


\begin{figure*}
\epsscale{1.0}
  \centering
  \begin{tabular}{l}
  \includegraphics[scale = 0.275, trim = 0mm 0mm 0mm 0mm, clip]{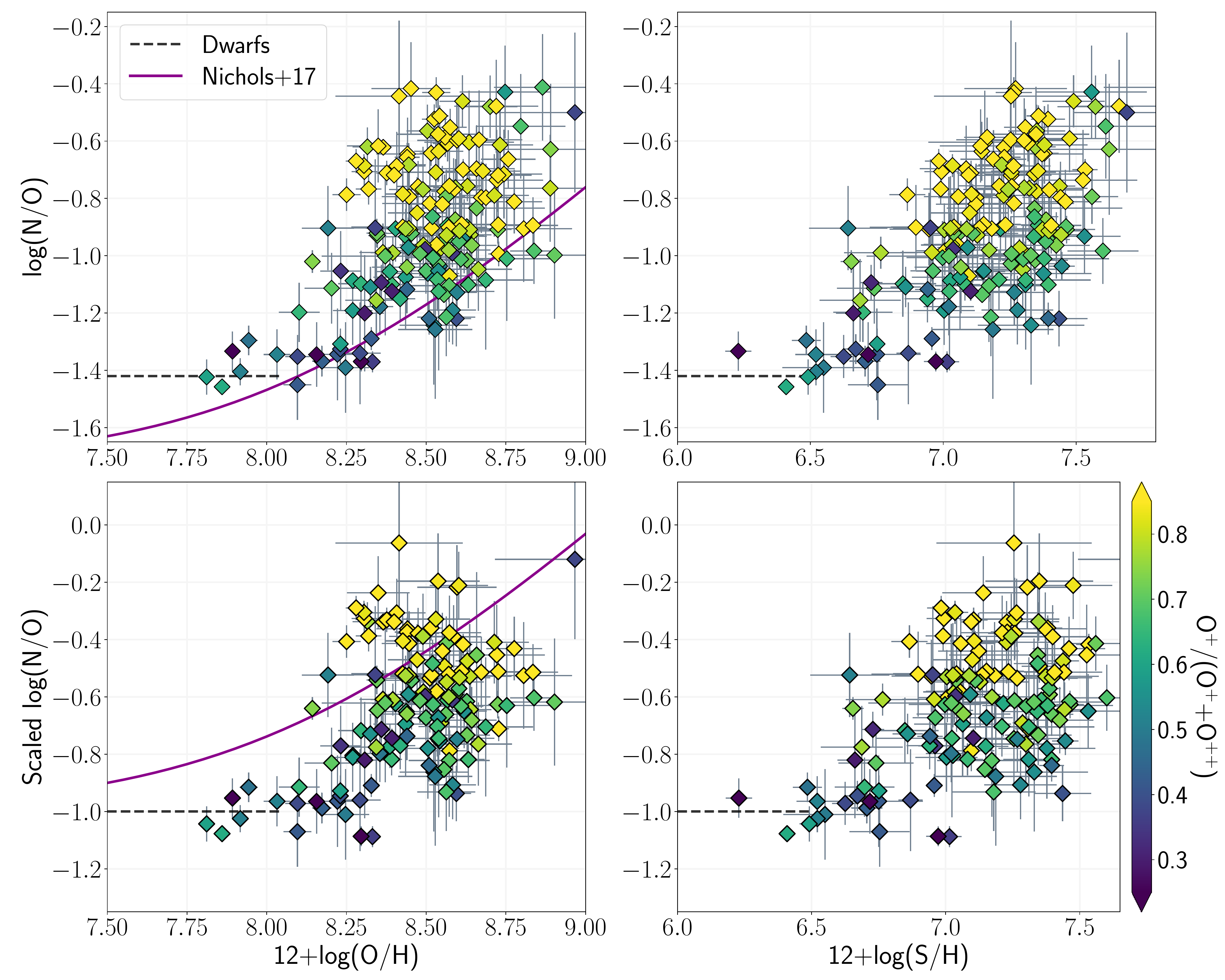}
  \end{tabular}
  \caption{ 
  {\it Top:} N/O relationships for the CHAOS sample relative to O/H ({\it left})
  and S/H ({\it right}).
  {\it Bottom:} scaled N/O trends relative to O/H ({\it left}) and S/H ({\it right}),
  where the differences in the star formation histories of individual galaxies are 
  removed by vertically shifting their primary N/O plateaus to align with log(N/O) $=-1.0$.
  For reference, the empirical stellar N/O--O/H trend from \citet{nicholls17} and 
  metal-poor N/O plateau for local dwarfs \citep{berg19} are also plotted.
  All plots are color-coded by the O$^+$/O ratio, or low-ionization fraction, and show
  a persistent trend of increasing N/O with O$^+$/O regardless of the N/O normalization.}
  \label{fig15}
\end{figure*}  


\subsection{The Scatter in the \\
N/O--O/H Relationship}
In Figure~\ref{fig13} we plotted the N/O--O/H trend of the CHAOS galaxies and 
found large observed scatter in N/O for a given O/H.
Given the tight correlations measured for the CHAOS N/O radial gradients 
(see Table~\ref{tbl2}), this scatter seems to be real.
Previous works have suggested that some of this scatter may be due to the time-dependent
nature of N/O production \citep[i.e., a N/O ``clock";][]{garnett90,pilyugin99,henry06}.
A directly observable effect of an aging ionizing stellar population
is an increasing fraction of low- to high-ionization gas in the \ion{H}{2} region
\citep[see, for example, how the shape of the ionizing continuum changes with age in][]{chisholm19}.

In Figure~\ref{fig15} we reproduce the N/O--O/H and N/O--S/H trends, color-coded
by the O$^+$/O ratio, or low-ionization fraction.
Interestingly, the overall trend of increasing N/O seems to be ordered
by ionization or age.
In the bottom panels of Figure~\ref{fig15}, we scale N/O (as was done in
Section~4.3) by shifting the vertical offsets in order to remove differences
in individual primary N/O plateaus and SFHs, yet the overall trend of
increasing N/O ordered by ionization remains.
Nearly all of the CHAOS points now have N/O abundances that are lower
relative to the scaled average stellar relationship of \citet{nicholls17},
suggesting that the physics of a recent burst of star formation has the
effect of shifting the N/O abundances downward, as expected for a recent
injection of newly synthesized oxygen.
The regions with the lowest N/O also have high ionization.
However, the standard N/O clock assumes regions with high N/O ratios
have experienced a burst of star formation followed by a long quiescent period 
that allowed their gas to be enriched with N from slow-evolving stars after a few 100 Myrs.
Given the fact that typical \ion{H}{2} regions are younger than $\sim10$ Myr,
the simple delayed-release N clock hypothesis fails to explain our observed
spread in N/O at a given O/H.

Alternatively, \citet{coziol99} suggested high N/O ratios in starburst
nucleus galaxies could result if N production occurs from a different, 
older population of intermediate-mass stars, such as would result from 
a sequence of bursts of star formation.
Similarly, \citet{berg19} used chemical evolution models of dwarf galaxies to 
show that N/O was elevated in regions experiencing an extended duration of star 
formation (continuous star formation) up to 0.4 Gyr.
Then, the overall effect of observing a large sample of \ion{H}{2} regions 
with a range of {\it luminosity-weighted} average stellar population ages
may be to produce the vertical spread in N/O at a given O/H seen in Figure~\ref{fig15}.

Perhaps another reason for the increased scatter of the N/O--O/H trends
relative to the N/O--$R_g$ relationships
is the possibility that N production is (or behaves as) a secondary function of the 
carbon abundance, rather than the typically assumed oxygen abundance \citep{henry00}.
Recently, \citet{groh19} investigated grids of stellar models at very low metallicities
and found that the ratio between nitrogen and carbon abundances (N/C) 
remains generally unchanged for non-rotating stellar models during their main sequence phase. 
However, the N/C production can increase by as much as 10--20$\times$ in rotating models 
at the end of the main sequence.
Thus, variations in stellar rotation speeds of different burst populations 
could result in significant effects on setting the low-metallicity stage.
Additionally, \citet{berg19} showed differential outflows of interstellar medium gas 
can affect the primary C/O and N/O ratios. 
Since O and S are produced on different timescales than N, newly synthesized O and S
may be preferentially lost in SNe winds and these outflows may have a 
greater probability of escape in the outer parts of the disk. \looseness=-2

At higher metallicities, where the effects of stellar winds become more important,
other authors have suggested that Wolf-Rayet stars can expel significant amounts 
of N resulting in local regions of N/O enrichment.
For the CHAOS sample, however, we do not find any correlation in the N/O 
dispersion with the Wolf-Rayet features sometimes seen in the optical spectra.

Another hypothesis is that the dispersion in N/O could be explained 
if we are consistently underestimating the O/H abundance in low-ionization nebula.
We have tested this hypothesis by looking at the offset in O/H abundance from the radial 
gradients relative to the secondary N/O radial gradient offsets and find some evidence of an 
anti-correlation, but it cannot explain all of the dispersion observed in N/O.

In summary, while we have observed a universal N/O gradient for the CHAOS galaxies
that seems to be tied to the nucleosynthetic yields of N, we also observe a large
dispersion when plotted relative to O/H. 
We have discussed several possible scenarios that could contribute to the N/O--O/H scatter, 
including extended star formation periods, differential outflows, and a secondary 
dependence on carbon abundance, but the importance of these contributions has not yet been determined.
At this time, the source of the scatter in the N/O--O/H relationship remains an open
question, but with several promising possibilities for future study.


\begin{deluxetable*}{llccccc}[H]
\tablecaption{Linear Fits to CHAOS Gradients}
\tablehead{
\CH{$y$}  & \CH{$x$}  & \CH{Galaxy} & \CH{$\#$ Reg.} & \CH{Equation} & \CH{$\sigma_{\rm{int.}}$} & \CH{$\sigma_{\rm{tot.}}$}}
\startdata
12+log(O/H) (dex)&$R_g$ ($R_{25}^{-1}$)&{NGC~0628} & 45 & $y=(8.71\pm0.06) - (0.40\pm0.11)\times x$ & 0.12 & 0.13 \\
                &                     & {NGC~5194} & 28 & $y=(8.75\pm0.09) - (0.27\pm0.15)\times x$ & 0.07 & 0.10 \\
                &                     & {NGC~5457} & 72 & $y=(8.78\pm0.04) - (0.90\pm0.07)\times x$ & 0.10 & 0.11 \\
                &                     & {NGC~3184} & 30 & $y=(8.74\pm0.16) - (0.48\pm0.28)\times x$ & 0.14 & 0.16 \\
\vspace{-1ex} \\              
                & $R_g$ ($R_e^{-1}$)  & {NGC~0628} & 45 & $y=(8.70\pm0.06) - (0.11\pm0.03)\times x$ & 0.12 & 0.13 \\
                &                     & {NGC~5194} & 28 & $y=(8.67\pm0.08) - (0.07\pm0.04)\times x$ & 0.07 & 0.10 \\
                &                     & {NGC~5457} & 72 & $y=(8.75\pm0.03) - (0.20\pm0.02)\times x$ & 0.10 & 0.11 \\
                &                     & {NGC~3184} & 30 & $y=(8.71\pm0.15) - (0.18\pm0.10)\times x$ & 0.14 & 0.16 \\
\vspace{-1ex} \\  
12+log(S/H) (dex) & $R_g$ ($R_e^{-1}$)& {NGC~0628} & 45 & $y=(7.60\pm0.06) - (0.19\pm0.03)\times x$ & 0.12 & 0.13 \\
                &                     & {NGC~5194} & 28 & $y=(7.51\pm0.11) - (0.10\pm0.05)\times x$ & 0.07 & 0.12 \\
                &                     & {NGC~5457} & 72 & $y=(7.40\pm0.05) - (0.23\pm0.03)\times x$ & 0.18 & 0.19 \\
                &                     & {NGC~3184} & 30 & $y=(7.59\pm0.15) - (0.34\pm0.11)\times x$ & 0.11 & 0.13 \\
\vspace{-1ex} \\          
log(N/O) (dex)  &$R_g$ ($R_{25}^{-1}$)& {NGC~0628} & 59 & $y=(-0.64\pm0.04) - (0.61\pm0.06)\times x$ & 0.10	& 0.11	\\
                &                     & {NGC~5194} & 28 & $y=(-0.34\pm0.09) - (0.44\pm0.16)\times x$ & 0.05	& 0.08	\\
                &                     & {NGC~5457} & 72 & $y=(-0.73\pm0.03) - (0.81\pm0.06)\times x$ & 0.07	& 0.10	\\
                &                     & {NGC~3184} & 30 & $y=(-0.30\pm0.13) - (0.83\pm0.22)\times x$ & 0.04	& 0.08  \\
\vspace{-1ex} \\          
                & $R_g$ ($R_e^{-1}$)  & {NGC~0628} & 59 & $y=(-0.65\pm0.03) - (0.18\pm0.02)\times x$ & 0.10	& 0.11	\\
                &                     & {NGC~5194} & 28 & $y=(-0.34\pm0.09) - (0.12\pm0.04)\times x$ & 0.05	& 0.08	\\
                &                     & {NGC~5457} & 72 & $y=(-0.74\pm0.03) - (0.18\pm0.01)\times x$ & 0.08	& 0.10	\\
                &                     & {NGC~3184} & 30 & $y=(-0.30\pm0.12) - (0.35\pm0.09)\times x$ & 0.05	& 0.08  \\
\vspace{-1ex} \\
log(N/O)$_{prim.}$ (dex)&             & {NGC~0628} & 11 & $y=-1.28$                                  &      & 0.13  \\
                &                     & {NGC~5194} & 4  & $y=-0.71$                                  &      & 0.03  \\
                &                     & {NGC~5457} & 15 & $y=-1.38$                                  &      & 0.13  \\
                &                     & {NGC~3184} & 0  & $y=-${\it 1.15}                            &      & \nodata \\
\vspace{-1ex} \\          
log(N/O)$_{sec.}$ (dex)&$R_g$ ($R_e^{-1}$)& 
                                        {NGC~0628} & 38 & $y=(-0.43\pm0.05) - (0.34\pm0.04)\times x$ & 0.06	& 0.07	\\
                &                     & {NGC~5194} & 20 & $y=(-0.27\pm0.18) - (0.17\pm0.11)\times x$ & 0.07	& 0.09	\\
                &                     & {NGC~5457} & 45 & $y=(-0.58\pm0.07) - (0.30\pm0.05)\times x$ & 0.06	& 0.08	\\
                &                     & {NGC~3184} & 30 & $y=(-0.30\pm0.12) - (0.35\pm0.09)\times x$ & 0.05	& 0.08  \\
\vspace{-1ex} \\
Scaled &&&&&& \\
log(N/O)$_{sec.}$ (dex)& $R_g$ ($R_e^{-1}$)& 
                                        {All Four} & 133 & $y=(-0.15\pm0.03) - (0.36\pm0.02)\times x$& 0.05	& 0.09	\\
                &                     &{Non-Inter.}& 113 & $y=(-0.16\pm0.03) - (0.34\pm0.02)\times x$& 0.05	& 0.08	
\enddata
\tablecomments{
Linear fits to trends in abundance versus radius for the four CHAOS galaxies.
The fits are determined using the Bayesian linear mixture model implemented in
the {\sc linmix} python code, which fits data with uncertainties on two variables, 
including explicit treatment of intrinsic scatter.
The $y$ and $x$ variables are given in the first two columns, with the number of 
associated \ion{H}{2} regions used in the fit listed in Column 4.
The resulting best fit is given in Column 5, with uncertainties on both the 
slope and $y-$intercept.
Columns 6 and 7 list the intrinsic and total uncertainties, $\sigma_{tot.}$ and $\sigma_{int.}$.
Note that the primary log(N/O) value given for NGC~3184 is italicized to indicate
that this quantity is an estimated value from the extrapolated secondary fit,
and not a measurement. }
\label{tbl2}
\end{deluxetable*}




\section {CONCLUSIONS}
This work is the fourth paper in a series presenting the on-going
results of the CHemical Abundances of Spirals survey \citep[CHAOS][]{berg15},
a project that is building a large database of direct abundance measurements 
spanning a large range in physical conditions in \ion{H}{2} regions across the
disks of nearby spiral galaxies.
Previous results for NGCC~628, NGC~5194, and NGC~5457 have been 
reported individually in papers {\sc i}--{\sc iii}.
Here we present new LBT/MODS spectra of 52 \ion{H}{2} regions in NGC~3184 to 
amass a high-quality, coherent sample of 175 direct-abundances from the 
first four CHAOS galaxies. \looseness = -2

Taking advantage of the direct $T_e$ measurements from one or more
auroral line detections in 190 individual \ion{H}{2} regions,
we confirm our previous results that $T_e$[\ion{S}{3}] and 
$T_e$[\ion{N}{2}] provide robust measures of electron temperature 
in the metal-rich \ion{H}{2} regions typical of spiral galaxies.
Specifically, the $T_e$[\ion{S}{3}]--$T_e$[\ion{N}{2}] trend, which characterizes
the intermediate- to low-ionization zone temperatures, is especially
tight for low-ionization \ion{H}{2} regions (low $F_{\lambda5007}/F_{\lambda3727}$)
with temperatures of $T_e \lesssim 8\times10^3$ K.
Unsurprisingly, we also find that the $T_e$[\ion{O}{3}]--$T_e$[\ion{S}{3}]
relationship is tightly correlated for high-ionization \ion{H}{2} regions
(high $F_{\lambda5007}/F_{\lambda3727}$).
Given the observed dichotomy in temperature dispersions with ionization of 
the nebulae, we recommend new ionization-based temperature priorities 
and apply them to abundance determinations for the four CHAOS galaxies.

Prioritizing temperatures derived from [\ion{O}{3}], [\ion{S}{3}] or [\ion{N}{2}]
depending on the average ionization of the observed nebula, we measure the relative 
and absolute abundance trends of O, N, S, Ar, and Ne for the CHAOS sample.
While the average $\alpha$/O abundances of the CHAOS sample are consistent within
the uncertainties of flat trends, we find evidence of systematic offsets that further depend 
on ionization and will likely require more sophisticated ionization correction factors to correct.
For O/H, we examine gradients normalized to both the isophotal radius ($R_{25}$) and the 
effective radius ($R_e$).
In contrast to some recent empirical abundance studies, we do not find a universal 
direct-O/H gradient when radius is plotted relative $R_e$, but rather measure 
unique slopes ranging from $-0.07$ to $-0.20$ dex/$R_e$.

Similarly, we examine the N/O gradient of our sample using both $R_{25}$ and $R_e$.
While each galaxy in our sample has a unique zero point offset, interpreted here
as different primary N/O plateaus set by differences in their star formation histories,
the secondary N/O slopes all appear to be the same.
We, therefore, determine the first measurement of a universal N/O gradient of 
$\alpha_{\rm N/O}=-0.33$ dex/$R_e$ for $R_g/R_e < 2.0$, where N is dominated by secondary 
production, and which can be used to constrain stellar yields.

As expected for two alpha elements, we find similar gradients for S/H and O/H for the CHAOS galaxies.
These trends suggest S/H can serve as a useful direct abundance diagnostic in the absence of O/H,
such as data sets lacking the blue wavelength coverage of [\ion{O}{3}] \W4363.
However, direct S/H abundances will generally be significantly more uncertain than direct
O/H abundances owing to the often large sulfur ICF uncertainties.
Given that the observable ionic states of S probe lower ionization and excitation energies
than O, S/H might be more appropriate for characterizing abundances in the
moderate- to metal-rich \ion{H}{2} regions of spiral galaxies.
Further work is needed to better constrain S ICFs and quantify their uncertainties
in order to improve S/H abundance determinations.


\acknowledgments
DAB is supported by the US National Science Foundation Grant AST-1715284.
We are grateful to the referee for detailed comments and thoughtful suggestions 
that greatly improved the scope and clarity of this paper.

This paper uses data taken with the MODS spectrographs built with funding from 
NSF grant AST-9987045 and the NSF Telescope System Instrumentation Program (TSIP), 
with additional funds from the Ohio Board of Regents and the Ohio State University Office of Research. 
This paper made use of the modsIDL spectral data reduction pipeline 
developed by KVC in part with funds provided by NSF Grant AST-1108693.
This work was based in part on observations made with the Large Binocular Telescope (LBT). 
The LBT is an international collaboration among institutions in the United States, Italy and Germany. 
The LBT Corporation partners are: the University of Arizona on behalf of the Arizona university system; 
the Istituto Nazionale di Astrofisica, Italy; the LBT Beteiligungsgesellschaft, Germany, representing the 
Max Planck Society, the Astrophysical Institute Potsdam, and Heidelberg University; the Ohio State 
University; and the Research Corporation, on behalf of the University of Notre Dame, the University 
of Minnesota, and the University of Virginia. 

We are grateful to D. Fanning, J.\,X. Prochaska, J. Hennawi, C. Markwardt, and M. Williams, and others 
who have developed the IDL libraries of which we have made use: coyote graphics, XIDL, idlutils, MPFIT, MPFITXY, and impro.  

This research has made use of the NASA/IPAC Extragalactic Database (NED) which is 
operated by the Jet Propulsion Laboratory, California Institute of Technology, 
under contract with the National Aeronautics and Space Administration.

\clearpage


\appendix


\section{CHAOS {\sc iv}: NGC~3184 Measurements}\label{sec:A1}


In Tables~\ref{tbl3}--\ref{tbl5} we present details for the CHAOS optical MODS/LBT spectroscopic observations 
of NGC~3184 used in this work, the measured emission line intensities, and the calculated 
ionic and total abundances.


\begin{deluxetable*}{lcccccc|c|c|c|c|c|c|c}[H]
	\tabletypesize{\scriptsize}
	\setlength{\tabcolsep}{0pt}
	\tablecaption{NGC\,3184 MODS/LBT Observations}
	\tablewidth{0pt}
	\tablehead{
\CH{H \ii} 				& \CH{R.A.}	& \CH{Dec.}	& \CH{R$_g$}	  & \CH{R/R$_{25}$} 	& \CH{R$_g$}	& \CH{Offset}	& \multicolumn{5}{c}{Auroral Line Detections} 					& \CH{Wolf}\\ \cline{8-12} 
\CH{Region}			& \CH{(2000)} 	& \CH{(2000)} 	& \CH{(arcsec)}	  & \CH{}			& \CH{(kpc)} & \CH{(arcsec)} & \CH{[O \iii]} & \CH{[N \ii]} & \CH{[S \iii]} & \CH{[O \ii]} & \CH{[S \ii]} 	& \CH{Rayet}}
\startdata																			
Total Detections:		& {}			& {}			& {}	 	 & {}		& {}		& {}		 	& 5		& 20		& 16		& 36		& 42		& 6		\\				
\hline
NGC3184+2.7-0.5		& 10:18:17.0	& 41:25:26.52 	& 2.10	 & 0.009	& 0.12	& +2.7,-0.5	&		& \CM	& 		& \CM	&		&		\\	
NGC3184+24.4-11.6		& 10:18:19.0	& 41:25:15.42	& 27.11	 & 0.122	& 1.54	& +24.4,-11.6	&		&		&		&		&		&		\\	
NGC3184+27.0-10.7		& 10:18:19.2	& 41:25:16.27	& 29.29	 & 0.132	& 1.66	& +27.0,-10.7	&		&		&		& 		&		&		\\	
NGC3184-2.7-47.5		& 10:18:16.6	& 41:24:39.55	& 47.18	 & 0.213	& 2.68	& -2.7,-47.5	&		&		&		& 		& \CM  	&		\\	

NGC3184-6.4-48.7		& 10:18:16.2	& 41:24:38.34	& 48.81	 & 0.220	& 2.77	 & -6.4,-48.7	&		&		&		& 		&		&		\\ 	
NGC3184-11.3-50.1 		& 10:18:15.8	& 41:24:36.94 	& 51.22	 & 0.231	& 2.91	 & -11.3,-50.1	&		&		&		& \CM	& \CM	&		\\ 	
NGC3184-1.1+60.6		& 10:18:16.7	& 41:26:27.58	& 61.02	 & 0.275	& 3.46	 & -1.1,+60.6	&		&		&		&		& \CM	&		\\
\enddata
	\label{tblA1}
	\tablecomments{Observing logs for H\ii\ regions observed in NGC 3184 using MODS on the LBT on the UT dates of March 24, 2012 and January 6, 2013. 
	Each field was observed over an integrated exposure time of 1200s on clear nights, with, on average $\sim$1\farcs00 seeing, and airmasses less than 1.3. 
	Slit ID, composed of the galaxy name and the offset in R.A. and Dec., in arcseconds, from the central position listed in Table \ref{tbl1} is listed in Column 1. 
	The right ascension and declination of the individual H\ii\ regions are given in units of hours, minutes, and seconds, and degrees, arcminutes, and arcseconds respectively in columns 2 and 3.  
	The de-projected distances of H\ii\ regions from the center of the galaxy in arcseconds, fraction of R$_{25}$, and in kpc are listed in the Columns 4-6.  
	Columns 7-11 highlight which regions have [O~\iii] $\lambda$4363, [N\,\ii] $\lambda$5755, [S\,\iii] $\lambda$6312, [O~\ii] $\lambda\lambda7320,7330$ 
	and [S\,\ii] $\lambda\lambda$4068,4076 auroral lines detections at the 3$\sigma$ significance level.  
	Finally, column 12 indicates which H\ii\ regions have Wolf-Rayet feature detections.}
	\label{tbl3}
\end{deluxetable*}

\clearpage

\LongTables
\begin{deluxetable*}{lccccccc}
\tabletypesize{\scriptsize}
\setlength{\tabcolsep}{0.03in} 
\tablecaption{Emission-Line Intensities and Equivalent Widths for LBT Observations of \ion{H}{2} regions in NGC~3184}
\tablewidth{0pt}
\tablehead{
\CH{} & \multicolumn{7}{c}{$I(\lambda)/I(\mbox{H}\beta)$} }
\startdata
{Ion}              & +2.7-0.5                & +24.4-11.6                & +27.0-10.7                & -2.7-47.5                 & -6.4-48.7                 & -11.3-50.1                & -1.1+60.6                 \\
\hline
{H14~\W3721}       &  0.015$\pm$0.001         &  0.018$\pm$0.004         &  0.024$\pm$0.002         &  0.016$\pm$0.002         &  0.019$\pm$0.001         &  0.021$\pm$0.001         &  0.009$\pm$0.001         \\
{[O~\ii]~\W3727}   &  0.374$\pm$0.007         &  0.844$\pm$0.019         &  0.839$\pm$0.014         &  0.745$\pm$0.137         &  0.672$\pm$0.014         &  0.678$\pm$0.011         &  0.566$\pm$0.010         \\
{H13~\W3734}       &  0.018$\pm$0.001         &  0.022$\pm$0.004         &  0.029$\pm$0.003         &  0.020$\pm$0.003         &  0.024$\pm$0.002         &  0.026$\pm$0.001         &  0.011$\pm$0.001         \\
{H12~\W3750}       &  0.003$\pm$0.005         &  0.021$\pm$0.004         &  0.030$\pm$0.008         &  0.050$\pm$0.007         &  0.024$\pm$0.005         &  0.043$\pm$0.002         &  0.005$\pm$0.002         \\
{H11~\W3770}       &  0.053$\pm$0.004         &  0.040$\pm$0.007         &  0.076$\pm$0.006         &  0.050$\pm$0.005         &  0.031$\pm$0.002         &  0.048$\pm$0.001         &  \nodata                 \\
{H10~\W3797}       &  0.039$\pm$0.003         &  0.048$\pm$0.010         &  0.062$\pm$0.006         &  0.043$\pm$0.006         &  0.051$\pm$0.003         &  0.056$\pm$0.003         &  0.025$\pm$0.002         \\
{He~I~\W3819}      &  0.012$\pm$0.002         &  0.036$\pm$0.007         &  0.017$\pm$0.006         &  0.040$\pm$0.007         &  0.005$\pm$0.002         &  0.010$\pm$0.001         &  \nodata                 \\
\enddata
\tablecomments{\ Reddening-corrected emission line intensities relative to H$\beta$. 
The last two rows list the $E(B-V)$ reddening and H$\beta$ line flux, given in units of
$10^{-16}$ ergs s$^{-1}$ cm$^{-2}$.
Note that previous CHAOS papers incorrectly reported the $E(B-V)$ quantities as $C_{H\beta}$, whereas 
the values listed here are actually $E(B-V)$.}
\label{tbl4}
\end{deluxetable*}


\clearpage 
\begin{deluxetable*}{lcccccc}
\tablecolumns{8}
\tablewidth{0pt}	
\tabletypesize{\scriptsize}
\setlength{\tabcolsep}{2pt} 
\tablecaption{Ionic and Total Abundances for LBT Observations of NGC~3184}
\startdata
\hline\hline \\
Property                   & +2.7-0.5           & -59.5-37.7                & +51.2+60.4                & -15.0-88.4                & -14.9-95.5                & -71.1+67.7                \\
\hline
$T_e$[S\ii]~(K)             &  \nodata           &  5700$\pm$400       &  8200$\pm$400       &  \nodata             &  14000$\pm$400       &  7600$\pm$400       \\
$T_e$[N\ii]~(K)             &  6000$\pm$1700       &  6300$\pm$1700       &  6500$\pm$1700       &  \nodata             &  6500$\pm$1700       &  \nodata             \\
$T_e$[O\ii]~(K)             &  8500$\pm$400       &  7800$\pm$400       &  12500$\pm$400       &  \nodata             &  6800$\pm$400       &  7600$\pm$400       \\
$T_e$[S\iii]~(K)            &  \nodata             &  \nodata             &  \nodata             &  6400$\pm$600       &  \nodata             &  6800$\pm$600       \\
T$_e$ [O~\iii] (K)          &  \nodata             &  \nodata             &  \nodata             &  \nodata             &  \nodata             &  \nodata             \\
$n_{e,m}$~(cm$^{-3}$)       &  210$\pm$10       &  60$\pm$10       &  50$\pm$10       &  29500$\pm$0       &  30$\pm$10       &  40$\pm$10       \\
\enddata 
\tablecomments{Electron temperatures and ionic and total abundances for objects with an 
[\ion{O}{3}] \W4363, [\ion{N}{2}] \W5755, or  [\ion{S}{3}] \W6312 
line signal to noise ratio of $3\sigma$ or greater. 
Electron temperatures for different ionization zones were calculated using the 
[\ion{O}{3}] ($\lambda4959 + \lambda5007$)/$\lambda4363$, [\ion{N}{2}] 
($\lambda6548 + \lambda6584$)/$\lambda5755$, or the [\ion{S}{3}] 
($\lambda9069 + \lambda9532$)/$\lambda6312$  diagnostic line ratios,
following the $T_e$ prioritization of \citetalias{berg15}.} 
\label{tbl5}
\end{deluxetable*}


\section{Re-derived Relative and Total Abudnances for CHAOS Galaxies}\label{sec:A2}

The gradients for NGC~5457 presented in \citetalias{croxall16} focused on 
abundances derived using $T_e$[\ion{O}{3}] for the purposed of comparing to 
previously reported trends in the literature that also used $T_e$[\ion{O}{3}] measurements.
In contrast, NGC~628 and NGC~5194 used the $T_e$ prioritization rules recommended by
\citetalias{berg15}.
Here, in Tables~\ref{tbl6}-\ref{tbl8}, we present recalculated ionic and total abundances 
for all three previously studied CHAOS galaxies: NGC~628, NGC~5194, and NGC~5457.
These updated values adopt the ionization-based temperature selection criteria 
proposed in this work in order
to form a uniform, coherent sample of 190 CHAOS \ion{H}{2} regions with 
direct electron temperature measurements.
This is the largest sample of its kind to date.

\clearpage


\begin{deluxetable*}{lccccc}
\tablewidth{0pt}	
\tabletypesize{\scriptsize}
\setlength{\tabcolsep}{2pt} 
\tablecaption{Updated Abundances for NGC~628 Using Ionization-Based $T_e$ Priorities:}
\tablehead{
\CH{H$\alpha$ Region}  & \CH{12+log(O/H)}	& \CH{log(N/O)} & \CH{log(S/O)} & \CH{log(Ar/O)} & \CH{log(Ne/O)}}
\startdata
    NGC628-35.9+57.7 & 8.52$\pm$0.04 & -0.64$\pm$0.05 & -1.14$\pm$0.07 & -1.90$\pm$0.06 &  \dots         \\ 
    NGC628+49.8+48.7 & 8.64$\pm$0.04 & -0.76$\pm$0.05 & -1.20$\pm$0.07 & -2.03$\pm$0.06 & -1.02$\pm$0.03 \\ 
    NGC628-73.1-27.3 & 8.37$\pm$0.02 & -0.62$\pm$0.02 & -1.14$\pm$0.03 & -2.03$\pm$0.05 & -0.73$\pm$0.04 \\ 
    NGC628-76.2+22.9 & 8.40$\pm$0.08 & -0.63$\pm$0.11 & -1.02$\pm$0.15 & -1.92$\pm$0.13 &  \dots         \\ 
    NGC628-36.8-73.4 & 8.43$\pm$0.04 & -0.65$\pm$0.05 & -1.21$\pm$0.07 & -2.15$\pm$0.06 &  \dots         \\ 
    NGC628+68.5+53.4 & 8.32$\pm$0.05 & -0.62$\pm$0.07 & -0.97$\pm$0.08 & -1.89$\pm$0.07 & -0.67$\pm$0.04 \\ 
    NGC628+81.6-32.3 & 8.62$\pm$0.03 & -0.70$\pm$0.04 & -1.08$\pm$0.06 & -1.89$\pm$0.06 &  \dots         \\ 
    NGC628-68.5+61.7 & 8.67$\pm$0.09 & -0.78$\pm$0.12 & -1.22$\pm$0.16 & -2.02$\pm$0.11 &  \dots         \\ 
    NGC628+76.9-49.6 & 8.74$\pm$0.14 & -0.74$\pm$0.17 & -1.21$\pm$0.23 & -1.96$\pm$0.16 &  \dots         \\ 
   NGC628-13.1+107.5 & 8.59$\pm$0.01 & -0.87$\pm$0.02 & -1.20$\pm$0.03 & -2.02$\pm$0.04 & -0.81$\pm$0.01 \\ 
   NGC628+53.5-104.0 & 8.47$\pm$0.13 & -0.78$\pm$0.17 & -1.23$\pm$0.23 & -2.09$\pm$0.18 &  \dots         \\ 
   NGC628-35.7+119.6 & 8.73$\pm$0.03 & -0.91$\pm$0.05 & -1.26$\pm$0.06 & -2.15$\pm$0.05 & -0.83$\pm$0.02 \\ 
   NGC628-20.3+124.6 & 8.63$\pm$0.01 & -0.90$\pm$0.02 & -1.17$\pm$0.03 & -2.07$\pm$0.04 & -0.87$\pm$0.01 \\ 
   NGC628-59.6-111.6 & 8.58$\pm$0.04 & -0.78$\pm$0.06 & -1.23$\pm$0.08 & -2.07$\pm$0.06 & -0.73$\pm$0.03 \\ 
   NGC628+61.2+113.5 & 8.73$\pm$0.03 & -0.99$\pm$0.04 & -1.42$\pm$0.06 & -2.13$\pm$0.05 & -1.03$\pm$0.03 \\ 
   NGC628+42.6-120.7 & 8.62$\pm$0.06 & -0.92$\pm$0.08 & -1.18$\pm$0.10 & -2.10$\pm$0.07 & -0.68$\pm$0.04 \\ 
   NGC628+131.9+18.5 & 8.56$\pm$0.02 & -0.89$\pm$0.03 & -1.20$\pm$0.03 & -2.19$\pm$0.05 & -0.65$\pm$0.02 \\ 
   NGC628+125.4-62.4 & 8.64$\pm$0.11 & -1.04$\pm$0.15 & -1.36$\pm$0.19 & -2.16$\pm$0.15 &  \dots         \\ 
   NGC628-130.9+71.8 & 8.55$\pm$0.02 & -0.97$\pm$0.03 & -1.26$\pm$0.04 & -2.13$\pm$0.05 & -0.82$\pm$0.02 \\ 
   NGC628+131.7-70.2 & 8.57$\pm$0.05 & -1.07$\pm$0.07 & -1.48$\pm$0.09 & -2.24$\pm$0.08 & -0.67$\pm$0.07 \\ 
   NGC628+151.0+22.3 & 8.61$\pm$0.07 & -0.93$\pm$0.12 & -1.09$\pm$0.14 & -2.10$\pm$0.10 & -0.87$\pm$0.09 \\ 
    NGC628-157.9-0.3 & 8.45$\pm$0.09 & -0.94$\pm$0.14 & -1.11$\pm$0.16 & -2.03$\pm$0.12 &  \dots         \\ 
   NGC628-24.5-155.6 & 8.62$\pm$0.06 & -1.01$\pm$0.09 & -1.29$\pm$0.12 & -2.17$\pm$0.10 & -0.75$\pm$0.10 \\ 
   NGC628-129.8+94.7 & 8.57$\pm$0.05 & -0.95$\pm$0.08 & -1.17$\pm$0.10 & -2.19$\pm$0.07 & -0.61$\pm$0.03 \\ 
   NGC628+140.3+82.0 & 8.35$\pm$0.03 & -0.93$\pm$0.04 & -1.29$\pm$0.06 & -2.16$\pm$0.06 & -0.99$\pm$0.04 \\ 
   NGC628-42.8-158.2 & 8.54$\pm$0.03 & -1.03$\pm$0.05 & -1.09$\pm$0.06 & -2.15$\pm$0.05 & -0.78$\pm$0.02 \\ 
   NGC628+147.9-71.8 & 8.55$\pm$0.11 & -1.03$\pm$0.15 & -1.29$\pm$0.20 & -2.19$\pm$0.15 & -0.75$\pm$0.19 \\ 
   NGC628+163.5+64.4 & 8.65$\pm$0.08 & -0.97$\pm$0.12 & -1.22$\pm$0.15 & -2.15$\pm$0.10 & -0.77$\pm$0.06 \\ 
    NGC628-4.5+185.6 & 8.39$\pm$0.03 & -1.10$\pm$0.06 & -1.31$\pm$0.07 & -2.21$\pm$0.05 & -0.84$\pm$0.02 \\ 
   NGC628+176.7-50.0 & 8.41$\pm$0.08 & -0.91$\pm$0.12 & -1.19$\pm$0.16 & -2.15$\pm$0.12 & -0.68$\pm$0.13 \\ 
   NGC628-76.2-171.8 & 8.63$\pm$0.05 & -1.10$\pm$0.08 & -1.32$\pm$0.10 & -2.30$\pm$0.09 & -0.52$\pm$0.09 \\ 
   NGC628+31.6-191.1 & 8.55$\pm$0.13 & -1.09$\pm$0.20 & -1.38$\pm$0.23 & -2.25$\pm$0.14 & -0.79$\pm$0.06 \\ 
    NGC628-200.6-4.2 & 8.53$\pm$0.10 & -1.07$\pm$0.17 & -1.15$\pm$0.19 & -2.11$\pm$0.12 & -0.85$\pm$0.11 \\ 
   NGC628-184.7+83.4 & 8.63$\pm$0.01 & -1.10$\pm$0.02 & -1.24$\pm$0.03 & -2.20$\pm$0.04 & -0.70$\pm$0.01 \\ 
   NGC628-206.5-25.7 & 8.60$\pm$0.05 & -1.22$\pm$0.11 & -1.16$\pm$0.12 & -2.14$\pm$0.10 & -0.61$\pm$0.09 \\ 
   NGC628-90.1+190.2 & 8.56$\pm$0.01 & -1.14$\pm$0.01 & -1.41$\pm$0.02 & -2.24$\pm$0.04 & -0.79$\pm$0.01 \\ 
  NGC628-168.2+150.8 & 8.27$\pm$0.01 & -1.09$\pm$0.02 & -1.24$\pm$0.02 & -2.23$\pm$0.04 & -0.76$\pm$0.02 \\ 
    NGC628+232.7+6.6 & 8.58$\pm$0.07 & -1.19$\pm$0.12 & -1.30$\pm$0.15 & -2.31$\pm$0.12 & -0.56$\pm$0.13 \\ 
    NGC628+237.6+3.0 & 8.58$\pm$0.10 & -1.23$\pm$0.16 & -1.39$\pm$0.20 & -2.36$\pm$0.16 & -0.57$\pm$0.16 \\ 
   NGC628+254.3-42.8 & 8.39$\pm$0.04 & -1.05$\pm$0.07 & -1.24$\pm$0.09 & -2.44$\pm$0.08 & -0.49$\pm$0.07 \\ 
   NGC628+252.1-92.1 & 8.24$\pm$0.06 & -1.05$\pm$0.14 & -1.26$\pm$0.09 & -2.17$\pm$0.10 & -0.61$\pm$0.10 \\ 
   NGC628+261.9-99.7 & 8.21$\pm$0.09 & -1.12$\pm$0.13 & -1.47$\pm$0.17 & -2.43$\pm$0.13 & -0.40$\pm$0.14 \\ 
  NGC628+265.2-102.2 & 8.10$\pm$0.07 & -1.20$\pm$0.11 & -1.41$\pm$0.13 & -2.26$\pm$0.11 &  \dots         \\ 
   NGC628+289.9-17.4 & 8.33$\pm$0.02 & -1.37$\pm$0.04 & -1.32$\pm$0.05 & -2.11$\pm$0.05 & -0.67$\pm$0.04 \\ 
   NGC628+298.4+12.3 & 8.30$\pm$0.02 & -1.37$\pm$0.04 & -1.32$\pm$0.05 & -1.99$\pm$0.05 & -0.64$\pm$0.03  
\enddata 
\label{tbl6}
\end{deluxetable*}

\clearpage

\begin{deluxetable*}{lccccc}
\tablewidth{0pt}	
\tabletypesize{\scriptsize}
\setlength{\tabcolsep}{2pt} 
\tablecaption{Updated Abundances for NGC~5194 Using Ionization-Based $T_e$ Priorities:}
\tablehead{
\CH{H$\alpha$ Region}  & \CH{12+log(O/H)}	& \CH{log(N/O)} & \CH{log(S/O)} & \CH{log(Ar/O)} & \CH{log(Ne/O)}}
\startdata
    NGC5194-4.3+63.3 & 8.73$\pm$0.14 & -0.43$\pm$0.16 & -1.19$\pm$0.24 & -2.56$\pm$0.21 &  \dots         \\ 
   NGC5194-33.2+58.0 & 8.89$\pm$0.16 & -0.42$\pm$0.19 & -1.02$\pm$0.28 & -2.06$\pm$0.24 &  \dots         \\ 
   NGC5194-62.2+50.3 & 8.79$\pm$0.13 & -0.55$\pm$0.17 & -1.19$\pm$0.24 & -2.08$\pm$0.20 &  \dots         \\ 
   NGC5194+75.5-28.7 & 8.47$\pm$0.12 & -0.42$\pm$0.15 & -1.18$\pm$0.21 & -2.04$\pm$0.17 &  \dots         \\ 
   NGC5194+96.1+16.8 & 8.70$\pm$0.09 & -0.48$\pm$0.11 & -1.13$\pm$0.16 & -2.06$\pm$0.14 &  \dots         \\ 
   NGC5194+91.0+69.0 & 8.72$\pm$0.13 & -0.48$\pm$0.16 & -1.07$\pm$0.21 & -1.87$\pm$0.15 &  \dots         \\ 
   NGC5194-86.5-79.4 & 8.54$\pm$0.04 & -0.43$\pm$0.05 & -1.27$\pm$0.08 & -2.24$\pm$0.08 &  \dots         \\ 
  NGC5194-22.5+122.8 & 8.77$\pm$0.15 & -0.58$\pm$0.20 & -1.25$\pm$0.27 & -2.15$\pm$0.22 &  \dots         \\ 
  NGC5194+112.7+37.7 & 8.61$\pm$0.08 & -0.46$\pm$0.10 & -1.12$\pm$0.14 & -2.06$\pm$0.12 &  \dots         \\ 
   NGC5194+76.6+96.3 & 8.77$\pm$0.10 & -0.63$\pm$0.14 & -1.36$\pm$0.18 & -2.30$\pm$0.16 &  \dots         \\ 
   NGC5194-97.0-78.4 & 8.52$\pm$0.03 & -0.52$\pm$0.05 & -1.13$\pm$0.07 & -1.98$\pm$0.06 &  \dots         \\ 
   NGC5194-3.0+131.9 & 8.64$\pm$0.11 & -0.69$\pm$0.14 & -1.55$\pm$0.19 & -2.19$\pm$0.14 & -0.70$\pm$0.12 \\ 
  NGC5194-57.2+118.2 & 8.53$\pm$0.08 & -0.63$\pm$0.10 & -1.38$\pm$0.14 & -2.17$\pm$0.11 & -0.76$\pm$0.11 \\ 
  NGC5194-78.9+107.4 & 8.87$\pm$0.15 & -0.76$\pm$0.19 & -1.56$\pm$0.25 & -2.27$\pm$0.21 &  \dots         \\ 
  NGC5194-82.0-102.7 & 8.59$\pm$0.12 & -0.59$\pm$0.15 & -1.47$\pm$0.21 & -2.68$\pm$0.18 &  \dots         \\ 
  NGC5194-66.6+122.9 & 8.69$\pm$0.09 & -0.80$\pm$0.12 & -1.56$\pm$0.16 & -2.09$\pm$0.13 &  \dots         \\ 
  NGC5194+56.8+126.5 & 8.68$\pm$0.14 & -0.60$\pm$0.17 & -1.42$\pm$0.23 & -2.28$\pm$0.20 &  \dots         \\ 
  NGC5194+30.8+139.0 & 8.75$\pm$0.09 & -0.66$\pm$0.10 & -1.57$\pm$0.15 & -2.37$\pm$0.11 & -0.69$\pm$0.10 \\ 
 NGC5194+104.1-105.5 & 8.56$\pm$0.04 & -0.60$\pm$0.06 & -1.24$\pm$0.08 & -2.04$\pm$0.07 &  \dots         \\ 
  NGC5194+98.1-113.8 & 8.54$\pm$0.03 & -0.51$\pm$0.04 & -1.19$\pm$0.05 & -2.10$\pm$0.05 & -0.86$\pm$0.04 \\ 
  NGC5194+71.2+135.9 & 8.56$\pm$0.05 & -0.59$\pm$0.06 & -1.41$\pm$0.09 & -2.24$\pm$0.08 & -0.78$\pm$0.08 \\ 
  NGC5194+83.4-133.1 & 8.57$\pm$0.06 & -0.55$\pm$0.07 & -1.26$\pm$0.10 & -1.95$\pm$0.08 &  \dots         \\ 
 NGC5194+109.9-121.4 & 8.51$\pm$0.07 & -0.57$\pm$0.09 & -1.16$\pm$0.12 & -2.08$\pm$0.10 & -0.70$\pm$0.10 \\ 
 NGC5194+112.2-126.6 & 8.75$\pm$0.12 & -0.72$\pm$0.16 & -1.40$\pm$0.22 & -2.17$\pm$0.18 &  \dots         \\ 
  NGC5194+150.6+99.0 & 8.67$\pm$0.11 & -0.70$\pm$0.15 & -1.31$\pm$0.19 & -2.11$\pm$0.14 &  \dots         \\ 
 NGC5194-159.5-116.4 & 8.73$\pm$0.09 & -0.72$\pm$0.11 & -1.47$\pm$0.16 & -2.14$\pm$0.12 &  \dots         \\ 
 NGC5194-135.4-181.4 & 8.61$\pm$0.07 & -0.77$\pm$0.10 & -1.30$\pm$0.13 & -2.18$\pm$0.10 & -0.65$\pm$0.08 \\ 
 NGC5194+114.5+230.8 & 8.46$\pm$0.06 & -0.69$\pm$0.08 & -1.25$\pm$0.11 & -2.08$\pm$0.08 & -0.83$\pm$0.06 
\enddata 
\label{tbl7}
\end{deluxetable*}

\clearpage

\begin{deluxetable*}{lccccc}
\tablewidth{10pt}	
\tabletypesize{\scriptsize}
\setlength{\tabcolsep}{2pt} 
\tablecaption{Updated Abundances for NGC~5457 Using Ionization-Based $T_e$ Priorities:}
\tablehead{
\CH{H$\alpha$ Region}  & \CH{12+log(O/H)}	& \CH{log(N/O)} & \CH{log(S/O)} & \CH{log(Ar/O)} & \CH{log(Ne/O)}}
\startdata
   NGC5457-75.0+29.3 & 8.65$\pm$0.14 & -0.62$\pm$0.16 & -1.34$\pm$0.22 & -2.21$\pm$0.16 &  \dots         \\ 
  NGC5457+22.1-102.1 & 8.73$\pm$0.11 & -0.80$\pm$0.15 & -1.49$\pm$0.20 & -2.28$\pm$0.16 & -0.49$\pm$0.20 \\ 
  NGC5457+47.9-103.2 & 8.79$\pm$0.09 & -0.82$\pm$0.11 & -1.32$\pm$0.15 & -2.07$\pm$0.13 &  \dots         \\ 
  NGC5457-12.0+139.0 & 8.53$\pm$0.13 & -0.71$\pm$0.17 & -1.28$\pm$0.22 & -2.21$\pm$0.18 & -0.58$\pm$0.21 \\ 
  NGC5457+138.9+30.6 & 8.50$\pm$0.07 & -0.77$\pm$0.09 & -1.24$\pm$0.11 & -2.05$\pm$0.08 & -0.61$\pm$0.06 \\ 
  NGC5457+134.4-58.8 & 8.62$\pm$0.10 & -0.91$\pm$0.13 & -1.66$\pm$0.17 & -2.36$\pm$0.13 & -0.66$\pm$0.10 \\ 
   NGC5457+164.6+9.9 & 8.60$\pm$0.03 & -0.95$\pm$0.05 & -1.22$\pm$0.05 & -2.12$\pm$0.05 & -0.88$\pm$0.01 \\ 
  NGC5457+89.3+149.7 & 8.83$\pm$0.16 & -0.92$\pm$0.22 & -1.64$\pm$0.26 & -2.35$\pm$0.19 & -0.74$\pm$0.16 \\ 
  NGC5457-70.2+162.2 & 8.62$\pm$0.13 & -0.90$\pm$0.18 & -1.24$\pm$0.22 & -2.13$\pm$0.14 & -1.06$\pm$0.07 \\ 
  NGC5457+166.4+86.3 & 8.42$\pm$0.08 & -0.78$\pm$0.11 & -1.35$\pm$0.14 & -2.21$\pm$0.10 &  \dots         \\ 
  NGC5457+177.2-42.8 & 8.45$\pm$0.06 & -0.91$\pm$0.09 & -1.46$\pm$0.11 & -2.22$\pm$0.08 & -0.95$\pm$0.05 \\ 
  NGC5457-159.9+89.6 & 8.58$\pm$0.08 & -0.77$\pm$0.11 & -1.19$\pm$0.15 & -2.03$\pm$0.11 & -0.61$\pm$0.09 \\ 
 NGC5457+133.1-126.8 & 8.58$\pm$0.14 & -0.91$\pm$0.17 & -1.70$\pm$0.22 & -2.42$\pm$0.16 & -0.72$\pm$0.12 \\ 
  NGC5457+177.2+76.1 & 8.61$\pm$0.09 & -0.89$\pm$0.12 & -1.44$\pm$0.16 & -2.18$\pm$0.11 & -0.97$\pm$0.07 \\ 
 NGC5457-120.2+146.9 & 8.57$\pm$0.06 & -0.80$\pm$0.10 & -1.02$\pm$0.12 & -2.04$\pm$0.08 & -0.97$\pm$0.05 \\ 
 NGC5457+130.2+157.4 & 8.55$\pm$0.10 & -0.82$\pm$0.13 & -1.42$\pm$0.17 & -2.20$\pm$0.12 & -0.75$\pm$0.08 \\ 
 NGC5457+129.2+161.7 & 8.30$\pm$0.04 & -0.71$\pm$0.05 & -1.31$\pm$0.07 & -2.17$\pm$0.07 & -1.05$\pm$0.06 \\ 
 NGC5457-145.1+146.8 & 8.87$\pm$0.08 & -1.00$\pm$0.12 & -1.25$\pm$0.16 & -2.17$\pm$0.13 & -0.77$\pm$0.13 \\ 
 NGC5457+103.5+192.6 & 8.46$\pm$0.08 & -0.85$\pm$0.11 & -1.35$\pm$0.16 & -2.11$\pm$0.13 & -0.98$\pm$0.14 \\ 
  NGC5457-205.4-98.2 & 8.57$\pm$0.09 & -0.90$\pm$0.13 & -1.17$\pm$0.17 & -2.28$\pm$0.11 & -1.05$\pm$0.08 \\ 
  NGC5457+17.3-235.4 & 8.62$\pm$0.07 & -1.02$\pm$0.12 & -1.32$\pm$0.14 & -2.26$\pm$0.09 & -0.97$\pm$0.04 \\ 
  NGC5457+36.8-233.4 & 8.45$\pm$0.03 & -0.99$\pm$0.06 & -1.50$\pm$0.06 & -2.27$\pm$0.06 & -0.86$\pm$0.06 \\ 
 NGC5457+139.0+200.7 & 8.53$\pm$0.08 & -0.96$\pm$0.10 & -1.48$\pm$0.14 & -2.26$\pm$0.11 & -0.75$\pm$0.13 \\ 
 NGC5457+189.2-136.3 & 8.58$\pm$0.02 & -0.99$\pm$0.05 & -1.59$\pm$0.05 & -2.27$\pm$0.05 & -0.81$\pm$0.01 \\ 
 NGC5457-183.9-179.0 & 8.68$\pm$0.14 & -1.08$\pm$0.23 & -1.36$\pm$0.25 & -2.25$\pm$0.15 & -0.67$\pm$0.07 \\ 
  NGC5457-249.4-51.3 & 8.51$\pm$0.06 & -0.90$\pm$0.08 & -1.38$\pm$0.11 & -2.28$\pm$0.09 & -0.54$\pm$0.08 \\ 
  NGC5457-250.8-52.0 & 8.59$\pm$0.11 & -0.94$\pm$0.15 & -1.45$\pm$0.20 & -2.46$\pm$0.16 & -0.28$\pm$0.16 \\ 
 NGC5457+225.6-124.1 & 8.49$\pm$0.05 & -1.05$\pm$0.08 & -1.53$\pm$0.10 & -2.35$\pm$0.07 & -0.81$\pm$0.03 \\ 
 NGC5457+117.9-235.0 & 8.30$\pm$0.07 & -1.10$\pm$0.11 & -1.45$\pm$0.14 & -2.24$\pm$0.11 & -0.95$\pm$0.11 \\ 
 NGC5457-208.0-180.7 & 8.45$\pm$0.10 & -0.92$\pm$0.17 & -1.34$\pm$0.19 & -2.29$\pm$0.11 & -0.65$\pm$0.05 \\ 
  NGC5457-12.3-271.1 & 8.49$\pm$0.08 & -0.99$\pm$0.10 & -1.49$\pm$0.14 & -2.06$\pm$0.12 & -0.99$\pm$0.12 \\ 
 NGC5457-200.3-193.6 & 8.60$\pm$0.06 & -1.13$\pm$0.11 & -1.33$\pm$0.12 & -2.32$\pm$0.10 & -0.72$\pm$0.10 \\ 
  NGC5457+96.7+266.9 & 8.49$\pm$0.08 & -0.96$\pm$0.11 & -1.46$\pm$0.14 & -2.32$\pm$0.09 & -0.70$\pm$0.06 \\ 
  NGC5457+67.5+277.0 & 8.52$\pm$0.05 & -1.04$\pm$0.10 & -1.37$\pm$0.11 & -2.27$\pm$0.07 & -0.73$\pm$0.05 \\ 
 NGC5457+252.2-109.8 & 8.53$\pm$0.08 & -1.01$\pm$0.13 & -1.52$\pm$0.15 & -2.48$\pm$0.10 & -0.77$\pm$0.06 \\ 
 NGC5457+254.6-107.2 & 8.50$\pm$0.01 & -0.98$\pm$0.02 & -1.46$\pm$0.04 & -2.18$\pm$0.04 & -0.77$\pm$0.01 \\ 
  NGC5457+281.4-71.8 & 8.42$\pm$0.05 & -1.15$\pm$0.08 & -1.48$\pm$0.09 & -2.28$\pm$0.08 & -0.89$\pm$0.07 \\ 
 NGC5457-243.0+159.6 & 8.49$\pm$0.06 & -1.00$\pm$0.09 & -1.23$\pm$0.12 & -2.20$\pm$0.11 & -0.92$\pm$0.12 \\ 
 NGC5457+249.3+201.9 & 8.42$\pm$0.09 & -1.06$\pm$0.16 & -1.34$\pm$0.18 & -2.31$\pm$0.10 & -0.71$\pm$0.05 \\ 
  NGC5457-297.7+87.1 & 8.45$\pm$0.10 & -1.04$\pm$0.15 & -1.38$\pm$0.19 & -2.25$\pm$0.13 & -0.93$\pm$0.11 \\ 
  NGC5457-309.4+56.9 & 8.35$\pm$0.03 & -0.92$\pm$0.05 & -1.31$\pm$0.06 & -2.41$\pm$0.07 & -0.78$\pm$0.06 \\ 
  NGC5457+354.1+71.2 & 8.51$\pm$0.10 & -1.23$\pm$0.19 & -1.17$\pm$0.20 & -2.18$\pm$0.11 & -0.70$\pm$0.03 \\ 
 NGC5457-164.9-333.9 & 8.53$\pm$0.03 & -1.23$\pm$0.08 & -1.33$\pm$0.07 & -2.27$\pm$0.06 & -0.89$\pm$0.05 \\ 
  NGC5457+360.9+75.3 & 8.51$\pm$0.02 & -1.22$\pm$0.05 & -1.11$\pm$0.05 & -2.16$\pm$0.05 & -0.69$\pm$0.03 \\ 
  NGC5457-377.9-64.9 & 8.52$\pm$0.06 & -1.08$\pm$0.10 & -1.33$\pm$0.11 & -2.33$\pm$0.08 & -0.62$\pm$0.07 \\ 
  NGC5457-99.6-388.0 & 8.39$\pm$0.01 & -1.12$\pm$0.03 & -1.29$\pm$0.04 & -2.08$\pm$0.04 & -0.71$\pm$0.01 \\ 
  NGC5457-397.4-71.7 & 8.33$\pm$0.04 & -1.06$\pm$0.07 & -1.46$\pm$0.08 & -2.64$\pm$0.07 & -0.60$\pm$0.07 \\ 
 NGC5457-226.9-366.4 & 8.28$\pm$0.06 & -1.20$\pm$0.12 & -1.28$\pm$0.13 & -2.22$\pm$0.08 & -0.93$\pm$0.05 \\ 
 NGC5457-405.5-157.7 & 8.14$\pm$0.02 & -1.02$\pm$0.04 & -1.49$\pm$0.04 & -2.26$\pm$0.05 & -0.84$\pm$0.04 \\ 
 NGC5457-345.5+273.8 & 8.35$\pm$0.04 & -1.17$\pm$0.09 & -1.33$\pm$0.09 & -2.44$\pm$0.09 & -0.53$\pm$0.07 \\ 
 NGC5457-410.3-206.3 & 8.32$\pm$0.09 & -1.14$\pm$0.14 & -1.62$\pm$0.16 & -2.31$\pm$0.11 & -0.94$\pm$0.08 \\ 
 NGC5457-371.1-280.0 & 8.33$\pm$0.03 & -1.11$\pm$0.07 & -1.48$\pm$0.06 & -2.36$\pm$0.05 & -0.67$\pm$0.01 \\ 
 NGC5457-368.3-285.6 & 8.45$\pm$0.02 & -1.12$\pm$0.05 & -1.51$\pm$0.05 & -2.41$\pm$0.05 & -0.70$\pm$0.01 \\ 
  NGC5457-455.7-55.8 & 8.18$\pm$0.03 & -1.37$\pm$0.05 & -1.47$\pm$0.06 & -2.43$\pm$0.06 & -0.61$\pm$0.05 \\ 
 NGC5457-392.0-270.1 & 8.36$\pm$0.02 & -1.09$\pm$0.06 & -1.63$\pm$0.04 & -2.27$\pm$0.04 & -0.70$\pm$0.01 \\ 
 NGC5457-414.1-253.6 & 8.28$\pm$0.03 & -1.15$\pm$0.10 & -1.60$\pm$0.06 & -2.15$\pm$0.07 & -0.67$\pm$0.06 \\ 
 NGC5457-464.7-131.0 & 8.16$\pm$0.01 & -1.34$\pm$0.05 & -1.53$\pm$0.02 & -2.46$\pm$0.05 & -0.63$\pm$0.02 \\ 
 NGC5457-466.1-128.2 & 8.01$\pm$0.04 & -1.34$\pm$0.08 & -1.50$\pm$0.08 & -2.39$\pm$0.07 & -0.69$\pm$0.06 \\ 
   NGC5457-479.7-3.9 & 8.15$\pm$0.01 & -0.90$\pm$0.07 & -1.61$\pm$0.02 & -2.53$\pm$0.05 & -0.70$\pm$0.02 \\ 
   NGC5457-481.4-0.5 & 7.95$\pm$0.03 & -1.30$\pm$0.05 & -1.46$\pm$0.06 & -2.47$\pm$0.06 & -0.57$\pm$0.05 \\ 
 NGC5457-453.8-191.8 & 8.24$\pm$0.06 & -1.38$\pm$0.13 & -1.68$\pm$0.14 & -2.60$\pm$0.07 & -0.64$\pm$0.02 \\ 
 NGC5457+331.9+401.0 & 8.23$\pm$0.01 & -1.33$\pm$0.05 & -1.56$\pm$0.04 & -2.38$\pm$0.05 & -0.69$\pm$0.03 \\ 
 NGC5457+324.5+415.8 & 8.23$\pm$0.02 & -1.31$\pm$0.05 & -1.48$\pm$0.03 & -2.49$\pm$0.06 & -0.70$\pm$0.04 \\ 
 NGC5457+315.3+434.4 & 8.33$\pm$0.01 & -1.29$\pm$0.04 & -1.37$\pm$0.02 & -2.42$\pm$0.05 & -0.62$\pm$0.04 \\ 
 NGC5457-540.5-149.9 & 7.89$\pm$0.01 & -1.33$\pm$0.07 & -1.66$\pm$0.05 & -2.26$\pm$0.05 & -0.63$\pm$0.03 \\ 
 NGC5457+509.5+264.1 & 8.29$\pm$0.06 & -1.34$\pm$0.17 & -1.43$\pm$0.08 & -2.27$\pm$0.08 & -0.64$\pm$0.02 \\ 
 NGC5457+266.0+534.1 & 8.18$\pm$0.03 & -1.37$\pm$0.05 & -1.47$\pm$0.06 & -2.43$\pm$0.06 & -0.61$\pm$0.05 \\ 
 NGC5457+667.9+174.1 & 8.16$\pm$0.02 & -1.34$\pm$0.11 & -1.43$\pm$0.05 & -2.17$\pm$0.05 & -0.72$\pm$0.02 \\ 
 NGC5457+650.1+270.7 & 8.09$\pm$0.04 & -1.35$\pm$0.08 & -1.47$\pm$0.06 &  \dots         & -0.58$\pm$0.10 \\ 
 NGC5457+692.1+272.9 & 8.10$\pm$0.04 & -1.45$\pm$0.12 & -1.35$\pm$0.12 & -2.33$\pm$0.09 & -0.67$\pm$0.07 \\ 
   NGC5457+1.0+885.8 & 7.86$\pm$0.01 & -1.46$\pm$0.02 & -1.45$\pm$0.02 & -2.69$\pm$0.06 & -0.68$\pm$0.05 \\ 
   NGC5457+6.6+886.3 & 7.92$\pm$0.01 & -1.40$\pm$0.05 & -1.40$\pm$0.02 & -2.65$\pm$0.05 & -0.67$\pm$0.03 \\ 
   NGC5457-8.5+886.7 & 7.81$\pm$0.02 & -1.42$\pm$0.06 & -1.32$\pm$0.04 & -2.79$\pm$0.10 & -0.68$\pm$0.10  
\enddata 
\label{tbl8}
\end{deluxetable*}


\section{CHAOS Surface Density Profiles}\label{sec:A3}

In order to test whether the abundance gradients in CHAOS galaxies correlate with 
their individual disk properties, we need to determine surface density properties
that correspond to our observed \ion{H}{2} regions.
We therefore examine the surface density profiles of the stellar mass, the gas mass,
and the SFR of our CHAOS sample. \looseness=-2

\subsection{Data and Profile Determinations}
Owing to the well-studied nature of the galaxies in our sample, there exists a
plethora of ancillary data to aid in this task.
Specifically, we use HERACLES CO(2--1) line integrated intensity (moment--0) maps 
\citep{leroy09} to trace the molecular gas, THINGS \ion{H}{1} 21cm line integrated 
intensity maps \citep{walter08} to trace the atomic gas, {\it Spitzer} IRAC 3.6$\mu$m 
images to trace stellar mass, and SFR surface density maps created in the $z=0$ 
Multi-wavelength Galaxy Synthesis project \citep[Z0MGS;][]{leroy19}.

The CO maps were converted into molecular gas surface density maps by assuming a 
standard Galactic CO-to-H$_2$ conversion factor of 
$\alpha_{\rm CO} = 4.35\rm\;M_\odot\,pc^{-2}\,(K\,km\,s^{-1})^{-1}$ 
\citep[including heavy element contribution;][]{bolatto13}, and a CO(2--1)/(1--0) 
line ratio of $R_{21} = 0.7$ \citep{leroy13,saintonge17}.
For the atomic gas surface density maps, \ion{H}{1} intensities were converted using
a standard conversion factor of $1.97\times10^{-2}\rm\;M_\odot\,pc^{-2}\,(K\,km\,s^{-1})^{-1}$, 
which includes heavy element contribution.
The stellar mass surface density distributions adopted a conversion factor of 
$420\rm\;M_\odot\,pc^{-2}\,(MJy\,sr^{-1})$, assuming a fixed mass-to-light ratio of 
$Y_{3.6} = 0.6\; M_{\odot}/L_{\odot,\,3.6}$ \citep{querejeta15}.
For all galaxies except NGC~5457, we also have dust-corrected IRAC 3.6$\mu$m images 
from S$^4$G \citep{sheth10,querejeta15}. 
The same conversion factors were used for these maps.
Finally, the SFR surface density maps were derived by combining background-subtracted, 
astrometry-matched, and resolution-matched GALEX FUV and WISE 24$\mu$m images, 
and converting the measured broad-band intensities to SFR surface density 
\citep{jarrett13,cluver17}.

Next, we built radial profiles from the mass and SFR surface density maps.
Using the galaxy parameters listed in Table~\ref{tbl1},
we determined the deprojected galactocentric radius for each pixel in each map.
Pixels were then assigned to a series of radial bins each having a width of 15\arcsec,
where this bin size was limited by the beam size of the Z0MGS SFR maps
(all other maps have smaller beam sizes).
Within each radial bin, we derived mean, median, and 16--84\% percentiles for 
each surface density tracer. 
For molecular gas surface density in particular, we also estimated the error on 
the mean value based on the published moment--0 uncertainty maps (Leroy et al. 2009).
The resulting derived molecular gas surface density profiles have S/N $\geq3$ in most of the bins,
however, in cases with lower S/N the 3$\sigma$ upper limit was provided. 
    
The Z0MGS WISE 1 maps (which traces stellar mass distribution) were also used to 
derive the effective radii, $R_e$, used throughout this work.
For this calculation, all foreground stars in the field of view were masked.
All pixels were put into a series of radial bins, where {\it masked}
pixels with $R_g < 0.4\times R_{25}$ and {\it all} pixels with $R_g > 0.4\times R_{25}$
were substituted for the median unmasked pixel value within the same radial bin.
The resulting maps were then integrated out to $1.5\times R_{25}$ to determine each 
galaxy's integrated flux, and the equi-radius contour encompassing half of this integrated
flux is the effective radius (see Table~\ref{tbl1}).


\begin{figure*} 
  \centering
  \includegraphics[scale=0.29,trim = 0mm 0mm 0mm 0mm, clip]{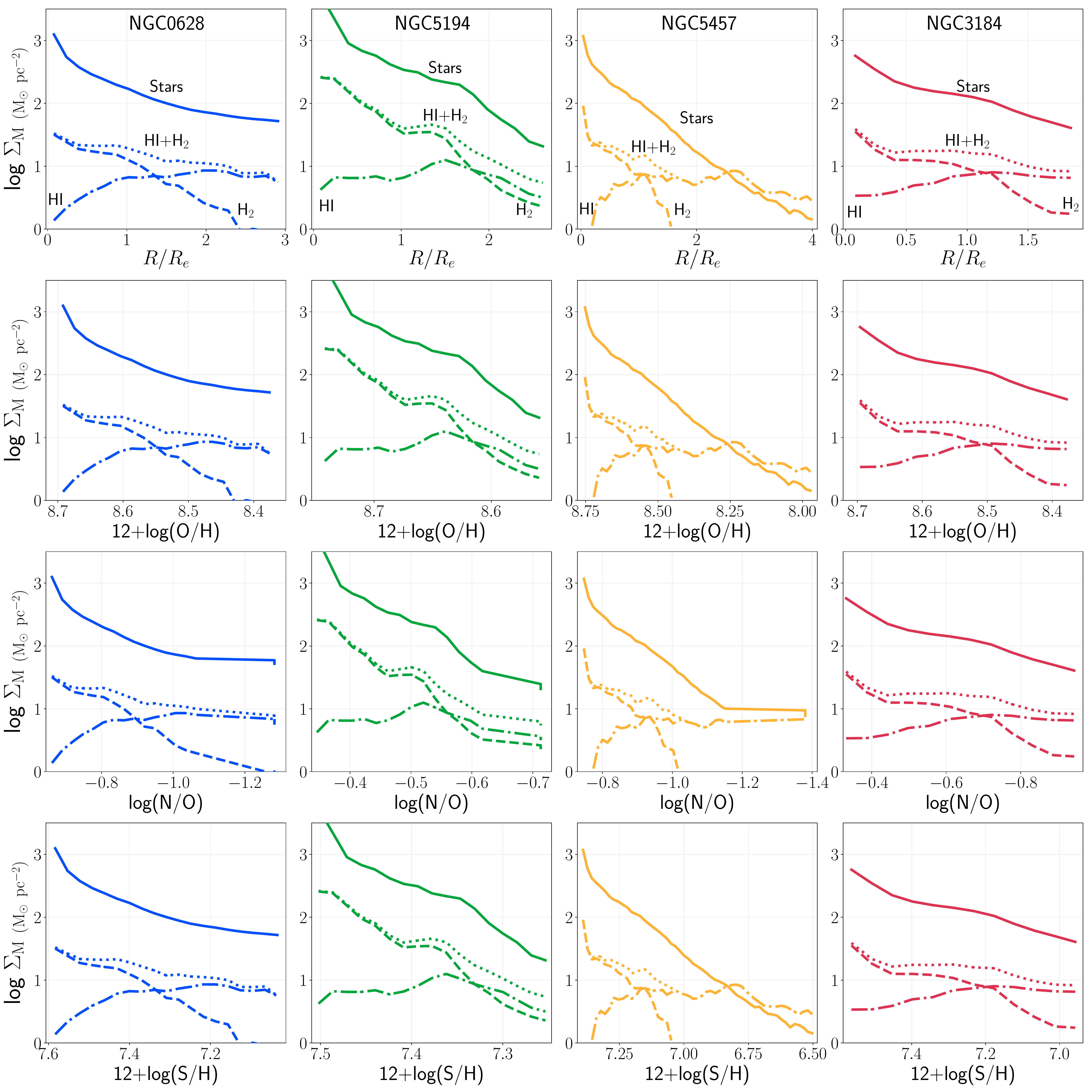} 
  \caption{Mass surface density profiles for different components of the first four CHAOS 
  galaxies versus galactocentric radius (top row), oxygen abundance (middle row), and
  N/O abundance (bottom row).
  Values for 12+log(O/H) and log(N/O) are from the linear fits plotted 
  in Figures~\ref{fig6} and \ref{fig9}.
  Stellar mass profiles are plotted as solid lines and decrease with increasing radius.
  Molecular H$_2$ gas profiles are plotted as dashed lines and also generally decrease outward.
  Atomic \ion{H}{1} gas profiles are plotted as dot-dashed lines, intersecting the
  H$_2$ trends at unique points in each galaxy.
  Total \ion{H}{1}$+$H$_2$ gas is shown by the dotted line.}
  \label{fig16}
\end{figure*}  


\begin{figure} 
  \centering
  \includegraphics[scale=0.25,trim = 0mm 0mm 0mm 0mm, clip]{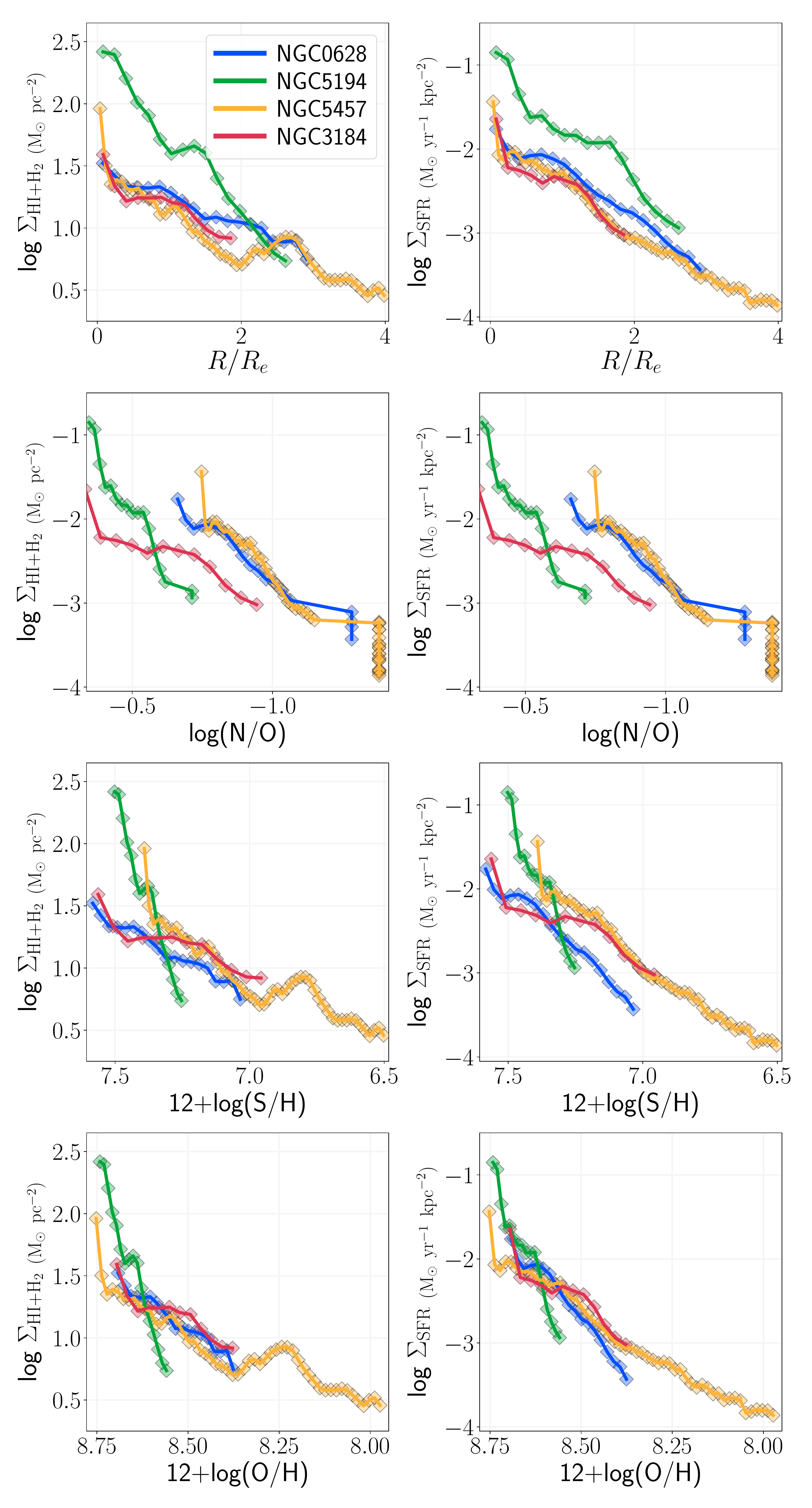}
  \caption{
  {\it Left:} Total gas mass surface density profiles for the four CHAOS galaxies as 
  scaled by disk effective radius ({\it top}), N/O abundance ({\it second}), sulfur 
  abundance ({\it third}), and oxygen abundance ({\it bottom}). 
  These profiles are dominated by the H$_2$ gas for most of the disk.
  {\it Right:} Total star formation rate surface density profiles versus radius and
  abundance trends.
  The right and left columns of panels show the observed increase in local star 
  formation rate with increasing H$_2$ mass surface density, which drives the
  increasing abundance trends. }
  \label{fig17}
\end{figure}


\subsection{Profile Comparisons for CHAOS Galaxies}
The derived mass surface densities for various galaxy components (i.e., stars, \ion{H}{1}, H$_2$)
are plotted in Figure~\ref{fig16} for the CHAOS sample.
As expected for an interacting galaxy, the H$_2$ profile of NGC~5194 is 
different from the other three galaxies as it both dominates the gas profile and 
makes up a larger fraction of the total galaxy mass.
We find that the stellar and gas mass surface density profiles of the other three 
non-interacting galaxies to look similar, with H$_2$ more prominent in the inner
$\sim 1 R_e$ of the disk, \ion{H}{1} dominating the outer disk, and the stellar mass
roughly following the total gas mass ($M_{gas}=M_{\rm HI}+M_{{\rm H}_2}$) for $R_g/R_e < 2$. 

We show the total gas mass surface density profiles, which are dominated by the H$_2$
gas for most of the disk, versus both radius and elemental abundances (O, N, and S )
in the left column of Figure~\ref{fig17}.
Interestingly, while we find the stellar mass and gas mass surface 
density profiles of individual galaxies to be offset from one another when plotted 
versus their N/O profiles, the shift is minimal for the O/H and S/H trends. 
Since the decline of H$_2$ gas mass with radius corresponds to a decreasing star 
formation rate (as shown in the right column of Figure~\ref{fig17}) and star 
formation efficiency, this could indicate that the H$_2$ mass surface density plays 
the leading role in the stellar and subsequent chemical evolution of these galaxies. \looseness=-2

\subsection{Local Scaling Relations}

\citet{rosales-ortega12}, using IFU spectroscopy from the PINGS \citep{rosales-ortega10} 
and CALIFA surveys, reported the first {\it local} mass-metallicity (M--Z) scaling 
relationship of \ion{H}{2} regions in spiral galaxies, with a secondary dependence 
on the equivalent width of H$\alpha$ (a proxy from SFR).
This local M--Z--EW(H$\alpha$) relationship is the logical product of 
inside-out disk growth and the dependence of SFR on mass.
While the more widely known fundamental M--Z--SFR {\it global} relationship \citep{mannucci10}
has been explained by galaxy growth via the accretion of cold gas that is altered 
by feedback of gaseous inflows and outflows,
the {\it local} M--Z--SFR relationship allows us to explore physical parameters 
that may be regulating the growth and chemical evolution {\it within} spiral disks.

\begin{figure*}
  \centering
  \includegraphics[scale=0.225,trim = 0mm 0mm 0mm 0mm, clip]{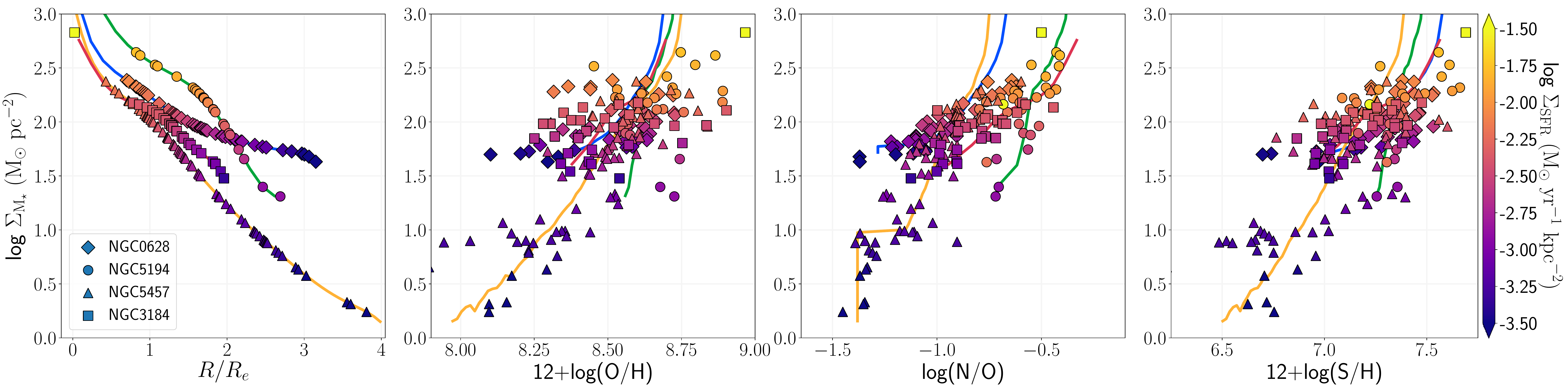} 
  \caption{
   {\it Left}: stellar mass surface densities for the CHAOS sample versus galactocentric radius
  where the colored lines represent the average profile of each galaxy.
  {\it Second:} the {\it local} M--Z--SFR relationship. 
  For each {\it observed} data point $x$, we plot the fit
  log $\Sigma {\rm M}_{\star} (R_x)$ versus the measured 12+log(O/H)$_x$, 
  color-coded by the its log $\Sigma {\rm SFR} (R_x)$ value.
  The large spread is due to the real scatter in observed O/H.
  {\it Third:} the {\it local} M--N/O--SFR relationship. 
  This panel shows the stellar mass surface density trend, similar to the second panel, but versus measured N/O,
  forming a tight correlation.
  {\it Right:} the {\it local} M--S/H--SFR relationship. }
   \label{fig18}
\end{figure*}

The stellar mass surface density ($\Sigma_{{\rm M}_\odot}$) radial profiles are reproduced 
for the four CHAOS galaxies in the first panel of Figure~\ref{fig18}.
We fit a polynomial to the $\Sigma_{{\rm M}_\odot}$--$R_e$ (Figure~\ref{fig18})
and $\Sigma_{\rm SFR}$--$R_e$ trend of each galaxy.
These fits are then used to plot the stellar mass surface densities corresponding to the
{\it observed} O/H and N/O abundances of the CHAOS \ion{H}{2} regions.
The $\Sigma_{{\rm M}_\odot}$--O/H and $\Sigma_{{\rm M}_\odot}$--N/O trends are 
plotted in the middle and right panels of Figure~\ref{fig18}, respectively, and
color-coded by SFR surface density ($\Sigma_{\rm SFR}$).
Since SFR is known to depend in stellar mass, the vertical color gradient seen
for the SFR in the $\Sigma_{{\rm M}_\odot}$--O/H is expected.
However, in the $\Sigma_{{\rm M}_\odot}$--N/O relationship, not only is the scatter
significantly reduced relative to the O/H trend, but SFR also appears to increase 
{\it along} the N/O gradient.

The metallicity-surface density relationship may reflect fundamental similarities in 
the evolution of non-barred, non-interacting spiral galaxies.
For example, \cite{ryder95} argues for a galaxy evolution model that includes self-regulating
star formation, where energy injected into the ISM by newly-formed stars inhibits 
further star formation.
These models were able to successfully reproduce the observed correlations between surface 
brightness and SFR \citep{dopita94} and surface mass density \citep[e.g.,][]{phillips91,ryder95,garnett97}.
The current work supports the idea that stellar mass surface density is a fundamental parameter 
governing spiral galaxy evolution, and is particularly important for the relative timescales
involved in N/O production.


\clearpage


\clearpage

\bibliography{mybib}{}

\clearpage


\end{document}